\documentclass[sigconf]{acmart}
\AtBeginDocument{%
  }
\renewcommand\footnotetextcopyrightpermission[1]{}

\usepackage{caption}
\usepackage{subcaption}
\usepackage{tcolorbox}
\usepackage{graphicx}
\usepackage{multirow}
\usepackage{tikz}
\usetikzlibrary{positioning,decorations.pathmorphing,arrows,shapes}
\usetikzlibrary{fit,calc}
\usepackage{bm}
\usepackage{xcolor}
\usepackage{colortbl}
\usepackage{comment}
\usepackage{booktabs}
\usepackage{setspace}

\usepackage[ruled,linesnumbered, vlined]{algorithm2e}
\makeatletter
\renewcommand{\tcp}[1]{%
  \unskip\hfill\textcolor{brown}{$\triangleright$~#1}\par%
}
\renewcommand{\tcc}[1]{%
  \unskip\textcolor{brown}{$\triangleright$~#1}\par%
}
\makeatother

\usepackage{enumitem}
\usepackage{multicol}
\usepackage{parcolumns}
\usepackage{balance}
\usepackage{makecell}

\usepackage{xspace}

\newcommand{\name}{\textsc{ProRepair}\xspace}

\usepackage{pifont}
\usepackage{amsmath,amsfonts}
\usepackage{amssymb}

\newtheorem{definition}{Definition}[section]
\newtheorem{theorem}{Theorem}

\newcommand*{\tikzmk}[1]{\tikz[remember picture,overlay,] \node (#1) {};\ignorespaces}
\newcommand{\boxit}[2]{\tikz[remember picture,overlay]{\node[yshift=3pt,xshift=3pt,fill=#1,opacity=.15,fit={(A)($(B)+(#2\linewidth,.8\baselineskip)$)}] {};}\ignorespaces}
\newcommand{\HightlightDef}[4]{
    \tikzmk{A}
    \Fn{#3}{\tikzmk{B} \boxit{#1}{#2}
        #4
    }
}
\newcommand{\DefHelperFunc}[2]{
    \HightlightDef{gray}{.865}{#1}{#2}
}

\usepackage{hyperref}
\hypersetup{
colorlinks=true,
citecolor=black,
linkcolor=blue,
urlcolor=blue
}

\copyrightyear{2025}
\acmYear{2025}
\setcopyright{acmlicensed}
\acmConference[CCS '25] {Proceedings of the 2025 ACM SIGSAC Conference on Computer and Communications Security}{ October 13--17, 2025}{Taipei, Taiwan.}
\acmBooktitle{Proceedings of the 2025 ACM SIGSAC Conference on Computer and Communications Security (CCS '25), October 13--17, 2025, Taipei, Taiwan}
\acmISBN{979-8-4007-1525-9/2025/10}
\acmDOI{10.1145/3719027.3765057}

\begin{document}


\title{Provable Repair of Deep Neural Network Defects by Preimage Synthesis and Property Refinement}

\author{Jianan Ma}
\authornote{Part of Jianan Ma's work was done when visiting Zhejiang University.}
\affiliation{%
  \institution{Hangzhou Dianzi University}
  \city{Hangzhou}
  \country{China}
}
\affiliation{%
  \institution{Zhejiang University}
  \city{Hangzhou}
  \country{China}
}
\email{majianannn@gmail.com}

\author{Jingyi Wang}
\authornote{Corresponding author: Jingyi Wang.}
\affiliation{%
  \institution{Zhejiang University}
  \city{Hangzhou}
  \country{China}
}
\email{wangjyee@zju.edu.cn}

\author{Qi Xuan}
\affiliation{%
  \institution{Institute of Cyberspace Security, \\ Zhejiang University of Technology}
  \city{Hangzhou}
  \country{China}
}
\affiliation{%
  \institution{Binjiang Institute of AI, ZJUT}
  \city{Hangzhou}
  \country{China}
}
\email{xuanqi@zjut.edu.cn}

\author{Zhen Wang}
\affiliation{%
  \institution{School of Cyberspace, \\ Hangzhou Dianzi University}
  \city{Hangzhou}
  \country{China}
}
\email{wangzhen@hdu.edu.cn}

\renewcommand{\shortauthors}{Jianan Ma, Jingyi Wang, Qi Xuan, and Zhen Wang}

\begin{abstract}
It is known that deep neural networks may exhibit dangerous behaviors under various security threats (e.g., backdoor attacks, adversarial attacks and safety property violation)
and there exists an ongoing arms race between attackers and defenders. 
In this work, we propose a complementary perspective to utilize recent progress on ``neural network repair'' to mitigate these security threats and repair various kinds of neural network defects (arising from different security threats) within a unified framework, offering a potential silver bullet solution to real-world scenarios.
To substantially push the boundary of existing repair techniques (suffering from limitations such as lack of guarantees, limited scalability, considerable overhead, etc) in addressing more practical contexts, we propose \name, a novel \emph{provable} neural network repair framework driven by formal preimage synthesis and property refinement.
The key intuitions are: (i) synthesizing a precise proxy box to characterize the feature space preimage, which can derive a bounded distance term sufficient to guide the subsequent repair step towards the correct outputs, and (ii) performing property refinement to enable surgical corrections and scale to more complex tasks.
We evaluate \name across four security threats repair tasks on six benchmarks and the results demonstrate it outperforms existing methods in effectiveness, efficiency and scalability.
For point-wise repair, \name corrects models while preserving performance and achieving significantly improved generalization, with a speed-up of 5$\times$ to 2000$\times$ over existing provable approaches.
In region-wise repair, \name successfully repairs all 36 safety property violation instances (compared to 8 by the best existing method), and can handle 18$\times$ higher dimensional spaces.
\end{abstract}




\maketitle

\begingroup
\renewcommand\thefootnote{} 
\footnotetext{This is the full version of the paper accepted by CCS 2025.}
\endgroup

\section{Introduction}
\label{sec:intro}

Numerous studies have demonstrated that deep neural networks (DNNs) are inherently vulnerable to a wide range of security threats, including adversarial attacks~\cite{goodfellow2014explaining, wan2023bounceattack}, natural corruption~\cite{hendrycks2019augmix, wang2021augmax}, backdoor attacks~\cite{liu2018trojaning, chen2017targeted, barni2019new}, and violations of safety properties~\cite{katz2017reluplex, marabou}, all of which may lead to dangerous and unpredictable results in safety-critical applications.
The evolving nature of security threats has triggered an ongoing arms race between attackers and defenders. 
Attackers continuously enhance their strategies to bypass existing defenses, while defenders respond by developing new techniques to address emerging risks, such as adversarial training~\cite{gehr2018ai2, tramer2019adversarial, levi2024splitting} and defense~\cite{xiang2021patchguard, zhu2023ai, zhang2024text, cullen2024s}, backdoor detection~\cite{wang2019neural, xu2021detecting, mo2024robust} and removal~\cite{zhu2023selective, gong2023redeem, wang2024mm}.
These defenses are typically tailored to specific threats and applied before model deployment, making them inadequate to address unforeseen failures that arise after the model is deployed. 
As a consequence, DNNs deployed in real-world environments remain vulnerable to unexpected failures.
This raises a crucial question: \textit{how to utilize the failures arising from
diverse security threats after the model is deployed to further enhance its safety and reliability?}

Recent advances in \textbf{neural network (NN) repair} offer a promising silver bullet solution to this critical problem. 
Unlike existing defenses, NN repair aims to handle post-deployment defects that may arise from various security threats through iterative error-driven updates.
This enables a unified and adaptive framework that is particularly suitable for real-world scenarios.
Building on its potential, a number of NN repair methods have been proposed, progressively improving effectiveness, efficiency and applicability across diverse security threats.
Formally, given a property $\phi$ that defines the desired behavior as a tuple of input set $\phi^{in}$ and output space $\phi^{out}$, NN repair aims to correct a buggy model $f$ to $\tilde{f}$ such that $\forall \bm{x} \in \phi^{in}, \; \tilde{f}(\bm{x} )\in \phi^{out}$.
For example, the property $\phi$ may encode specifications like robustness to bounded perturbations, where $\phi^{in}=\{\bm{x} \mid \|\bm{x} - \bm{x}_0\| \le \delta\}$ defines a local space and $\phi^{out}=\{\bm{y} \mid \bm{y}_{true} \ge \bm{y}_{other}\}$ ensures correct classification.

Existing solutions to this problem fall into two categories:
heuristic repair based on fault localization and provable repair with theoretical guarantees. 
The first category focuses on identifying the key units in the model that are responsible for errors, which typically leverage gradients~\cite{usman2021nn}, activation values~\cite{sohn2023arachne}, or causal models~\cite{sun2022causality} to pinpoint faulty neurons or parameters.
However, they lack theoretical guarantees (the repaired NN may still violate $\phi$) and cannot handle region-wise properties (where $\phi^{in}$ is an infinite set).
To address these issues, recent \textit{provable repair} techniques~\cite{sotoudeh2021provable,reassure,tao2023architecture} leverage the piecewise-linear nature of ReLU NNs: if the activation status (active or inactive for ReLU) of all neurons are fixed, the entire model can be expressed as a linear function of the inputs or parameters, 
allowing the repair to be converted into a linear programming (LP), which can be solved by constraint solvers.

\textbf{Research gap} – While prior provable works have been proven effective in certain cases, they still face several key limitations when applied to more practical contexts:
1) \textit{Limited scalability}. 
Existing methods enforce fixed activation status throughout the input space $\phi^{in}$. 
This rigid requirement severely limits their capability to find feasible solutions, especially when there are multiple properties to repair or for high-dimensional spaces. 
Additionally, the dependence on activation status restricts their applicability to NNs with piecewise-linear activation functions.
2) \textit{High computational cost}.
To ensure the model behaves linearly in $\phi^{in}$, they must enumerate either all activation status or all vertices of the space, which results in exponential complexity with respect to its dimensionality.
This produces extremely large LP requiring massive computational resources.
For example, ~\cite{tao2023architecture} (SOTA) builds an LP with over a billion coefficients when repairing in a 16D space, which takes over 10,000 seconds and 200 GB of memory to solve.
3) \textit{Unstable Fidelity preservation}.
Since modifying earlier layers would cause uncontrolled activation status in all subsequent layers, they are limited to repairing only the final few layers. 
This restriction compels the LP solution to make excessive modifications on these layers, which may significantly degrade the model's original performance.

\textbf{Insight} - Instead of modifying the final few layers (for decision making), we propose an inverse perspective that focuses on repairing the early layers (for feature extraction).
The intuition is that overwhelming modifications to later layers cause dramatic shifts on the feature-space preimage (the set of features that produce desired outputs when fed to subsequent layers), thereby compromising the model performance. 
This motivates us to characterize the feature space preimage, which can guide the correction of the early layers while leaving the preimage itself unchanged.
To further address the repair cost and scalability issues, we propose an adaptive space refinement approach which employs NN verification techniques to eliminate activation status dependency while providing formal repair guarantees. 
By adaptively refining the input space, the approximation errors induced by verification are effectively reduced, with each subspace receiving more surgical corrections that collectively enable the repair of more complex tasks.

\begin{figure*}[!t]
\centering
\includegraphics[scale=0.5]{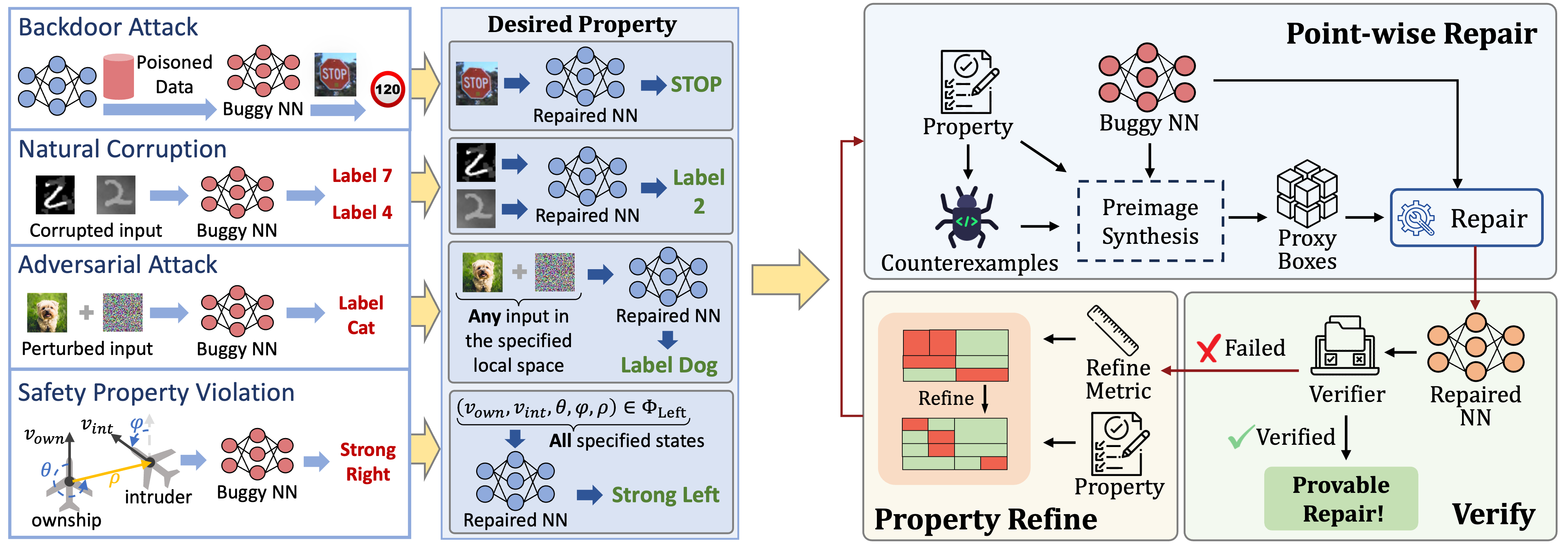}
\caption{The overview of \name framework.}
\label{fig:overview}
\end{figure*}

\textbf{Solution} - Based on the above insights, we design and implement \name, a novel provable NN repair framework to mitigate diverse kinds of security threats.
As shown in Fig.~\ref{fig:overview}, \name first generates counterexamples for the given desired properties, which are then used to synthesize preimages.
Since the numerous nonlinear units in the NN make exact synthesis impractical, \name proposes to employ linear relaxation with iterative center shift to generate a small proxy box to precisely and efficiently characterize the preimage.
The geometric nature of proxy boxes further enables us to derive a bounded distance measure carefully crafted for repair. 
After one repair iteration, \name selects a critical dimension to split for each space that fails verification, allowing space refinement for surgical repair.

\begin{table}[!t]
\centering
\footnotesize
\scriptsize
\tiny
\fontsize{7pt}{9pt}\selectfont
\caption{A comparison of different NN repair methods. \(\blacksquare\) and \(\square\) denote the method supports the attribute or not. \(\vartriangle\): REASSURE only supports ReLU. \(\blacktriangle\): PRDNN theoretically supports any activation function for point-wise repair and piece-wise linear functions for region-wise repair, but the implementation only supports ReLU. \(\lozenge\): APRNN supports activation functions that have linear pieces such as ReLU. -, +, and ++ denotes low, medium, and high efficiency, respectively.}
\setlength{\tabcolsep}{2.2pt}
\label{tab:comparison}
\begin{tabular}{lcccccc}
\hline
Method & Provable & \makecell{No Extra \\ Data} & \makecell{Architecture \\ Preserving} & \makecell{Region \\ Repair} & \makecell{Complex 
\\ Activation \\Functions} & Efficiency \\
\hline
NNREPAIR~\cite{usman2021nn} &  \(\square\) &  \(\square\)  &  \(\blacksquare\)  & \(\square\)  &  \(\square\) & - \\
Arachne~\cite{sohn2023arachne} & \(\square\) &  \(\square\)  &  \(\blacksquare\)  & \(\square\)  &  \(\blacksquare\) & +\\
CARE~\cite{sun2022causality}     &  \(\square\) &  \(\square\) &  \(\blacksquare\) &  \(\square\) &  \(\blacksquare\) & +\\
I-REPAIR~\cite{henriksen2022repairing} &  \(\square\) &  \(\square\) & \(\blacksquare\)  & \(\square\) & \(\blacksquare\) & ++\\
VeRe~\cite{ma2024vere}     &  \(\square\) &  \(\square\) &  \(\blacksquare\) &  \(\square\) &  \(\blacksquare\) & +\\
AI-Lancet~\cite{zhao2021ai} &  \(\square\) & \(\blacksquare\)  & \(\blacksquare\) & \(\square\) & \(\blacksquare\) & +\\
REASSURE~\cite{reassure} &  \(\blacksquare\) & \(\blacksquare\) & \(\square\) & \(\vartriangle\) & \(\square\) & -\\
PRDNN~\cite{sotoudeh2021provable}   & \(\blacksquare\) & \(\blacksquare\) & \(\square\) & \(\blacktriangle\) & \(\blacktriangle\) & -\\
APRNN~\cite{tao2023architecture}  & \(\blacksquare\) &\(\blacksquare\) & \(\blacksquare\) & \(\lozenge\) &\(\lozenge\) & -\\
This work  &  \(\blacksquare\) & \(\blacksquare\) & \(\blacksquare\) & \(\blacksquare\) & \(\blacksquare\) & ++ \\
\hline
\end{tabular}
\end{table}

We have implemented \name as a self-contained toolkit and evaluated it across four security threats repair tasks using various benchmarks and models.
The results demonstrate its effectiveness and efficiency, with significant improvements in region-wise repair capability.
For point-wise repair, \name preserves model fidelity while achieving remarkable generalization, showcasing a speed improvement of 5$\times$ to 2000$\times$ over existing provable methods. 
For region-wise repair, it successfully repairs all 36 safety property violation instances (compared to only 8 repairs by the state-of-the-art method) and can handle 18$\times$ higher-dimensional spaces.
A comprehensive comparison with existing approaches is presented in Table~\ref{tab:comparison}.
\noindent In summary, our main contributions are:
\begin{itemize}[left=0pt]
    \item We devise a new repair perspective that focuses on the feature extractor, enabled by a novel proxy box synthesis method to capture a meaningful subset of the preimage.
    \item We propose \name, a novel provable NN repair framework that integrates a bounded distance measure with adaptive property refinement to correct NNs with provable guarantees in more practical contexts.
    \item Extensive experiments on 4 repair tasks and 76 models across 6 benchmark datasets demonstrate that our approach significantly outperforms existing provable repair methods in terms of effectiveness, efficiency and scalability.
    \item We release \name at this \href{https://github.com/nninjn/ProRepair}{\textbf{repository}} along with all the codes, datasets and models to facilitate future studies in this area.
\end{itemize}

\section{Preliminaries}

\subsection{Deep neural networks} 
\label{sec:define-property}
A DNN can be defined as a function $f:\mathbb R^m \to \mathbb R^n$ maps a high-dimensional input $\bm{x} \in \mathbb R^m$ to an output $f(\bm{x}) \in \mathbb R^n$, where $n$ is the total number of classes.
A typical DNN includes multiple hidden layers $f_{\mathrm{e}}$ to extract feature representations, followed by fewer layers $f_{\mathrm{c}}$ to classify based on the extracted features, represented as $f=f_{\mathrm{c}} \circ f_{\mathrm{e}}$.
Finally, it chooses the dimension with the highest score as the classification result, i.e., $\arg \max_{1 \le i \le n } f(\bm{x})_i$. 
We further define the specification of the desirable property that DNNs should satisfy.

\begin{definition}[\textbf{Desired Property Specification}]
For a neural network $f:\mathbb{R}^m\rightarrow\mathbb{R}^n$, a desired property $\phi$ is a triple $(\phi^{in}, \phi^{cons}, \phi^{out})$, where $\phi^{in}\subseteq\mathbb{R}^m$ defines the input space of interest.
The output specification is governed by the constraint set $\phi^{cons}=\{\psi^1, \dots, \psi^q\}$ where each linear constraint $\psi$ enforces $\bm{c}_{\psi}^\top f(\cdot)+d_{\psi}\geq0$ with $\bm{c}_{\psi}\in\mathbb{R}^n$ and $d_{\psi}\in\mathbb{R}$. 
This induces the valid output space $\phi^{out}=\{\bm{y}\in\mathbb{R}^n\mid\bigwedge_{\psi \in \phi^{cons}} (\bm{c}_{\psi}^\top\bm{y}+d_{\psi}\geq0)\}$.
\end{definition}

In this work, we consider $\phi^{in}$ to be a bounded box $\{\bm{x}\in\mathbb{R}^m\mid\forall i\in[1,m]: \bm{l}^\phi_i\leq\bm{x}_i\leq\bm{u}^\phi_i\}$ defined by lower and upper bounds $\bm{l}^\phi,\bm{u}^\phi\in\mathbb{R}^m$, which is commonly used in NN repair~\cite{sotoudeh2021provable, reassure, tao2023architecture}.
Note that this includes the special case of a single input point when $\bm{l}^\phi=\bm{u}^\phi$. 
We write $f\models\phi$ when $f(\bm{x})\in\phi^{out}$ holds for all $\bm{x}\in\phi^{in}$, and extend this notation to the property set $\mathcal{P}$ with $f\models\mathcal{P}$ meaning $\forall\phi\in\mathcal{P}:f\models\phi$. A repair is classified as \emph{region-wise} if $\exists\phi\in\mathcal{P}$ where $\phi^{in}$ contains infinitely many points, otherwise \emph{point-wise}.


\subsection{Neural network verification}
Neural network verification aims to rigorously prove or disprove whether a given network satisfies certain properties, such as correctness, robustness and fairness. 
Recently, numerous verification frameworks have been proposed, which can be broadly categorized into \textit{complete} and \textit{incomplete} verification. Complete verification typically based on constraint solving~\cite{narodytska2018verifying,marabou,ehlers2017formal,huang2017safety}, 
which can give a definite `yes/no' result in sufficient time.
However, verifying NNs with both soundness and completeness is a NP-hard problem~\cite{katz2017reluplex}, making it challenging to terminate in a reasonable time.
As a result, such solver-based verification can only scale to DNNs with hundreds of neurons. 
In practice, incomplete verification strikes a balance between precision and efficiency, typically based on linear relaxation~\cite{julian2019verifying, xu2020automatic,fastlin,batten2021efficient, xu2020fast} and abstract interpretation~\cite{gehr2018ai2, singh2019abstract, deepsrgrextended, singh2018fast}. 
This category of approaches can scale up to medium size models, which contain around \(10^5\) neurons.

Here we use auto-LiRPA~\cite{xu2020automatic} (one of the most widely used verifiers) as an example to briefly illustrate how linear relaxation works.
The key idea is to over-approximate the outputs of non-linear activation units with two linear functions as lower/upper bound.
For example, an unstable ReLU neuron $z'=\mathrm{ReLU}(z)$ with its input range $[l,u]$ ($l<0<u$) can be relaxed as:
\(k \cdot z \le z' \le \frac{u(z-l)}{u-l}, \;  k \in [0, 1]\).
Then the numerical lower bound of $z'$ can be calculated by substituting $z$ in the term $k \cdot z$ with its affine bounds.
This process is iteratively applied, with each newly introduced variable is replaced by its affine bounds, until all variables within the expression are reduced to input variables.
In this way, the output of $f$ over a given space $\mathcal{B} \subset \mathbb R^m$  can be linearly relaxed as: $\forall \bm{x} \in \mathcal{B}, \, \underline{\bm{w}} \cdot \bm{x}   +\underline{\bm{b}} \leq  f(\bm{x}) \leq \overline{\bm{w}}   \cdot \bm{x}+\overline{\bm{b}}$.
This recursive process of propagating affine bounds layer by layer is commonly referred to as \emph{bound propagation}, and forms the basis of many incomplete verifiers such as auto-LiRPA and Deeppoly~\cite{singh2019abstract}.

\subsection{Preimage}
Given an input space \( \mathcal{I} \) and an output space \( \mathcal{O} \), the preimage is the set of all inputs $\bm{x} \in \mathcal{I}$ that are mapped to an element in $\mathcal{O} $ by a function \(f\).
Formally, we define the preimage for NNs as:
\begin{definition}[\textbf{Neural network preimage}]
\label{def:preimage}
Given a DNN $f: \mathbb{R}^{m} \rightarrow \mathbb{R}^{n}$, an input space \( \mathcal{I}\subset \mathbb{R}^{m}  \) and an output space \( \mathcal{O} \subset \mathbb{R}^{n}\), the preimage is defined as: $f^{-1}(\mathcal{I}, \mathcal{O}) = \left\{ \bm{x} \in \mathcal{I}\mid 
f\left(\bm{x}\right) \in \mathcal{O}
\right\} \subset \mathbb{R}^{m}$.

\end{definition}
\noindent Generating the exact preimage for DNNs~\cite{matoba2020exact} is intractable due to the presence of numerous nonlinear units.
Recently, some works have focused on under- and over-approximating the preimage. For instance,~\cite{kotha2023provably} utilizes
lagrangian dual optimization for preimage over-approximations while~\cite{zhang2024provable} presents an anytime method to produce under-approximation of preimage.
Overall, they yield a convex approximation of the preimage, i.e., 
\(\mathcal{U} \subseteq f^{-1}(\mathcal{I}, \mathcal{O}) \subseteq \mathcal{T}\), where \(\mathcal{U}\) and \(\mathcal{T}\) are convex sets.

\section{Threat Model} 
We consider a post-deployment scenario where the defender has deployed a model which exhibits failures in operation, potentially triggered by the following security threats (illustrated in Fig.~\ref{fig:overview}, left).

\noindent \textbf{Threat 1: Adversarial Perturbation.} 
For predetermined inputs (e.g., specific traffic signs), adversaries can craft imperceptible perturbations under $\ell_\infty$-norm bounds to force misclassification, which are particularly hazardous when targeting safety-sensitive inputs (e.g., perturbing stop signs to be misclassified as yield signs)~\cite{eykholt2018robust}.

\noindent \textbf{Threat 2: Natural Corruption.}
Deployed models frequently encounter reliability degradation due to real-world environmental disturbances that deviate from training conditions~\cite{xie2020adversarial, guo2022improving}. 
These disturbances manifest as input corruptions like weather effects (fog, rain) and sensor noise (motion blur, brightness variations), revealing the fragility of DNNs trained on idealized datasets~\cite{hendrycks2019benchmarking}.

\noindent \textbf{Threat 3: Backdoor Attack.}
Due to the growing popularity of online machine learning platforms~\cite{mo2024robust}, backdoor attacks have become a significant concern.
In line with the classic threat model from existing backdoor learning studies~\cite{xu2021detecting, zhu2023selective, gong2023redeem}, we assume the defender deploys a third-party model implanted with malicious triggers of arbitrary shape/size.

\noindent \textbf{Threat 4: Safety Property Violation.}
In this case, we consider security-critical systems that rely on DNNs for decision making, such as ACAS Xu~\cite{marston2015acas}, an airborne collision avoidance system for unmanned aircraft. In these systems, violations of established safety properties~\cite{marabou, chi2025patch}, like failure to prevent a collision, pose significant security risks. We follow the general threat model from the neural network verification domain, assuming the defender deploys a model that may potentially violate safety properties.

\noindent \textbf{Defender's Capabilities.}
We assume the defender has white-box access to the target model (architecture and parameters) and can derive desired properties from multiple sources. 
For runtime-observed failures (e.g., a backdoor-triggered stop sign misclassification), the defender can extract the concrete counterexample $\phi^{in} = \{\bm{x}_0\}$ and formulate constraints $\phi^{cons} = \{\psi^j \mid \psi^j := f(\bm{x})_{y_0} - f(\bm{x})_j \geq 0, \forall j\}$, where $y_0$ is the class ``STOP''.  
Another common source is developer-specified requirements in safety-critical systems, where the defender can encode domain knowledge as formal properties.
With ACAS Xu as a canonical example, the defender may require the model to output "Clear-of-Conflict" (COC) decisions when the intruder's relative distance exceeds safety threshold $d_{min}$ and its velocity is below danger threshold $v_{max}$, formally expressed as
 $\phi^{in} = \{\bm{x} \mid \bm{x}_{distance} \geq d_{min} \land \bm{x}_{velocity} \leq v_{max} \}$ and $\phi^{cons} = \{\psi^j \mid \psi^j := f(\bm{x})_{\mathrm{COC}} - f(\bm{x})_j \geq 0, \forall j\}$.
These formulations accommodate both reactive repairs and proactive safety enforcement without requiring additional training data, aligning with established repair frameworks~\cite{tao2023architecture,reassure,sotoudeh2021provable}.

\noindent \textbf{Goals.} 
Given a model $f$ and a set of desired properties $\mathcal{P}$ where $f \not\models \mathcal{P}$, the defender aims to repair it to $\tilde{f}$ such that $\tilde{f} \models \mathcal{P}$. 
We also aim for the repair to be as efficient as possible while preserving the model's original performance.

\section{Related Work}
\label{sec:motivation}
Recent NN repair methods can be categorized into non-provable and provable repair depending on whether they guarantee that the repaired network satisfies the desired properties.
Tab.~\ref{tab:comparison} summarizes the comparison of our method and existing repair frameworks.

\textbf{Non-provable repair.} Most of these methods start with fault localization, aiming to identify neurons~\cite{sun2022causality, chen2024isolation} or parameters~\cite{usman2021nn, henriksen2022repairing, sohn2023arachne, chen2024interpretability} that are strongly correlated with the bugs. This is typically followed by a repair step, where heuristic algorithms modify the identified neurons or parameters. 
As they are generally statistics-based, their effectiveness heavily depends on the availability of large amounts of data. 
More recently, \cite{ma2024vere} introduced a verification-guided synthesis framework, which improves localization precision when the available data is limited. 
However, it still requires additional clean data for effective repair. 
Although these methods generally demonstrate high efficiency and scalability, they are unable to provide theoretical guarantees and handle region-wise property.

\textbf{Provable repair.}
Several approaches use formal methods such as constraint solving to ensure that the repaired NN rigorously satisfies the desired properties, providing guarantees for repair.
Notably, PRDNN~\cite{sotoudeh2021provable}, REASSURE~\cite{reassure}
and APRNN~\cite{tao2023architecture} are recently proposed representative methods. 
The core challenge faced by these methods is how to transform the repair to linear programming, since directly modifying the parameters in hidden layers introduces nonlinear effects on the model's outputs that existing solvers struggle to handle.
Their key idea involves first stabilizing the model's activation status on the given input space so that the model's output can be formalized as a linear transformation of \(\Delta \theta\), where \(\Delta \theta\) denotes parameter modifications in few layers of \(f_{\mathrm{c}}\) (PRDNN and APRNN) or the parameters of the patch network (REASSURE). 
Notably, they adjust the parameters of \(f_{\mathrm{c}}\) because this does not affect the neuron activation status in \(f_{\mathrm{e}}\), while modifying \(f_{\mathrm{e}}\) would simultaneously alter the activation status of neurons in \(f_{\mathrm{c}}\).
Finally, they invoke a solver to calculate $\Delta \theta$ to correct the model.

\begin{figure}[t]
    \centering
    \begin{subfigure}[b]{0.335\linewidth}
        \centering
        \captionsetup{skip=3pt}
        \includegraphics[width=\linewidth]{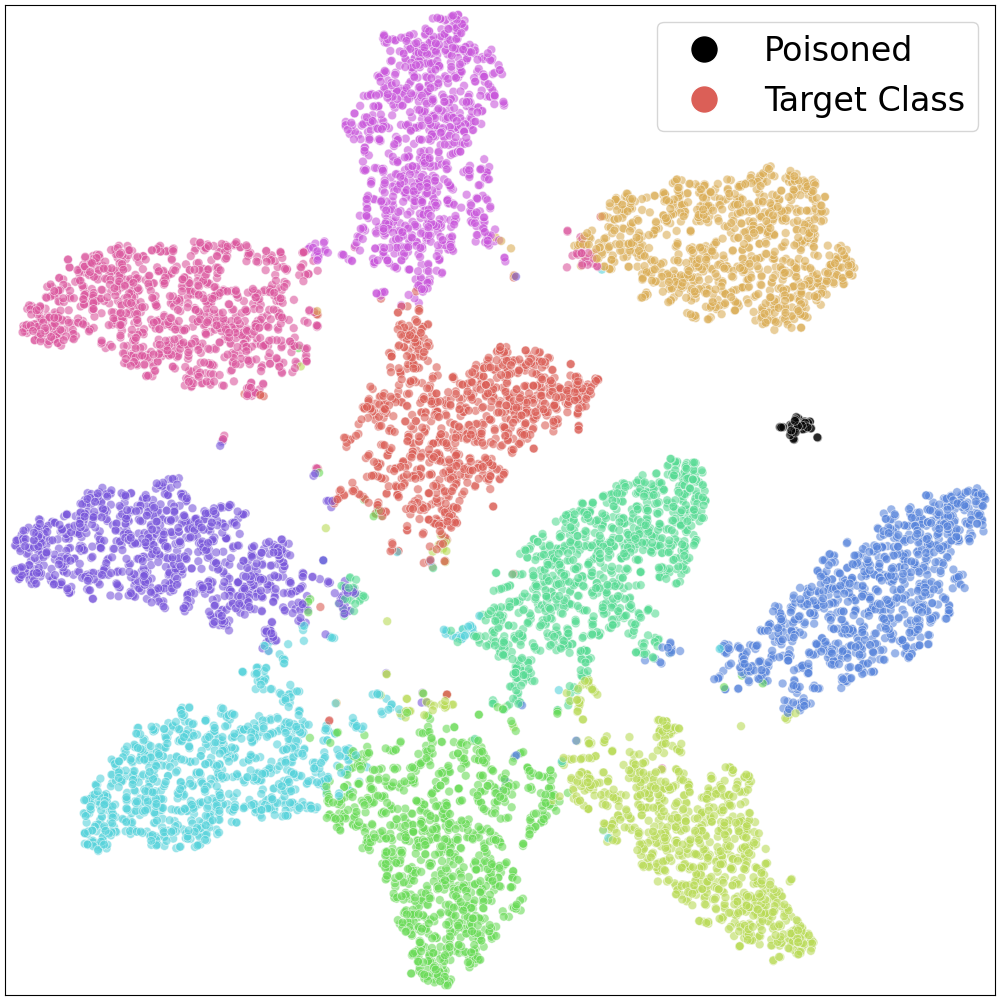} 
        \caption{Original}
    \end{subfigure}\hspace{-0.5em}
    \begin{subfigure}[b]{0.335\linewidth}
        \centering
        \captionsetup{skip=3pt}
        \includegraphics[width=\linewidth]{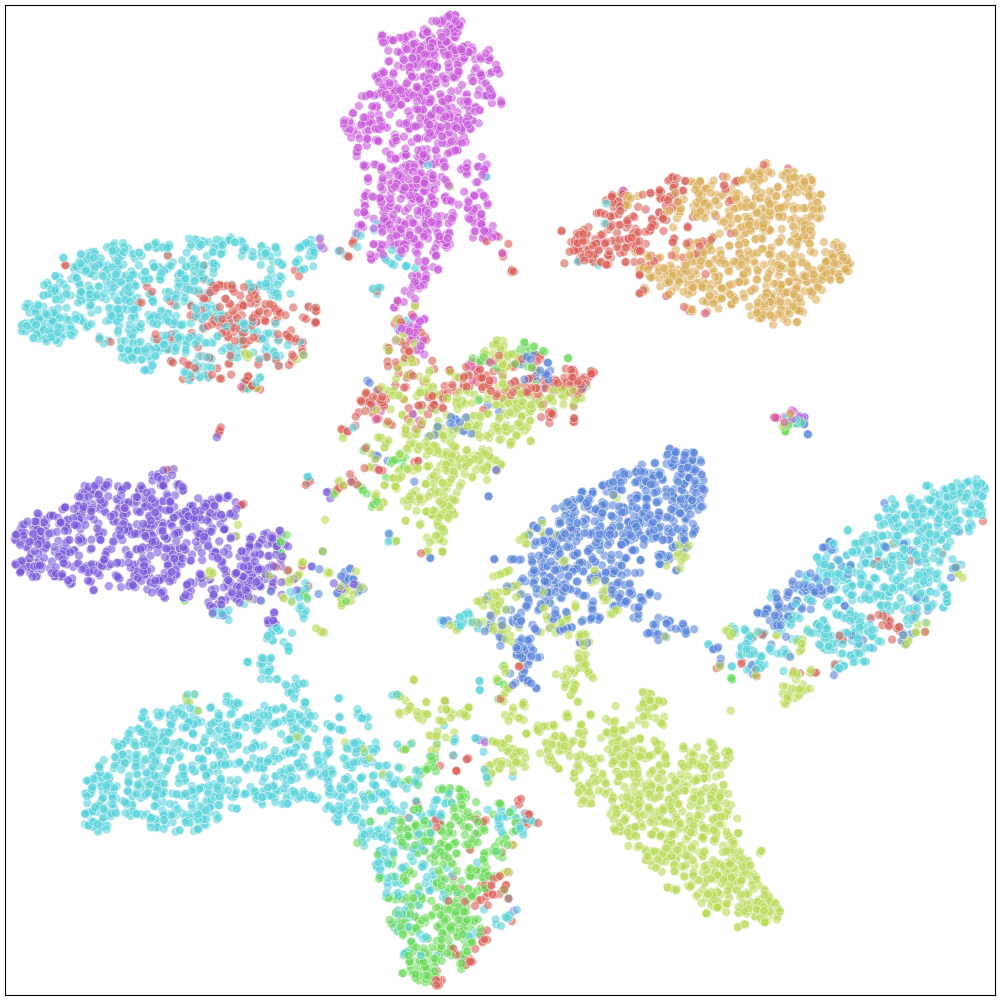} 
        \caption{APRNN}
        \label{fig:tsne-aprnn}
    \end{subfigure}\hspace{-0.5em}
    \begin{subfigure}[b]{0.335\linewidth}
        \centering
        \captionsetup{skip=3pt}
        \includegraphics[width=\linewidth]{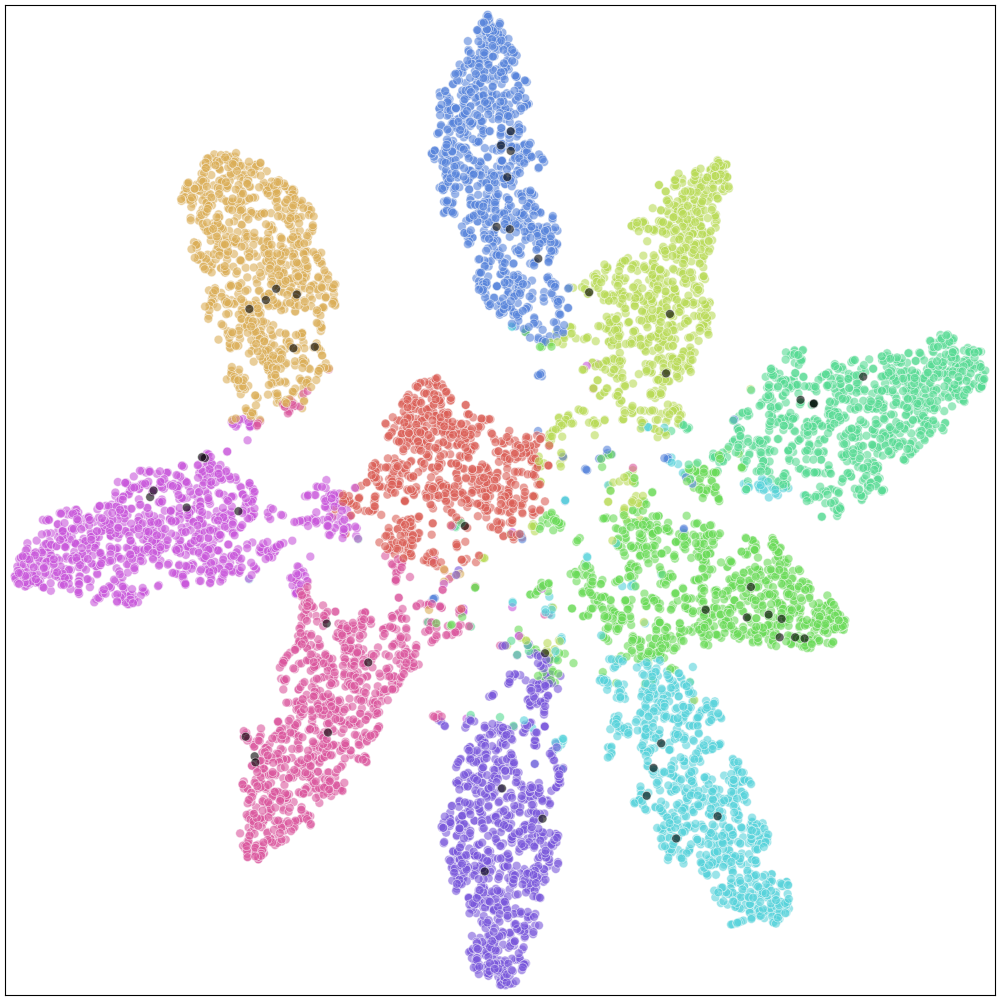} 
        \caption{Ours}
        \label{fig:tsne-our}
    \end{subfigure}
    \caption{t-SNE visualization of feature representations where each group of colored points can be approximately represented as the feature space preimage (set of all features mapping to correct outputs) for the corresponding class. (a) Original network shows well-separated clusters indicating high accuracy; (b) APRNN-repaired net exhibits merged clusters (40\% accuracy drop) despite identical feature extraction; (c) Our method repairs the feature extractor while keeping the well-separated preimages unchanged (1.5\% accuracy drop).}
    \label{fig:tsne}
\end{figure}

However, existing provable methods inherently depend on activation status, which limits applicability to NNs with piecewise linear activation functions and causes scalability issues in more challenging tasks.
Specifically, the rigid requirement of identical activation status across the entire input space not only severely limits their capability to find feasible solutions, but also forces vertex enumeration for region-wise repair, which leads to prohibitive computational costs. 
Consequently, even state-of-the-art methods like APRNN require over 10,000 seconds and 200 GB of memory to repair a single 16D input region. 
Moreover, focusing on modifying the parameters of $f_{\mathrm{c}}$ to satisfy all the constraints in the constructed LP often significantly compromises model's performance. 
We conducted a backdoor repair case study to explain this phenomenon based on preimage analysis. 
As shown in Fig.~\ref{fig:tsne}, while APRNN keeps the extracted features unchanged, it drastically distorts the feature space preimages, resulting in nearly 40\% accuracy degradation.

\section{Methodology}
Motivated by the preimage analysis in Sec.~\ref{sec:motivation}, we devise a new perspective on repair: instead of modifying the parameters of \(f_{\mathrm{c}}\) (which changes the preimages in the feature space), we focus on correcting \(f_{\mathrm{e}}\), to ensure that the extracted features directly fall into the preimage of the corresponding property.
Fig.~\ref{fig:tsne-our} shows an example (we continue to mark the features of poisoned inputs in black for clarity). 
We can see that after repair, the features extracted by \(f_{\mathrm{e}}\) for the given backdoored inputs are all located within the correct preimages (clusters). 
Note that the positions of the clusters change due to t-SNE adjusting the layout based on the input features, but the preimages in the feature space remain unchanged.


\subsection{Point-wise Repair}
\label{sec:point-wise-repair}
In point-wise repair, the input space \(\phi^{in}\) of each property \(\phi \in \mathcal{P}\) consists of a single point \(\bm{x}_{\phi}\). 
As described in Sec.~\ref{sec:motivation}, we expect that the feature of \(\bm{x}_{\phi}\) extracted by the repaired model to be positioned within the feature space preimage, which can be accomplished in two steps: (1) \textit{calculate the preimage of property \(\phi\)} and $f_{\mathrm{c}}$, and (2) \textit{modify the feature extractor \(f_{\mathrm{e}}\) to ensure \(f_{\mathrm{e}}(\bm{x}_{\phi})\) falls within the preimage.}
According to the definition~\ref{def:preimage}, we can directly define the exact feature space preimage based on  \(\phi^{out}\) and  \(f_{\mathrm{c}}\) as:
\begin{equation}
f_{\mathrm{c}}^{-1}(\mathbb{R}^{\mathrm{s}}, \phi^{out}) = \left\{ \bm{h} \in \mathbb{R}^{\mathrm{s}} \mid f_{\mathrm{c}}\left(\bm{h}\right) \in \phi^{out}
\right\} 
\end{equation}
where \(\mathrm{s}\) denotes the dimensionality of the feature space.
Since exactly calculating the preimage is intractable, a natural solution here for step (1) is to utilize existing tools like~\cite{zhang2024provable, kotha2023provably} to generate the under-approximation \(\mathcal{U}\) of the preimage \(f_{\mathrm{c}}^{-1}(\mathbb{R}^{\mathrm{s}}, \phi^{out})\), which can further provide valuable information for step (2). 
For instance, such under-approximate preimage can be leveraged to optimize \(f_{\mathrm{e}}\) to minimize the distance between \(\mathcal{U}\) and \(f_{\mathrm{e}}(\bm{x}_\phi)\), i.e., 
\begin{equation}
\label{eq:opt-func}
\underset{f_{\mathrm{e}}}{\min}  \left\|  f_{\mathrm{e}}(\bm{x}_\phi) - \Pi_{\mathcal{U}}f_{\mathrm{e}}(\bm{x}_\phi) \right\|
\end{equation}
where \(\Pi\) is the projection operator onto \(\mathcal{U}\).

However, existing preimage approximation frameworks typically require the input space to be a bounded box. 
In our scenario, the input space is the entire unbounded feature space \(\mathbb{R}^{\mathrm{s}}\), which makes it impractical to directly apply these frameworks. 
In addition, the under-approximation \(\mathcal{U}\) obtained from these frameworks is a general convex set defined by multiple linear constraints. This necessitates solving a linear programming for projection operations in the problem~\eqref{eq:opt-func}, which is computationally expensive.

\noindent \textbf{Proxy Boxes Synthesis.} To address above issues, rather than generating approximations of the whole preimage for each property \(\phi \in \mathcal{P}\), we concentrate on synthesizing a proxy box \(\mathcal{B}_\phi \subset f_{\mathrm{c}}^{-1}(\mathbb{R}^{\mathrm{s}}, \phi^{out})\).
The key idea is that a small box located within the preimage and in close proximity to the original feature \(f_{\mathrm{e}}(\bm{x}_\phi)\) is sufficient to provide the necessary information for the repair.
To obtain such box, we propose an algorithm based on linear relaxation and iterative center shifting, detailed in Algorithm~\ref{algorithm:point-wise} (lines 1-7).
It starts by constructing the initial box centered around \(f_{\mathrm{e}}(\bm{x}_\phi)\), and then iteratively searches the current box for the point with minimal constraints violation, which then serves as the center for the next box. 
Specifically, for a given box centered at \(\bm{h}_{\phi}\) with a radius \(r\), defined as \(
    \mathcal{B}(\bm{h}_{\phi}, r) = \left\{ \bm{h} \in \mathbb{R}^{\mathrm{s}} \mid \left\|\bm{h}-\bm{h_{\phi}}\right\|_{\infty} \leq r \right\}
\), 
we aim to find the point \(\bm{h}_{\phi}^{\prime}\) within this box that minimize the degree of constraints violation as follows:
\begin{equation}
\label{eq:box-opt}
\bm{h}_{\phi}^{\prime} = \arg \underset{\bm{h} \in  \mathcal{B}(\bm{h}_{\phi}, r)}{\min}
\underset{\psi \in \phi^{cons}}{\sum} - \min\left(\bm{c}_{\psi}^\top \cdot f_{\mathrm{c}}(\bm{h})+d_{\psi}, 0\right)
\end{equation}

\begin{algorithm}[t]
\caption{Point-wise Repair}
\label{algorithm:point-wise}
\DontPrintSemicolon
\SetKwProg{Fn}{\textbf{Function}}{:}{}
\SetKwFunction{main}{Point-wise Repair}
\SetKwFunction{boxg}{Proxy Box Synthesis}\label{func:proxy-box}
\KwIn {DNN \( f = f_{\mathrm{c}} \circ f_{\mathrm{e}} \), property set \( \mathcal{P} \), box radius \( r \), max iterations for box synthesis \( T_\text{b} \), 
max iterations for repair \( T_\text{r} \)}
\KwOut{repaired model \( \tilde{f} \) that \( \tilde{f} \models \mathcal{P} \) or failure \(\bot\)}
\vspace{0.5ex}

\DefHelperFunc{\boxg{\(f_{\mathrm{c}}, \bm{h}_{\phi}, \phi,  r, T_\text{b}\)}}{
    \( \bm{\alpha}, lb \gets \textsc{LinearBounds}(f_{\mathrm{c}}, \mathcal{B}(\bm{h}_{\phi}, r), \phi^{cons}) \)\tcp{\(\bm{\alpha} : \{\bm{\alpha}_{\psi} \in \mathbb{R}^{s} \mid \psi \in \phi^{cons}\}, lb : \{lb_{\psi} \in \mathbb{R} \mid \psi \in \phi^{cons}\}\)}

    \For{$1$ \KwTo $T_\text{b}$}{
    
        \lIf{$\min_{\psi \in \phi^{cons}} lb_{\psi} \ge 0$}{    
            \Return \( \bm{h}_{\phi} \)
        }
        
        \( \bm{h}_{\phi} \gets \arg \underset{\bm{h} \in \mathcal{B}(\bm{h}_{\phi}, r)}{\max} \sum_{\psi \in \phi^{cons}} \left( \mathcal{I}_{lb_\psi < 0} \cdot \bm{\alpha}_\psi^\top \right) \cdot \bm{h} \)
        
        \( \bm{\alpha}, lb \gets \textsc{LinearBounds}(f_{\mathrm{c}}, \mathcal{B}(\bm{h}_{\phi}, r), \phi^{cons}) \)
    
    }
    
    \Return \(\bot\)
}

\DefHelperFunc{\main{\(f, \mathcal{P}, r, T_\text{b}, T_\text{r}\)}}{
    \( \mathcal{C} \gets \{\} \)
    
    \ForEach{\( \phi \in \mathcal{P} \)}{     
    \tcc{ $\phi^{in}$ consists of a single point $\bm{x}_{\phi}$}
    
        \( \bm{h}_{\phi}^* \gets \boxg(f_{\mathrm{c}}, f_{\mathrm{e}}(\bm{x}_{\phi}), \phi, r, T_\text{b}) \)
        
        \lIf{\( \bm{h}_{\phi}^* = \bot \)}{        
            \Return \(\bot\)
        }
        
        \( \mathcal{C} \gets \mathcal{C} \cup \left\{ \left(\bm{x}_{\phi}, \bm{h}_{\phi}^* \right) \right\} \)
    
    }
    
    \For{$1$ \KwTo $T_\text{r}$}{
    
        \lIf{\( \bigwedge_{\phi \in \mathcal{P}} f \models \phi \)}{    
            \Return \( f_{\mathrm{c}} \circ f_{\mathrm{e}} \)
        }
        
        \( \mathcal{L} \gets \frac{1}{|\mathcal{C}|} \sum_{(\bm{x}_{\phi}, \bm{h}_{\phi}^*) \in \mathcal{C}} \left\| f_{\mathrm{e}}(\bm{x}_{\phi}) - \bm{h}_{\phi}^* \right\|_2 \)
        
        \( f_{\mathrm{e}} \gets \mathrm{Adam}(f_{\mathrm{e}}; \min \mathcal{L}) \)
    
    }
    
    \Return \(\bot\)
}
\end{algorithm}

However, precisely solving this optimization problem is challenging due to two inherent complexities: 1) the non-convex activation functions within \( f_{\mathrm{c}} \), and 2) the inner \emph{min} operator). To address the former, we take the box \(\mathcal{B}(\bm{h}_{\phi}, r)\) as the input space and conduct linear relaxation on the subnetwork \(f_{\mathrm{c}}\).
Specifically, we employ auto-LiRPA~\cite{xu2020automatic} to obtain the linear lower bound on the \(f_{\mathrm{c}}\)'s output over all constraints \(\psi \in \phi^{cons}\) as follows:
\begin{equation}
\label{eq:verify-box}
\bm{\alpha}_\psi^\top \cdot \bm{h} + \beta_\psi \le  \bm{c}_{\psi}^\top \cdot f_{\mathrm{c}}(\bm{h}) +d_{\psi}
 \quad \forall \bm{h} \in \mathcal{B}(\bm{h}_{\phi}, r) , \,  \psi \in \phi^{cons}
\end{equation}
where \(\bm{\alpha}_\psi \in \mathbb{R}^{\mathrm{s}}\) and \( \beta_\psi \in \mathbb{R}\) are the coefficients and bias. 
For the inner \emph{min} operator in Eq.~\eqref{eq:box-opt}, we note that its inclusion is necessary since different points within the box may not uniformly satisfy (or violate) the constraint \(\psi\); otherwise, the \emph{min} function could be omitted.
To simplify it, we propose a satisfaction approximation strategy.
We first approximately infer whether the entire box satisfies the constraint \(\psi\), and further generalize this inference to all points within the box. Given a constraint \(\psi\), the property satisfaction for the entire box can be determined by minimizing the lower bound $lb_\psi = \underset{\bm{h} \in \mathcal{B}(\bm{h}_{\phi}, r)}{\min} \bm{\alpha}_\psi^{\top} \cdot \bm{h} + \beta_\psi$
, where \(lb_\psi \geq 0\) implies that the entire box satisfies the constraint \(\psi\), i.e.,  \(\min (\bm{c}_{\psi}^\top \cdot f_{\mathrm{c}}(\bm{h})+d_{\psi}, 0)\) equals 0.
We filter out these already satisfied constraints and generalize the satisfaction state of the entire box to all points for the remaining constraints. Consequently, the problem Eq.~\eqref{eq:box-opt} can be simplified to a linear optimization problem as: 
\begin{equation}
\label{eq:linear-shift}
\bm{h}_{\phi}^{\prime} = \arg \underset{\bm{h} \in  \mathcal{B}(\bm{h}_{\phi}, r)}{\min} -
\underset{\psi \in \phi^{cons}}{\sum} \left(
\bm{\alpha}_\psi^\top \cdot \bm{h} + \beta_\psi
\right) \cdot \mathcal{I}_{lb_\psi < 0}
\end{equation}
where \(\mathcal{I}_{(\cdot)}\) represents the indicator function. 
Since the feasible region \(\mathcal{B}(\bm{h}_{\phi}, r)\) is an axis-aligned box, problem~\eqref{eq:linear-shift} can be directly solved in a closed form. 
As described in Alg.~\ref{algorithm:point-wise} (lines 3-6 and 10-14), 
for each property \(\phi \in \mathcal{P}\), 
we iteratively calculate \(\bm{h}_{\phi}^{\prime}\) and shift the center of the current box to \(\bm{h}_{\phi}^{\prime}\) to synthesize the next box until all constraints are satisfied within the box.
This box synthesis process analyzes the model’s behavior over the entire box \(\mathcal{B}\) at each step, thus preventing certain feature dimensions from getting stuck in local optima where the gradients become negligible. We validate this advantage by comparing our method with gradient-based optimization and varying the radius of the box (Sec.~\ref{sec:ablation}).

Upon obtaining the proxy box $\mathcal{B}_{\phi}$, 
using it to replace $\mathcal{U}$ in Eq.~\eqref{eq:opt-func} can bypass the costly LP solving needed for projection.
However, a critical issue arises if we directly minimize the distance $\mathcal{L}_{\mathrm{Proj}}:=\|  f_{\mathrm{e}}(\bm{x}_\phi) - \Pi_{\mathcal{B}_{\phi}}f_{\mathrm{e}}(\bm{x}_\phi) \|_p$: as $f_{\mathrm{e}}(\bm{x}_\phi)$ approaches the surface of $\mathcal{B}_{\phi}$, the distance (and its gradient w.r.t. $f_{\mathrm{e}}$) diminishes rapidly, causing the repair process to prematurely converge before $f_{\mathrm{e}}(\bm{x}_\phi)$ properly enters the box.
To address this issue, we exploit the proxy box's geometric nature to derive a bounded distance measure. 
Specifically, we establish the following theorem (proof in Appendix~\ref{appendix:proof-1}):
\begin{theorem}
\label{thm:point}
Let $\phi$ be a point-wise property with $\phi^{in} = \{\bm{x}_{\phi}\}$, and $\mathcal{B}_{\phi}$ be the proxy box of radius $r$ centered at $\bm{h}_{\phi}^*$.
Given a repaired model $\tilde{f}:=f_\mathrm{c} \circ \tilde{f}_{\mathrm{e}}$, 
the distance $\mathcal{L} := \lVert \tilde{f}_{\mathrm{e}}(\bm{x}_\phi) - \bm{h}_{\phi}^* \rVert_{p}$ satisfies:
\begin{equation}
\label{eq:distance-bound}
\max \left(\lVert \bm{h}_{\phi}^* - \Pi_{\mathcal{B}_{\phi}}\tilde{f}_{\mathrm{e}}(\bm{x}_\phi) \rVert_p \, , \, \mathcal{L}_{\mathrm{Proj}}\right) \leq \mathcal{L} 
\leq \mathcal{L}_{\mathrm{Proj}} + r\cdot \mathrm{s}^{1/p}
\end{equation}
where $p \ge 1$ and $\mathrm{s}$ is the feature space dimension.
We further have that if $\mathcal{L}\leq r$ holds, then $f_{\mathrm{c}}(\tilde{f}_{\mathrm{e}}(\bm{x}_\phi)) \in \phi^{out}$, which implies $\tilde{f} \models \phi$.
\end{theorem}

Theorem~\ref{thm:point} states that the original objective $\mathcal{L}_{\mathrm{Proj}}$ can be reformulated as the $\ell_p$-norm distance $\mathcal{L}$, which is bounded by $\mathcal{L}_{\mathrm{Proj}}$ and $\mathcal{L}_{\mathrm{Proj}} + r\cdot \mathrm{s}^{1/p}$. 
Crucially, before $f_{\mathrm{e}}(\bm{x}_\phi)$ enters $\mathcal{B}_{\phi}$, the inequality $r \leq \lVert \bm{h}_{\phi}^* - \Pi_{\mathcal{B}_{\phi}}\tilde{f}_{\mathrm{e}}(\bm{x}_\phi) \rVert_p \leq \mathcal{L}$ always holds, thereby avoiding the premature convergence issue. 
The overall algorithm for point-wise repair is summarized in Alg.~\ref{algorithm:point-wise}. 
For all properties \(\phi \in \mathcal{P}\), we synthesize the proxy box and optimize \(f_{\mathrm{e}}\) to minimize $\mathcal{L}$ until all properties are satisfied. 
For simplicity, we use the Euclidean distance ($p=2$) in our implementation. 
Notably, the calculation of the proxy box for each property (lines 10-14 of Alg.~\ref{algorithm:point-wise}) can be parallelized.

\subsection{Region-wise Repair}
Now we present how \name handles region-wise repair, which is more challenging than point-wise task.
The core difficulty stems from the input space $\phi^{in}$ consisting of an infinite number of points, making it necessary to characterize the model's behavior across the entire space and perform targeted adjustments.
For this, prior works combine the activation status with constraint solving, e.g., enumerating all vertices of $\phi^{in}$ and enforcing the model to exhibit identical activation status on all vertices to induce linear behavior throughout the space. 
Based on this idea, the state-of-the-art method APRNN can handle spaces up to 16 dimensions (with $2^{16}$ vertices). 
However, such activation-driven methods suffer from two inherent limitations: the exponential complexity of vertex enumeration confines the scalability to higher dimensional spaces, and the rigid requirement of uniform activation status across the entire space severely limits the feasible solution space.

To tackle these problems, we transform the region-wise repair into an iterative process, where we let counterexample generation take the lead, and employ property refinement for more surgical repair.
The core insight is that correcting the model's erroneous output at point $\bm{x}$ could simultaneously rectify erroneous behaviors over the neighborhood due to the model's continuity.

\noindent \textbf{Counterexample Generation.} 
Following the above idea, the first step is to identify a counterexample (an input point \(\bm{x} \in \phi^{in}\) that \(f(\bm{x}) \notin \phi^{out}\)) for each property.
Intuitively, repairing the input point with the most constraints violation within $\phi^{in}$ is more effective in correcting the model's erroneous behavior over the entire space. 
Inspired by this, we first identify the ``worst'' input \(\bm{x}_{\phi}\) for each property \(\phi\), which can be formalized as:
\begin{equation}
\label{eq:worst-input}
\bm{x}_{\phi} = \arg\underset{\bm{x} \in \phi^{in}}{\max}
\underset{\psi \in \phi^{cons}}{\sum} - \min\left(\bm{c}_{\psi}^\top \cdot f(\bm{x})+d_{\psi}, 0\right)
\end{equation}
We note that problem~\eqref{eq:worst-input} has the same structure as Eq.~\eqref{eq:box-opt} except for the inverted objective and different feasible space.
Thus, we conduct linear relaxation on the whole model $f$ with the space $\phi^{in}$ to derive the approximation of the objective in Eq.~\eqref{eq:worst-input}:
\begin{equation}
\label{eq:worst-input-approx}
\begin{aligned}
& \underset{\psi \in \phi^{cons}}{\sum} - \min\left(\bm{c}_{\psi}^\top  f(\bm{x})+d_{\psi}, 0\right) \\
\le  &
\underset{\psi \in \phi^{cons}}{\sum} - \min\left( \underline{\bm{w}}_\psi^\top  \bm{x}+b_{\psi}, 0\right) 
=\underset{\psi \in \mathrm{Vio}(\phi)}{\sum} - \min\left( \underline{\bm{w}}_\psi^\top  \bm{x}+b_{\psi}, 0\right) 
\end{aligned}
\end{equation}
where \(\underline{\bm{w}}_\psi \in \mathbb{R}^m \) and \(\underline{b}_\psi \in \mathbb{R} \) are the coefficients and bias computed via auto-LiRPA such that $\underline{\bm{w}}_\psi^\top \cdot \bm{x} + \underline{b}_\psi \le  \bm{c}_{\psi}^\top \cdot f(\bm{x}) +d_{\psi}$ holds, \(\mathrm{Vio}(\phi)\)  denotes the set of constraints that may not be satisfied, i.e., \(\{ \psi \in \phi^{cons} \mid lb_\psi = \min_{\bm{x} \in  \phi^{in} } \underline{\bm{w}}_\psi^{\top} \cdot \bm{x} + \underline{b}_\psi < 0\}\). 
We then utilize the satisfaction approximation strategy proposed in Sec.~\ref{sec:point-wise-repair} to linearize the objective as 
$\underset{\bm{x} \in \phi^{in}}{\max} \underset{\psi \in \mathrm{Vio}(\phi)}{\sum} - \left( \underline{\bm{w}}_\psi^\top \bm{x}+b_{\psi}\right)$.
The above simplification yields a linear objective, enabling us to compute the optimal solution directly.
After generating the counterexample for each property, we collect all the real counterexamples and formulate a point-wise repair task, which can be handled by Alg.~\ref{algorithm:point-wise}.

\begin{figure}[t]
\includegraphics[width=1.0\linewidth]{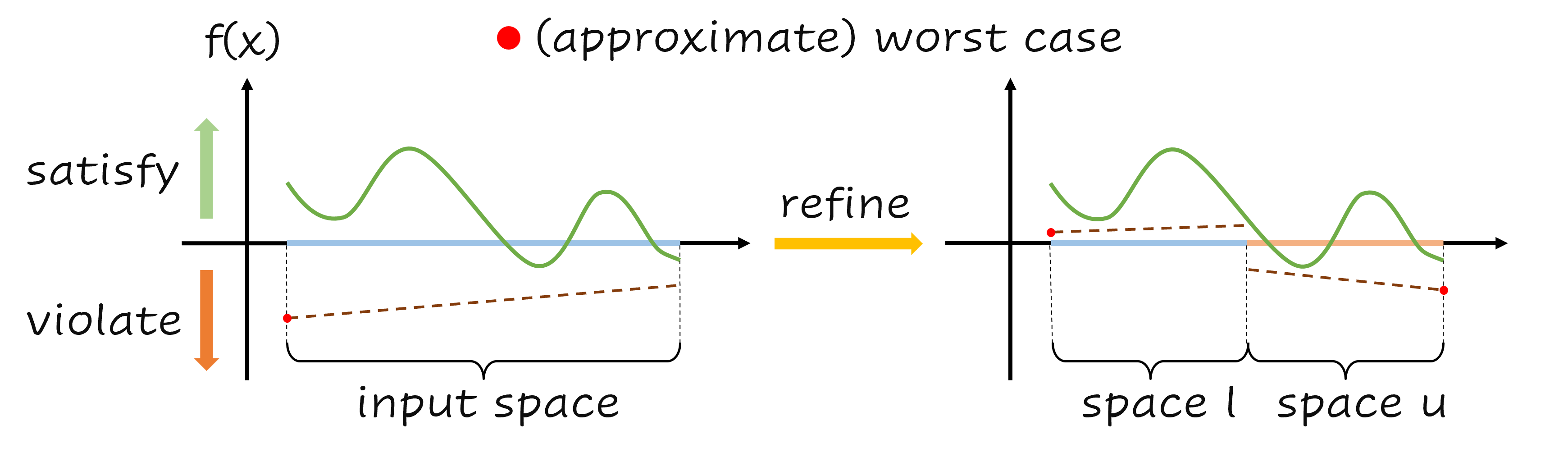}
\centering
\vspace{-0.5em}
\caption{Illustration of refinement: the approximate input does not violate constraints before refinement, while a valid counterexample is captured by splitting the input space.}
\label{fig:refine}
\end{figure}

However, the above counterexample generation process might not always produce valid counterexamples. 
This issue arises from the potentially large input space $\phi^{in}$ (high-dimensional with wide-ranging values), which can lead to excessive approximation errors in Eq.~\eqref{eq:worst-input-approx},  especially since, unlike the proxy box synthesis step, it needs to approximate a more complex model ($f$ rather than $f_{\mathrm{c}}$).

\noindent \textbf{Property Refinement.} 
To mitigate this problem, we propose to refine the unsatisfied properties after one iteration point-wise repair.
As illustrated in Fig.~\ref{fig:refine}, through partitioning the input space into subspaces, we aim to more precisely capture the model's outputs, thereby facilitating counterexample generation and enabling more targeted model repair.
Note that splitting all dimensions of the input space is impractical, as a single refinement step can result in an exponential number of sub-properties. 
Hence, selecting the appropriate dimensions to split is crucial.
To this end, we devise a significance metric called \textit{Refine Score} (\(\mathrm{RS}\)) to assess the splitting importance of each dimension $i$, considering both its input range magnitude and impact on the model's output:
\begin{equation}
    \; \mathrm{RS}(\phi)_i = \left( \bm{u}^{\phi}_i - \bm{l}^{\phi}_i \right) \odot \sum_{\psi \in \mathrm{Vio}(\phi)} \left |\underline{\bm{w}}_\psi \right |_i, \; i \in \left[1, m\right]
\end{equation}
where \(\odot\) is the element-wise multiplication, and \(\sum_{\psi \in \mathrm{Vio}(\phi)} |\underline{\bm{w}}_\psi |_i\) represents the estimated impact of the dimension $i$ on the model's output (considering the unsatisfied constraints). 
We select the dimension with the maximum \(\mathrm{RS}\) value to bisect the input space \(\phi^{in}\) into \(\phi_l^{in}\) and \(\phi_u^{in}\).

\begin{algorithm}[!t]
\caption{Region-wise Repair}
\label{algorithm:region-wise}
\DontPrintSemicolon
\SetInd{0.5em}{0.6em}
\SetKwProg{Fn}{\textbf{{Function}}}{:}{}
\SetKwFunction{main}{Region-wise Repair}
\SetKwFunction{verify}{Counterexample Generation}
\SetKwFunction{refine}{Property Refinement}
\KwIn {DNN $f=f_{\mathrm{c}} \circ f_{\mathrm{e}}$, property set $\mathcal{P}$, box radius $r$ and property budget $\mathbf{T}=10000$}
\KwOut {repaired model $\tilde{f}$ that $\tilde{f}\models \mathcal{P}$, or $\bot$}
\vspace{0.5ex}

\DefHelperFunc{\verify{\(f, \mathcal{P}\)}}{
    $\mathcal{Q} \gets \{  \}$\tcp{Store generated point-wise properties}
    
    \ForEach{$\phi \in \mathcal{P}$}{
        $\underline{\bm{w}}, lb \gets \textsc{LinearBounds}\left(f, \phi^{in}, \phi^{cons} \right)$

        $\mathrm{Vio}(\phi) \gets  \left\{ \psi \in \phi^{cons} \mid lb_\psi < 0\right\}$

        $\bm{x}_{\phi} \gets \arg \underset{\bm{x} \in \phi^{in}}{\min} \sum_{\psi \in \mathrm{Vio}(\phi)} \bm{w}_\psi^{\top} \cdot \bm{x}$

        \If{$f(\bm{x}_{\phi}) \notin \phi^{out}$}{
                
                
                $\mathcal{Q} \gets \mathcal{Q} \cup \left\{ \left( \left\{\bm{x}_{\phi}\right\}, \phi^{cons}, \phi^{out}\right) \right\}$
            }

        $\mathrm{RS}(\phi) \gets  \left( \bm{u}^{\phi} - \bm{l}^{\phi} \right)^{\top} \odot \underset{\psi \in \mathrm{Vio}(\phi)}{\sum} \left |\bm{w}_\psi \right|$\tcp{$\mathrm{RS}(\phi) \in \mathbb{R}^m$}
    }
    \Return $\mathrm{Vio}, \mathcal{Q}, \mathrm{RS}$
}

\DefHelperFunc{\refine{\(\mathcal{P}, \mathrm{Vio}, \mathrm{RS}\)}     \vspace{0.2em}}{
    $\mathcal{P}^{\prime} \gets \{\phi \in \mathcal{P}\mid \mathrm{Vio}(\phi) = \emptyset \}$

    \ForEach{$\phi \in \{\phi \mid \phi \in \mathcal{P} \wedge \mathrm{Vio}(\phi) \ne \emptyset \}$}{
            \vspace{0.1em}
            
            $d \gets \arg \underset{1 \leq i \leq m}{\max} \mathrm{RS}(\phi)_i$
            
            $\phi_{l} \gets \phi, \;\phi_l^{in} \gets \phi^{in} \cap \left\{ \bm{x} \mid  \bm{x}_d \leq (\bm{l}^{\phi}_d + \bm{u}^{\phi}_d) / 2 \right\}$
            
            $\phi_{u} \gets \phi, \;\phi_u^{in} \gets \phi^{in} \cap \left\{ \bm{x}  \mid  \bm{x}_{d} \geq (\bm{l}^{\phi}_d + \bm{u}^{\phi}_d) / 2 \right\}$
            
            $\mathcal{P}^{\prime} \gets \mathcal{P}^{\prime} \cup \{ \phi_{l},\ \phi_{u} \}$
            
            $\mathrm{Vio}(\phi_{l}) \gets \mathrm{Vio}(\phi),\: 
            \mathrm{Vio}(\phi_{u}) \gets \mathrm{Vio}(\phi)$
    }
    
    \Return $\mathcal{P}^{\prime}$   
}

\DefHelperFunc{\main{\(f\), \(\mathcal{P}\), \(r\)}}{
    $\mathrm{Vio}, \mathcal{Q} ,\mathrm{RS} \gets \verify(f, \mathcal{P})$

    \While{$|\mathcal{P}| \le \mathbf{T}$}{

        $f \gets \textsc{Point-Wise Repair}(f, \mathcal{Q}, r, T_\text{b}, T_\text{r})$\tcp{Alg.~\ref{algorithm:point-wise}}

        \lIf{$f = \bot$}{\Return $\bot$}
        
        $\mathrm{Vio}, \mathcal{Q} ,\mathrm{RS} \gets \verify(f, \mathcal{P})$

        \lIf{$\bigwedge_{\phi \in \mathcal{P}} \left( \mathrm{Vio}(\phi) = \emptyset \right)$}{
        \Return $f$
        }
        
        $\mathcal{P} \gets \refine(\mathcal{P}, \mathrm{Vio}, \mathrm{RS})$
    }
   
    \Return $\bot$

}
\end{algorithm}

The main loop of our region-wise repair method is summarized in lines 20-28 of Algorithm~\ref{algorithm:region-wise}, which is coupled with two key functions: counterexample generation (lines 1-10) and property refinement (lines 11-19). 
It first generates the counterexamples and then constructs a point-wise task, followed by a property refinement step.
When the set $\mathrm{Vio}(\phi)$ for all properties is empty, we return a repaired model. 
Otherwise, the algorithm continues to iterate until the property budget $\mathbf{T}$ is exhausted.
The following theorem shows the soundness of our method (proof in Appendix~\ref{appendix:proof-2}).
\begin{theorem}
\label{thm:region}
Given a property set $\mathcal{P}$ and a model $f$, Algorithm~\ref{algorithm:region-wise} returns a repaired model $\tilde{f}$ only if $\tilde{f} \models \mathcal{P}$ holds.
\end{theorem}

\noindent \textbf{Asymptotic complexity analysis.} 
The core components of \name include box synthesis, repair, counterexample generation and space refinement. 
Let the respective costs be denoted as $C_{\text{box}}$, $C_{\text{repair}}$, $C_{\text{counter}}$, and $C_{\text{refine}}$. 
With $T_{\text{b}}$, $T_{\text{r}}$ and $\mathbf{T}$ denoting the iteration budgets for box synthesis, repair and refinement, and assuming each layer contains at most $K$ neurons, the costs of these components are shown in Tab.~\ref{tab:complexity}.
The overall cost is $\mathbf{T}*(T_{\text{b}}*L_{\text{c}}^2*K^3+T_{\text{r}}*L_{\text{e}}^2*K^3+(L_{\text{e}}+L_{\text{c}})^2*K^3+m*q)$. 
Ignoring constant factors, this can be simplified to 
$\mathcal{O}((L_{\text{e}}+L_{\text{c}})^2*K^3)$, 
which grows quadratically with the total number of layers and cubically with the maximum number of neurons per layer.

\begin{table}[!t]
\centering
\caption{Complexity of each component in \name.}
\label{tab:complexity}
\begin{tabular}{l|p{4.6cm}|l}
\toprule
 & \textbf{Description} & \textbf{Cost} \\
\midrule
$C_{\text{box}}$ & Bound propagation over $f_{\text{c}}$ across $L_{\text{c}}$ layers, where each neuron contributes up to $L_{\text{c}}*K^2$ terms & $T_{\text{b}}* L_{\text{c}}^2* K^3$ \\
\midrule
$C_{\text{repair}}$ & Gradient descent optimization on $f_{\text{e}}$ ($L_{\text{e}}*K$ neurons with $L_{\text{e}} *K^2$ terms per neuron) & $T_{\text{r}} *L_{\text{e}}^2 *K^3$\\
\midrule
$C_{\text{counter}}$ & Bound propagation on the model $f$ & $(L_{\text{e}} + L_{\text{c}})^2* K^3$ \\
\midrule
$C_{\text{refine}}$ & Refinement score computation over an $m$-D space with $q$ constraints & $m *q$ \\
\bottomrule
\end{tabular}
\end{table}

\section{Evaluation}
We apply \name to four different repair tasks, including point-wise repair for natural corruption and backdoor attacks, as well as region-wise repair for adversarial attacks and violation of safety properties.
We briefly introduce the setup below.
More details on the models, datasets, and the construction of the desired property set are provided in Appendix~\ref{appendix:setup}.

\subsection{Experimental Setup}
\label{sec:setup}

\subsubsection{Repair Task}
\begin{itemize}[left=0pt]
\item \textbf{Backdoor.} This task aims to repair the buggy NN so that it correctly classify the backdoored data, which is stronger than merely defending against attacks (i.e., ensuring the classification is not the target label).
We conduct extensive evaluations on multiple datasets, attack strategies, and model architectures. 
More setup for this task are detailed in the Appendix~\ref{appendix:setup-backdoor} and Appendix~\ref{appendix:attacks}.
\item \textbf{Corruption.} This task is widely adopted in provable repair ~\cite{sotoudeh2021provable, tao2023architecture, reassure} to evaluate the effectiveness and efficiency of repair methods on small-scale datasets and models.
We use various NNs from~\cite{erangithub} on MNIST-C~\cite{mu2019mnist} dataset with all corruptions. 

\item \textbf{Adversarial attack.} This task aims to correct a buggy model so that it is provably robust within the given local spaces.
We conduct evaluations on the MNIST, GTRSB and CIFAR-10 datasets. The corresponding networks are 3x100, CNN-3 and CNN-4. 
We evaluate our method and APRNN by varying the perturbation dimensions and the number of desired properties.
\item \textbf{Safety Property Violation.} 
This task focuses on correcting the network so that it satisfies the property that may be defined on a global input space rather than a local neighborhood.
Specifically, we conduct the evaluation on the ACAS Xu benchmark that is designed for aircraft collision detection. 
As reported in~\cite{dong2021towards}, more than 30 ACAS Xu models violate Property-2 (detailed in Appendix~\ref{appendix:setup-safety}). 
We perform an evaluation of all these models.
\end{itemize}

\subsubsection{Baselines and Configurations}
To the best of our knowledge, APRNN, PRDNN, and REASSURE are the only three provable repair methods that scale to medium-sized and larger DNNs. We conduct the evaluation by comparing our method with these SOTA methods. For the backdoor repair task, we include two additional methods: SEAM~\cite{zhu2023selective} and I-REPAIR~\cite{henriksen2022repairing}. SEAM is the latest SOTA unlearning method for backdoor removal, while I-REPAIR is a repair framework based on fault localization and fine-tune. They do not provide guarantees and require additional clean data, but typically exhibit notable scalability and efficiency. 

For \name, we set the radius \(r\) of the box to 0.1 for all scenarios, except for backdoor repair, where it is set to 0.5 due to the larger model size. We analyze the impact of this parameter in Sec.~\ref{sec:ablation-radius}. Other setups for all methods are detailed in Appendix~\ref{appendix:baseline}.

\subsubsection{Evaluation metrics}
There are three metrics involved in the evaluation: \textit{PSR} (Property Satisfaction Rate), \textit{Acc} (Accuracy), and \textit{Gene} (Generalization). 
\textit{PSR} measures the proportion of given desired properties that are satisfied by the repaired model, i.e., $\mathrm{PSR}= \frac{ |\{\,\phi \in \mathcal{P} \mid \tilde{f} \models \phi\,\}|}{\mathcal{|P|}}, \:$
For provable repairs, the \textit{PSR} should reach 100\%.
\textit{Acc} measures the ratio of normal inputs from the test set that can be correctly classified. 
\textit{Gene} measures the proportion of inputs from the generalization set for which the model's outputs satisfy the desired properties.
Details on the construction of the desired property set \(\mathcal{P}\) and the generalization set are provided in Appendix~\ref{appendix:setup}.

\subsection{How effective and efficient is \name?}
\label{sec:main-exp}

\subsubsection{Point-wise Backdoor Repair}
Considering the limited availability of poisoned data in practice, we randomly selected 50 buggy data (nearly 0.1\% of the training dataset) to construct the desired property set. We first conducted evaluations under relatively common attack strategies, i.e., dirty label attack on single class. REASSURE was excluded from the evaluation, as it does not directly support such large scale models. 
For SEAM and I-REPAIR, we provide an additional clean set containing 1\% of the training data.

As shown in Tab.~\ref{table:backdoor}, both PRDNN and APRNN achieve provable repair (i.e., 100\% PSR). Nevertheless, they encounter failures in some cases: PRDNN requires significant memory, leading to out-of-memory issues, while APRNN encountered some unsolvable instances, as the constructed LP has no feasible solution. 
This is because they differ from data-driven approaches, as more properties to repair introduces additional constraints, making the task more challenging. 
By comparison, \name successfully corrects the model across all settings.
In Sec.~\ref{sec:number-repair}, we conduct further evaluation with varying amounts of properties to repair.

\begin{table*}[t]
\setlength{\tabcolsep}{3.4pt}
\caption{Point-wise backdoor repair on different models, datasets, and attacks. The best values are in \textbf{bold}, and the second-best values are \underline{underlined}. PSR: Property Satisfaction Rate (\%), Acc: Accuracy (\%), Gene: Generalization (\%), T: Time (seconds).}
\label{table:backdoor}
\centering
\fontsize{7pt}{8pt}\selectfont
\begin{tabular}{ccc|cc|>{\columncolor{green!6}}c>{\columncolor{green!6}}c>{\columncolor{green!6}}c>{\columncolor{green!6}}l|cccl|cccl|cccl|cccl}
\toprule
 &  & \multirow{2}{*}{Attack}  & \multicolumn{2}{c|}{Original} & \multicolumn{4}{>{\columncolor{green!6}}c|}{\name} & \multicolumn{4}{c|}{PRDNN\cite{sotoudeh2021provable}} & \multicolumn{4}{c|}{APRNN\cite{tao2023architecture}} & \multicolumn{4}{c|}{SEAM\cite{zhu2023selective}} & \multicolumn{4}{c}{I-REPAIR\cite{henriksen2022repairing}}  \\ 
&& & Acc &  Gene &  PSR & Acc &  Gene & T & PSR & Acc &  Gene  & T & PSR & Acc &   Gene  & T  & PSR & Acc &  Gene & T  & PSR & Acc &  Gene & T \\
\midrule
 \multirow{12}{*}{\rotatebox{90}{ResNet18}} & \multirow{3}{*}{\rotatebox{90}{SVHN}} & {BNA2O} & 95.9 & 8.4 & \textbf{100} & \textbf{94.91} & \textbf{94.63} & \textbf{1.2} & \textbf{100} & \underline{94.60} & 19.51 & 4107 & \textbf{100} & 94.37 & 18.84 & 913 & 96 & 62.91 & \underline{58.81} & \underline{5.1} & 74 & 94.53 & 52.68 & \textbf{20.2} \\
& & {Blend} & 95.8 & 6.4 & \textbf{100} & \underline{94.13} & \textbf{78.57} & \textbf{1.9} & \textbf{100} & 83.01 & 6.84 & 2822 & \textbf{100} & 85.64 & 8.90 & 1028 & 90 & 71.12 & \underline{55.65} & \underline{2.2} & 0 & \textbf{94.48} & 6.36 & \textbf{24.9} \\
& & {TrojanNN} & 96.1 & 6.7 & \textbf{100} & \textbf{95.27} & \textbf{72.76} & \textbf{1.2} & \textbf{100} & 84.58 & 9.51 & 3549 & \textbf{100} & 80.49 & 12.60 & 1198 & 84 & 68.05 & \underline{34.08} & \underline{2.6} & 18 & \underline{92.88} & 7.61 & \textbf{23.5} \\
\cmidrule{2-25}
  & \multirow{3}{*}{\rotatebox{90}{GTSRB}} & {BNA2O} & 98.6 & 7.4 & \textbf{100} & \textbf{98.55} & \textbf{98.44} & \textbf{1.2} & \textbf{100} & 84.60 & 23.02 & 23059 & \textbf{100} & 92.41 & 18.97 & 1313 & 98 & 80.09 & \underline{79.31} & \underline{1.9} & 84 & \underline{97.19} & 51.11 & \textbf{3.8} \\
& & {Blend} & 98.5 & 1.2 & \textbf{100} & \textbf{97.16} & \textbf{89.77} & \textbf{1.3} & \textbf{100} & 62.00 & 9.83 & 19076 & \textbf{100} & 58.45 & 6.55 & 1409 & 94 & 77.82 & \underline{59.21} & \underline{1.9} & 20 & \underline{93.02} & 7.17 & \textbf{4.1} \\
& & {TrojanNN} & 98.9 & 0.5 & \textbf{100} & \textbf{97.60} & \textbf{91.94} & \textbf{1.2} & \textbf{100} & 64.39 & 9.24 & 20162 & \textbf{100} & 77.12 & 12.79 & 1657 & 82 & 80.57 & 20.19 & \underline{1.4} & 92 & \underline{91.17} & \underline{39.48} & \textbf{3.4} \\
\cmidrule{2-25}
  & \multirow{3}{*}{\rotatebox{90}{\scriptsize CIFAR10}} & {BNA2O} & 93.6 & 12.1 & \textbf{100} & 90.50 & \textbf{89.84} & \textbf{1.2} & \textbf{100} & \underline{92.44} & 36.58 & 5617 & \textbf{100} & \textbf{92.87} & 30.20 & 1748 & 96 & 44.04 & 41.27 & \textbf{6.6} & 78 & 84.81 & \underline{57.02} & \underline{3.6} \\
& & {Blend} & 94.8 & 10.0 & \textbf{100} & \textbf{90.41} & \textbf{67.98} & \textbf{1.3} & \textbf{100} & 80.72 & 16.44 & 3691 & \textbf{100} & 77.11 & 16.67 & 672 & 82 & 65.49 & \underline{28.71} & \underline{1.5} & 0 & \underline{88.05} & 10.00 & \textbf{4.2} \\
& & {TrojanNN} & 94.4 & 10.0 & \textbf{100} & 92.13 & \textbf{62.36} & \textbf{1.3} & \textbf{100} & 91.50 & 18.70 & 6419 & \textbf{100} & \textbf{92.76} & 15.04 & 919 & 50 & 60.22 & \underline{21.24} & \underline{1.4} & 0 & \underline{92.48} & 10.04 & \textbf{4.3} \\
\cmidrule{2-25}
  & \multirow{3}{*}{\rotatebox{90}{\scriptsize CIFAR100}} & {BNA2O} & 76.0 & 5.5 & \textbf{100} & \textbf{73.50} & \textbf{64.26} & \textbf{1.6} & \multicolumn{4}{c|}{\multirow{3}{*}{\footnotesize  Out of Memory\footnotemark[1]}} & \textbf{100} & \underline{73.10} & 8.06 & 897 & 92 & 8.05 & 7.39 & \textbf{6.7} & 62 & 62.20 & \underline{12.01} & \underline{3.9} \\
& & {Blend} & 75.5 & 1.0 & \textbf{100} & \textbf{70.57} & \textbf{32.79} & \textbf{1.8} & & & & & \textbf{100} & 58.75 & 2.11 & 814 & 72 & 8.11 & \underline{5.14} & \textbf{6.0} & 4 & \underline{63.94} & 1.19 & \underline{4.3} \\
& & {TrojanNN} & 75.7 & 1.3 & \textbf{100} & \textbf{73.40} & \textbf{49.38} & \textbf{1.6} & & & & & \textbf{100} & 63.83 & 2.69 & 902 & 70 & 7.16 & \underline{3.64} & \textbf{4.7} & 24 & \underline{65.34} & 2.80 & \underline{4.3} \\
\cmidrule{2-25}
 \multirow{12}{*}{\rotatebox{90}{VGG11\footnotemark[2]}} & \multirow{3}{*}{\rotatebox{90}{SVHN}} & {BNA2O} & 95.2 & 8.7 & \textbf{100} & \textbf{94.68} & \textbf{93.86} & \textbf{1.4} & \textbf{100} & 42.19 & 38.08 & 1527 & \textbf{100} & 50.46 & 37.06 & 142 & 92 & 69.69 & \underline{68.16} & \textbf{3.3} & 12 & \underline{94.52} & 13.83 & \underline{2.2} \\
& & {Blend} & 94.8 & 6.4 & \textbf{100} & \textbf{94.23} & \textbf{82.90} & \textbf{1.6} & \textbf{100} & 50.39 & 34.88 & 2314 & \textbf{100} & 32.37 & 41.25 & 161 & 96 & 69.50 & \underline{63.77} & \textbf{3.4} & 36 & \underline{89.35} & 23.88 & \underline{2.3} \\
& & {TrojanNN} & 95.2 & 6.7 & \textbf{100} & \textbf{94.54} & \textbf{72.94} & \underline{1.3} & \textbf{100} & 48.03 & 17.23 & 1017 & \multicolumn{4}{c|}{Repair Failed}& 60 & 79.84 & \underline{27.92} & \textbf{0.9} & 12 & \underline{93.65} & 7.32 & \textbf{2.5} \\
\cmidrule{2-25}
  & \multirow{3}{*}{\rotatebox{90}{GTSRB}} & {BNA2O} & 97.5 & 7.6 & \textbf{100} & \textbf{97.32} & \textbf{95.15} & \underline{1.4} & \textbf{100} & 23.03 & 15.60 & 5765 & \textbf{100} & 19.79 & 18.28 & 219 & 96 & 64.12 & 50.32 & \textbf{1.6} & \textbf{100} & \underline{96.84} & \underline{67.09} & \textbf{1.2} \\
& & {Blend} & 97.6 & 1.2 & \textbf{100} & \textbf{96.70} & \textbf{88.83} & \underline{1.5} & \textbf{100} & 8.42 & 10.57 & 6018 & \textbf{100} & 5.42 & 11.75 & 342 & 90 & 80.48 & \underline{55.51} & \textbf{0.8} & 6 & \underline{96.55} & 1.90 & \textbf{2.1} \\
& & {TrojanNN} & 97.6 & 0.6 & \textbf{100} & \underline{96.13} & \textbf{73.83} & \underline{1.4} & \textbf{100} & 16.72 & 10.73 & 6420 & \textbf{100} & 8.76 & 11.55 & 665 & 84 & 81.92 & 36.06 & \textbf{0.4} & 98 & \textbf{96.32} & \underline{42.73} & \textbf{1.5} \\
\cmidrule{2-25}
  & \multirow{3}{*}{\rotatebox{90}{\scriptsize CIFAR10}} & {BNA2O} & 91.5 & 12.4 & \textbf{100} & \textbf{89.68} & \textbf{87.18} & \textbf{2.3} & \textbf{100} & 36.56 & 35.98 & 1016 & \multicolumn{4}{c|}{Repair Failed}& 70 & 66.51 & \underline{55.32} & \textbf{0.4} & 2 & \underline{87.11} & 13.36 & \underline{1.4} \\
& & {Blend} & 91.5 & 10.0 & \textbf{100} & \textbf{88.76} & \textbf{74.30} & \textbf{2.1} & \textbf{100} & 13.78 & 22.71 & 875 & \textbf{100} & 42.57 & 23.90 & 173 & 86 & 65.99 & \underline{29.34} & \textbf{0.7} & 0 & \underline{85.76} & 10.00 & \underline{1.8} \\
& & {TrojanNN} & 91.7 & 10.0 & \textbf{100} & \textbf{89.79} & \textbf{54.72} & \underline{1.4} & \textbf{100} & 23.79 & 20.08 & 1955 & \textbf{100} & 52.70 & 24.54 & 393 & 72 & 57.85 & \underline{28.33} & \textbf{1.2} & 0 & \underline{88.87} & 10.07 & \textbf{2.2} \\
\cmidrule{2-25}
  & \multirow{3}{*}{\rotatebox{90}{\scriptsize CIFAR100}} & {BNA2O} & 70.7 & 5.9 & \textbf{100} & \textbf{67.62} & \textbf{61.84} & \underline{2.1} & \textbf{100} & 16.01 & 13.34 & 13956 & \textbf{100} & 14.31 & \underline{15.69} & 271 & 24 & 26.63 & 8.56 & \textbf{1.1} & 58 & \underline{49.47} & 13.74 & \textbf{2.6} \\
& & {Blend} & 71.8 & 1.0 & \textbf{100} & \textbf{69.04} & \textbf{29.46} & \underline{1.6} & \textbf{100} & 14.63 & 4.80 & 15797 & \textbf{100} & 9.03 & \underline{5.92} & 253 & 34 & 40.49 & 3.32 & \textbf{0.6} & 18 & \underline{43.34} & 2.61 & \textbf{2.9} \\
& & {TrojanNN} & 71.1 & 1.2 & \textbf{100} & \textbf{68.01} & \textbf{38.99} & \underline{1.7} & \textbf{100} & 14.02 & 5.73 & 14592 & \textbf{100} & 16.06 & \underline{6.54} & 281 & 12 & 37.12 & 3.10 & \textbf{0.4} & 38 & \underline{49.80} & 4.07 & \textbf{2.6} \\
\bottomrule
 
\end{tabular}
    \footnotesize
    1. PRDNN required more than 256GB of memory in these settings, resulting in out-of-memory errors. \; 2. We employed VGG16 for CIFAR-100 and VGG11 for other datasets.

\end{table*}

\begin{table}[t]
\setlength{\tabcolsep}{4.5pt}
\caption{Backdoor repair against advanced attacks.}
\label{table:backdoor-other-att}
\centering
\footnotesize
\begin{tabular}{cc|cccc|cccc}
\toprule
 &  \multirow{2}{*}{Method} & \multicolumn{4}{c|}{Badnets All to All}  & \multicolumn{4}{c}{SIG}  \\ 
& & PSR & Acc &  Gene & T  & PSR & Acc & Gene & T  \\
\midrule
 \multirow{6}{*}{\rotatebox{90}{ResNet18}} &  Original & 0 & 94.94 & 2.97 & -& 0 & 98.72 & 0.73 & -\\
& {PRDNN} & 98\footnotemark[1] & \underline{94.83} & 25.18 & 5443 & \textbf{100} & 37.32 & 6.72 & 16683 \\
& {APRNN} & \textbf{100} & \textbf{94.87} & 15.61 & 1144 & \textbf{100} & 32.23 & 9.36 & 1426 \\
& {SEAM} & 88 & 52.86 & 45.21 & \underline{2.5} & 84 & \underline{72.01} & 32.66 & \underline{2.9} \\
& {IREPAIR} & \textbf{100} & 93.14 & \underline{77.63} & 6.3 & 60 & 71.16 & \underline{33.48} & 3.8 \\
  & \cellcolor{green!6}{Ours} & \cellcolor{green!6}\textbf{100} & \cellcolor{green!6}93.41 & \cellcolor{green!6}\textbf{86.74} & \cellcolor{green!6}\textbf{0.7} & \cellcolor{green!6}\textbf{100} & \cellcolor{green!6}\textbf{94.05} & \cellcolor{green!6}\textbf{69.50} & \cellcolor{green!6}\textbf{1.5} \\
\cmidrule{2-10}

 \multirow{6}{*}{\rotatebox{90}{VGG11}} &  Original & 0 & 91.87 & 4.18 & -& 0 & 97.83 & 3.42 & -\\
& {PRDNN} & \textbf{100} & \textbf{91.59} & 32.96 & 1008 & \textbf{100} & 12.42 & 18.58 & 6071 \\
& {APRNN} & \multicolumn{4}{c|}{Repair Failed}& \textbf{100} & 19.07 & 17.32 & 196.2 \\
& {SEAM} & 74 & 72.29 & \underline{65.97} & \textbf{0.8} & 60 & 86.30 & 31.03 & \textbf{0.8} \\
& {IREPAIR} & 52 & \underline{90.97} & 51.51 & 6.4 & 70 & \underline{93.44} & \underline{39.55} & 1.9 \\
 &  \cellcolor{green!6}{Ours} & \cellcolor{green!6}\textbf{100} & \cellcolor{green!6}90.06 & \cellcolor{green!6}\textbf{74.07} & \cellcolor{green!6}\underline{1.0} & \cellcolor{green!6}\textbf{100} & \cellcolor{green!6}\textbf{96.59} & \cellcolor{green!6}\textbf{59.75} & \cellcolor{green!6}\underline{1.1} \\
\bottomrule
\end{tabular}
\end{table}

\begin{table*}[t]
\setlength{\tabcolsep}{4.7pt}
\renewcommand{\arraystretch}{0.93}
\caption{Results of point-wise corruption repair on the 6×100 model under different corruptions.}
\label{table:corruption}
\centering
\small
\fontsize{8pt}{10.4pt}\selectfont
\begin{tabular}{c|cc|>{\columncolor{green!6}}c>{\columncolor{green!6}}c>{\columncolor{green!6}}c>{\columncolor{green!6}}c|llll|llll|llll}
\toprule
\multirow{2}{*}{Corruption}  & \multicolumn{2}{c|}{Original} & \multicolumn{4}{>{\columncolor{green!6}}c|}{\name} & \multicolumn{4}{c|}{PRDNN\cite{sotoudeh2021provable}}  & \multicolumn{4}{c|}{APRNN\cite{tao2023architecture}} & \multicolumn{4}{c}{REASSURE\textdagger\cite{reassure}} \\ 
& Acc & Gene & PSR & Acc & Gene & T & PSR & Acc & Gene & T & PSR & Acc & Gene & T & PSR & Acc & Gene & T  \\
\midrule
brightness & 96.29 & 39.89 & \textbf{100} & \underline{95.31} & \textbf{83.49} & \textbf{1.5} & \textbf{100} & 94.84 & 45.31 & \underline{4.4} & \textbf{100} & 93.60 & \underline{65.78} & 144.4 & \textbf{100} & \textbf{96.29} & 39.89 & 1498.1 \\
  canny edges & 96.29 & 61.10 & \textbf{100} & \underline{94.93} & \textbf{72.50} & \textbf{1.5} & \textbf{100} & 92.54 & 61.62 & \underline{5.4} & \textbf{100} & 93.32 & \underline{65.02} & 133.8 & \textbf{100} & \textbf{96.29} & 61.10 & 2056.0 \\
  dotted line & 96.29 & 94.21 & \textbf{100} & \textbf{96.58} & \textbf{94.67} & \textbf{1.6} & \textbf{100} & 95.56 & 93.49 & \underline{4.8} & \textbf{100} & 95.37 & 93.02 & 12.3 & \textbf{100} & \underline{96.29} & \underline{94.52} & 1857.6 \\
  fog & 96.29 & 23.20 & \textbf{100} & \underline{94.54} & \textbf{81.93} & \textbf{1.4} & \textbf{100} & 89.41 & 30.03 & \underline{5.3} & \textbf{100} & 93.65 & \underline{56.21} & 158.6 & \textbf{100} & \textbf{96.29} & 23.20 & 2115.3 \\
  glass blur & 96.29 & 92.67 & \textbf{100} & 94.42 & \textbf{92.88} & \textbf{1.6} & \textbf{100} & \underline{94.77} & 91.22 & \underline{5.0} & \textbf{100} & 93.66 & 90.90 & 131.5 & \textbf{100} & \textbf{96.29} & \underline{92.69} & 1805.6 \\
  identity & 96.29 & 96.32 & \textbf{100} & \textbf{97.27} & \textbf{97.06} & \textbf{1.9} & \textbf{100} & 96.39 & 96.21 & \underline{4.9} & \textbf{100} & \underline{96.74} & 96.58 & 113.0 & \textbf{100} & 96.29 & \underline{96.99} & 1845.3 \\
  impulse noise & 96.29 & 88.07 & \textbf{100} & \underline{96.20} & \textbf{90.57} & \textbf{1.5} & \textbf{100} & 95.40 & 87.81 & \underline{4.8} & \textbf{100} & 90.86 & 84.12 & 124.8 & \textbf{100} & \textbf{96.29} & \underline{88.07} & 1895.1 \\
  motion blur & 96.29 & 73.41 & \textbf{100} & \underline{93.71} & \textbf{87.76} & \textbf{1.6} & \textbf{100} & 88.31 & 73.91 & \underline{5.4} & \textbf{100} & 91.24 & \underline{78.59} & 119.0 & \textbf{100} & \textbf{96.29} & 73.41 & 2047.1 \\
  rotate & 96.29 & 81.41 & \textbf{100} & \underline{95.71} & \textbf{85.41} & \textbf{1.6} & \textbf{100} & 94.22 & \underline{82.08} & \underline{5.2} & \textbf{100} & 90.60 & 77.77 & 14.5 & \textbf{100} & \textbf{96.29} & 81.41 & 1971.8 \\
  scale & 96.29 & 60.18 & \textbf{100} & \underline{94.14} & \textbf{85.17} & \textbf{1.6} & \textbf{100} & 90.55 & 68.53 & \underline{5.7} & \textbf{100} & 91.81 & \underline{71.02} & 143.0 & \textbf{100} & \textbf{96.29} & 60.18 & 2102.3 \\
  shear & 96.29 & 91.07 & \textbf{100} & 95.41 & \textbf{91.48} & \textbf{1.9} & \textbf{100} & \underline{95.45} & 90.19 & \underline{5.3} & \textbf{100} & 91.70 & 86.12 & 140.6 & \textbf{100} & \textbf{96.29} & \underline{91.07} & 1967.3 \\
  shot noise & 96.29 & 95.20 & \textbf{100} & \textbf{97.14} & \textbf{95.71} & \textbf{1.7} & \textbf{100} & 95.70 & 94.74 & \underline{5.2} & \textbf{100} & 95.69 & 94.04 & 14.4 & \textbf{100} & \underline{96.29} & \underline{95.67} & 2020.5 \\
  spatter & 96.29 & 94.03 & \textbf{100} & \textbf{97.03} & \textbf{94.77} & \textbf{1.6} & \textbf{100} & 94.84 & 92.16 & \underline{5.2} & \textbf{100} & 96.03 & 93.88 & 13.9 & \textbf{100} & \underline{96.29} & \underline{94.51} & 1907.4 \\
  stripe & 96.29 & 54.46 & \textbf{100} & \underline{95.76} & \textbf{78.72} & \textbf{1.3} & \textbf{100} & 94.99 & 60.12 & \underline{5.4} & \textbf{100} & 92.73 & \underline{67.44} & 97.4 & \textbf{100} & \textbf{96.29} & 54.46 & 1871.8 \\
  zigzag & 96.29 & 83.64 & \textbf{100} & \underline{95.78} & \textbf{89.12} & \textbf{1.7} & \textbf{100} & 93.46 & 83.27 & \underline{4.9} & \textbf{100} & 94.53 & 83.23 & 135.6 & \textbf{100} & \textbf{96.29} & \underline{83.64} & 1855.7 \\
  \midrule
  Average & 96.29 & 75.26& \textbf{100} & \underline{95.60} & \textbf{88.08} & \textbf{1.6} & \textbf{100} & 93.76 & 76.71 & \underline{5.1} & \textbf{100} & 93.44 & \underline{80.25} & 99.79 & \textbf{100} & 96.29 & 75.39 & 1921.1 \\
  \bottomrule
\end{tabular}
\end{table*}

In terms of accuracy preservation and generalization, \name outperforms in the majority of settings: the average accuracy drop and generalization are below 2\% and about 70\%, respectively. 
In contrast, the baseline with the best accuracy preservation, I-REPAIR, shows a performance loss of about 6\%, while SEAM exhibits the best generalization among all baselines, at around 35\%. 
PRDNN and APRNN do not consistently protect model performance: they only perform well under few settings.
Their generalization is also unsatisfactory, possibly because they require that activation patterns remain unchanged before and after repair, as some studies~\cite{zheng2022pre, liu2019abs} reveal that backdoored models often show unusually high activation in certain neurons on poisoned samples.
On the other hand, the performance of SEAM and I-REPAIR is highly influenced by the dataset, due to their data-driven nature. On more complex datasets (e.g., CIFAR-100), they exhibit limited generalization and significant accuracy degradation, as these more complex scenarios require larger amounts of data. We evaluate their performance under varying sample (including both buggy and clean data) sizes in Sec.~\ref{sec:number-repair}. 
Nevertheless, they still demonstrate decent efficiency.
The other two provable repair methods often require several hundred to thousands of seconds, with PRDNN's overheads in some scenarios exceeding 10\,000 seconds, even surpassing the time needed to train a model from scratch.
Overall, our method performs the best in this regard, with repair overheads never exceeding 2 seconds (a speed-up of 160$\times$ to 2000$\times$ over existing provable methods).

\noindent\textbf{Against advanced attack.} We further consider two more advanced attacks: Badnets All-to-All and SIG on CIFAR-10 and GTSRB, respectively. The former involves dirty label attacks across multiple classes, while the latter is an invisible clean label attack. 
The desired property set is constructed from 50 randomly selected buggy data.
As shown in the Tab.~\ref{table:backdoor-other-att}, our method remains robust, with an average accuracy drop of only 2.3\%, and a generalization improvement that exceeds 70\%.
By comparison, other baselines exhibit unstable results. While I-REPAIR performs better than the others, it still falls significantly behind our method in terms of generalization and accuracy preservation.

\subsubsection{Point-wise Corruption Repair} 
The desired property set for this task is constructed using 100 corrupted inputs from the repair set.
We consider all 15 types of corruption from the MNIST-C dataset and report the full results for the 6x100 model in Tab.~\ref{table:corruption}.
Results for the remaining models are shown in Fig.~\ref{fig:corr-9100200} in the Appendix.

Considering all corruption types and model architectures, most methods achieved provable repair (i.e., 100\% PSR) in the majority of settings.
However, APRNN failed to find feasible solution in certain cases (e.g., 9x200 model with \textit{scale} corruption).
In terms of accuracy, APRNN and PRDNN showed varying degrees of drop: APRNN decreased by an average of 8.97\%, while PRDNN decreased by 10.39\%. 
In comparison, our method exhibited a much smaller average decrease of only 1.76\%. 
On the other hand, REASSURE did not show any decline, as it constructs a patch network and a support network for each buggy data, which essentially acts as a look-up table. 
However, this also leads to a  nearly 0\% generalization.
PRDNN and APRNN achieved better generalization of 67.2\% and 71.8\%, respectively.
Our method consistently achieved the highest generalization across all settings, with an average of 85\%.
The improvement in generalization is especially significant under challenging corruptions such as \textit{fog} and \textit{brightness}. 
For simpler corruptions, the gain is naturally smaller, as the models already exhibit strong generalization, sometimes approaching the accuracy on clean samples.

\noindent \textbf{Time Cost.}
REASSURE had a significant time overhead (more than 4\,000 seconds on average) due to its need to calculate the activation patterns and linear regions of the network for each buggy data, resulting in a substantial increase in time cost as the number of neurons increases. The average repair times for APRNN and PRDNN were 269s and 9s, respectively. 
Our method demonstrated the best efficiency with an average repair time of 1.8s, outperforming existing provable methods by 5$\times$ to 2000$\times$ in speed.

\begin{figure*}[t]
\includegraphics[width=1.0
\textwidth]{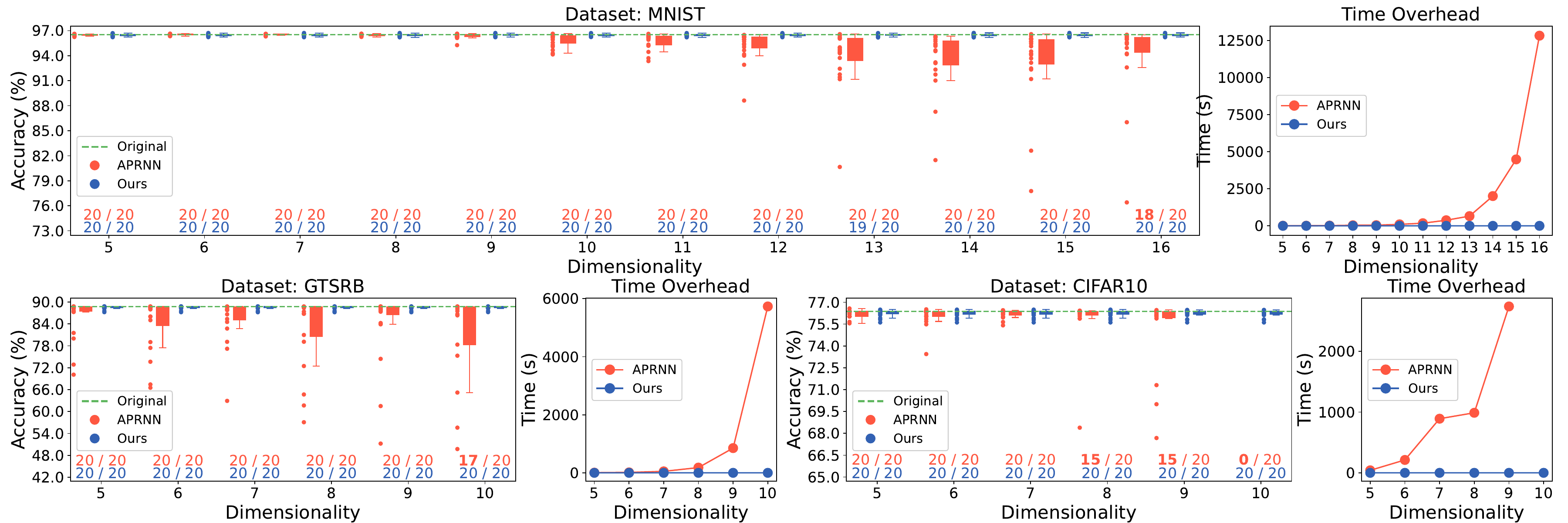}
\centering
\caption{Results of adversarial attack repair. The x-axis represents the dimensionality of the space defined in desired properties. The scatter points on box sides represent the accuracy after each repair task. We conducted the experiments 20 times, each with randomly selected data to construct the desired properties. 
Values above the x-axis show the number of successful/total repairs, with \textcolor[rgb]{0.996, 0.341, 0.255}{red} for APRNN and \textcolor[rgb]{0.196, 0.380, 0.702}{blue} for \name.}
\label{fig:robustness}
\end{figure*}

\begin{tcolorbox}[fonttitle = \bfseries, boxsep=1mm, top=1mm, bottom=1mm, left=1mm, right=1mm]
  \textbf{Remark 1:} 
    Compared to state-of-the-art repair methods, \name demonstrates superior effectiveness and efficiency (with \textbf{5$\times$-2000$\times$ speedup} against provable repair techniques) while maintaining model accuracy in both point-wise corruption and backdoor repair tasks across various settings.
\end{tcolorbox}

\subsubsection{Region-wise Adversarial Attack repair}
\label{sec:main-exp-rob}
We compare our method with APRNN, the SOTA region-wise repair approach.
Here, we focus on repairing a single local space (\(\left|\mathcal{P}\right|=1\)), same as the setup in APRNN (repairing with multiple properties is evaluated in Sec.~\ref{sec:number-repair}). 

Fig.~\ref{fig:robustness} shows that for MNIST, both APRNN and \name performed well when the dimensionality of the input space is less than 10, with a maximum accuracy drop of 1.28\% and 0.41\%, respectively.
However, as dimensionality increases, APRNN exhibits significant accuracy degradation, e.g., with a 16D space, the maximum accuracy drop reaches 20.13\%.
Additionally, among the 20 repair tasks, there was one instance of solving failure due to an infeasible LP and one instance of memory out.
This is because for the region-wise task, it enumerates the vertices of the input space, whose number grows exponentially with the dimensionality (e.g., \(2^{16}\) vertices for the 16D space). 
This overwhelming increase necessitates excessive modifications to the model parameters, leading to poor performance and scalability, and even infeasible instances.
The time costs also increase exponentially, averaging over 12\,000 seconds. 
By comparison, our method demonstrates significantly better scalability, consistently maintaining accuracy and completing repairs in less than 1 second.
On more complex datasets like CIFAR-10 and GTSRB, APRNN struggled further, with repairs failing at higher dimensions and accuracy drops of 7.99\% and 8.70\% at 8 and 9 dimensions, respectively. In comparison, \name leveraged NN verification to capture the model's outputs, thereby avoiding the constraints on activation status at all vertices, which enhances scalability. 
It consistently completes repairs in less than 1 second for CIFAR-10 (1.08\% drop) and in 3 seconds for GTSRB (2\% drop), outperforming APRNN in both time cost and accuracy retention.

\begin{table}[h]
\setlength{\tabcolsep}{2.2pt}
\caption{Results of global safety property repair. \#P represents the number of sub-properties contained in the desired property set \(\mathcal{P}\) after repair.}
\label{table:safety-property}
\centering
\scriptsize
\fontsize{7pt}{8.5pt}\selectfont
\begin{tabular}{c|cr|ccr|c|cr|ccr}
\toprule
& APRNN & Time & Ours & \#P & Time & & APRNN & Time & Ours & \#P & Time \\
\midrule
$N_{2,1}$ & \cellcolor[HTML]{FFCCD0} Fail &\cellcolor[HTML]{FFCCD0} - &\cellcolor{green!20} Succ &\cellcolor{green!20} 112 &\cellcolor{green!20} \textbf{2.99}&
$N_{2,2}$ & \cellcolor[HTML]{FFCCD0} Fail &\cellcolor[HTML]{FFCCD0} - &\cellcolor{green!20} Succ &\cellcolor{green!20} 114 &\cellcolor{green!20} \textbf{3.01}\\
$N_{2,3}$ & \cellcolor[HTML]{FFCCD0} Fail &\cellcolor[HTML]{FFCCD0} - &\cellcolor{green!20} Succ &\cellcolor{green!20} 119 &\cellcolor{green!20} \textbf{3.86}&
$N_{2,4}$ & \cellcolor{green!20} Succ &\cellcolor{green!20} 22.97 &\cellcolor{green!20} Succ &\cellcolor{green!20} 108 &\cellcolor{green!20} \textbf{3.18}\\
$N_{2,5}$ & \cellcolor[HTML]{FFCCD0} Fail &\cellcolor[HTML]{FFCCD0} - &\cellcolor{green!20} Succ &\cellcolor{green!20} 235 &\cellcolor{green!20} \textbf{4.14}&
$N_{2,6}$ & \cellcolor[HTML]{FFCCD0} Fail &\cellcolor[HTML]{FFCCD0} - &\cellcolor{green!20} Succ &\cellcolor{green!20} 22 &\cellcolor{green!20} \textbf{2.26}\\
$N_{2,7}$ & \cellcolor[HTML]{FFCCD0} Fail &\cellcolor[HTML]{FFCCD0} - &\cellcolor{green!20} Succ &\cellcolor{green!20} 114 &\cellcolor{green!20} \textbf{15.91}&
$N_{2,8}$ & \cellcolor[HTML]{FFCCD0} Fail &\cellcolor[HTML]{FFCCD0} - &\cellcolor{green!20} Succ &\cellcolor{green!20} 202 &\cellcolor{green!20} \textbf{3.42}\\
$N_{2,9}$ & \cellcolor[HTML]{FFCCD0} Fail &\cellcolor[HTML]{FFCCD0} - &\cellcolor{green!20} Succ &\cellcolor{green!20} 424 &\cellcolor{green!20} \textbf{7.80}&
$N_{3,1}$ & \cellcolor{green!20} Succ &\cellcolor{green!20} 19.66 &\cellcolor{green!20} Succ &\cellcolor{green!20} 268 &\cellcolor{green!20} \textbf{3.87}\\
$N_{3,2}$ & \cellcolor[HTML]{FFCCD0} Fail &\cellcolor[HTML]{FFCCD0} - &\cellcolor{green!20} Succ &\cellcolor{green!20} 129 &\cellcolor{green!20} \textbf{3.01}&
$N_{3,3}$ & \cellcolor{green!20} Succ &\cellcolor{green!20} 26.05 &\cellcolor{green!20} Succ &\cellcolor{green!20} 102 &\cellcolor{green!20} \textbf{3.16}\\
$N_{3,4}$ & \cellcolor[HTML]{FFCCD0} Fail &\cellcolor[HTML]{FFCCD0} - &\cellcolor{green!20} Succ &\cellcolor{green!20} 32 &\cellcolor{green!20} \textbf{0.90}&
$N_{3,5}$ & \cellcolor[HTML]{FFCCD0} Fail &\cellcolor[HTML]{FFCCD0} - &\cellcolor{green!20} Succ &\cellcolor{green!20} 134 &\cellcolor{green!20} \textbf{3.83}\\
$N_{3,6}$ & \cellcolor[HTML]{FFCCD0} Fail &\cellcolor[HTML]{FFCCD0} - &\cellcolor{green!20} Succ &\cellcolor{green!20} 126 &\cellcolor{green!20} \textbf{3.82}&
$N_{3,7}$ & \cellcolor[HTML]{FFCCD0} Fail &\cellcolor[HTML]{FFCCD0} - &\cellcolor{green!20} Succ &\cellcolor{green!20} 105 &\cellcolor{green!20} \textbf{2.89}\\
$N_{3,8}$ & \cellcolor[HTML]{FFCCD0} Fail &\cellcolor[HTML]{FFCCD0} - &\cellcolor{green!20} Succ &\cellcolor{green!20} 142 &\cellcolor{green!20} \textbf{3.57}&
$N_{3,9}$ & \cellcolor[HTML]{FFCCD0} Fail &\cellcolor[HTML]{FFCCD0} - &\cellcolor{green!20} Succ &\cellcolor{green!20} 445 &\cellcolor{green!20} \textbf{9.78}\\
$N_{4,1}$ & \cellcolor{green!20} Succ &\cellcolor{green!20} 20.67 &\cellcolor{green!20} Succ &\cellcolor{green!20} 139 &\cellcolor{green!20} \textbf{3.24}&
$N_{4,2}$ & \cellcolor[HTML]{FFCCD0} Fail &\cellcolor[HTML]{FFCCD0} - &\cellcolor{green!20} Succ &\cellcolor{green!20} 111 &\cellcolor{green!20} \textbf{3.33}\\
$N_{4,3}$ & \cellcolor[HTML]{FFCCD0} Fail &\cellcolor[HTML]{FFCCD0} - &\cellcolor{green!20} Succ &\cellcolor{green!20} 110 &\cellcolor{green!20} \textbf{4.16}&
$N_{4,4}$ & \cellcolor[HTML]{FFCCD0} Fail &\cellcolor[HTML]{FFCCD0} - &\cellcolor{green!20} Succ &\cellcolor{green!20} 105 &\cellcolor{green!20} \textbf{3.48}\\
$N_{4,5}$ & \cellcolor[HTML]{FFCCD0} Fail &\cellcolor[HTML]{FFCCD0} - &\cellcolor{green!20} Succ &\cellcolor{green!20} 244 &\cellcolor{green!20} \textbf{3.69}&
$N_{4,6}$ & \cellcolor[HTML]{FFCCD0} Fail &\cellcolor[HTML]{FFCCD0} - &\cellcolor{green!20} Succ &\cellcolor{green!20} 117 &\cellcolor{green!20} \textbf{3.52}\\
$N_{4,7}$ & \cellcolor[HTML]{FFCCD0} Fail &\cellcolor[HTML]{FFCCD0} - &\cellcolor{green!20} Succ &\cellcolor{green!20} 112 &\cellcolor{green!20} \textbf{4.15}&
$N_{4,8}$ & \cellcolor{green!20} Succ &\cellcolor{green!20} 22.66 &\cellcolor{green!20} Succ &\cellcolor{green!20} 219 &\cellcolor{green!20} \textbf{3.42}\\
$N_{4,9}$ & \cellcolor{green!20} Succ &\cellcolor{green!20} 22.90 &\cellcolor{green!20} Succ &\cellcolor{green!20} 413 &\cellcolor{green!20} \textbf{20.94}&
$N_{5,1}$ & \cellcolor[HTML]{FFCCD0} Fail &\cellcolor[HTML]{FFCCD0} - &\cellcolor{green!20} Succ &\cellcolor{green!20} 111 &\cellcolor{green!20} \textbf{3.37}\\
$N_{5,2}$ & \cellcolor{green!20} Succ &\cellcolor{green!20} 21.51 &\cellcolor{green!20} Succ &\cellcolor{green!20} 62 &\cellcolor{green!20} \textbf{1.17}&
$N_{5,3}$ & \cellcolor{green!20} Succ &\cellcolor{green!20} 24.65 &\cellcolor{green!20} Succ &\cellcolor{green!20} 37 &\cellcolor{green!20} \textbf{0.94}\\
$N_{5,4}$ & \cellcolor[HTML]{FFCCD0} Fail &\cellcolor[HTML]{FFCCD0} - &\cellcolor{green!20} Succ &\cellcolor{green!20} 125 &\cellcolor{green!20} \textbf{4.13}&
$N_{5,5}$ & \cellcolor[HTML]{FFCCD0} Fail &\cellcolor[HTML]{FFCCD0} - &\cellcolor{green!20} Succ &\cellcolor{green!20} 112 &\cellcolor{green!20} \textbf{3.47}\\
$N_{5,6}$ & \cellcolor[HTML]{FFCCD0} Fail &\cellcolor[HTML]{FFCCD0} - &\cellcolor{green!20} Succ &\cellcolor{green!20} 120 &\cellcolor{green!20} \textbf{3.07}&
$N_{5,7}$ & \cellcolor[HTML]{FFCCD0} Fail &\cellcolor[HTML]{FFCCD0} - &\cellcolor{green!20} Succ &\cellcolor{green!20} 101 &\cellcolor{green!20} \textbf{3.46}\\
$N_{5,8}$ & \cellcolor[HTML]{FFCCD0} Fail &\cellcolor[HTML]{FFCCD0} - &\cellcolor{green!20} Succ &\cellcolor{green!20} 115 &\cellcolor{green!20} \textbf{3.03}&
$N_{5,9}$ & \cellcolor[HTML]{FFCCD0} Fail &\cellcolor[HTML]{FFCCD0} - &\cellcolor{green!20} Succ &\cellcolor{green!20} 115 &\cellcolor{green!20} \textbf{4.67}\\\bottomrule
\end{tabular}
\end{table}

\subsubsection{Region-wise Global Safety Property Repair}
This task aims to repair a network so that the global property is satisfied. 
Note that non-complete verifiers face challenges with certain hard cases in verifying these global properties.
Thus, we invoke a complete verifier~\cite{wang2021beta} whenever the total number of sub-properties doubles, starting with an initial threshold of 100.
We compare APRNN with our method in terms of the number of successful repairs and repair costs. We also record the number of sub-properties refined during the repair process, as the training and test sets of the ACAS Xu network are not publicly available.
Notably, APRNN manually partitions the space into more than 500 subspaces before repair, as enforcing consistent activation status across all vertices in the global space is impractical.
The results shown in Tab.~\ref{table:safety-property} highlight the remarkable improvement on repair capability of our method, as it successfully repairs all 36 instances, significantly outperforming APRNN, which only managed to repair 8 models.
Moreover, \name demonstrates advantages in both the number of refinements and the overhead. 
This is attributed to its targeted counterexample generation and adaptive space refinement during the repair process, so that each sub-space can receive more surgical correction.

\begin{tcolorbox}[fonttitle = \bfseries, boxsep=1.2mm, top=1mm, bottom=1mm, left=2mm, right=2mm]
  \textbf{Remark 2:} \name outperforms current provable methods in handling region-wise tasks, achieving notable improvements in repair capability  and computational efficiency.
\end{tcolorbox}

\subsection{How does the number of properties impact?}
\label{sec:number-repair}
In this section, we investigate how all repair methods perform with a different number of desired properties (i.e., \(\left|\mathcal{P}\right|\)).

\begin{figure}[h]
\includegraphics[width=1.0\columnwidth]{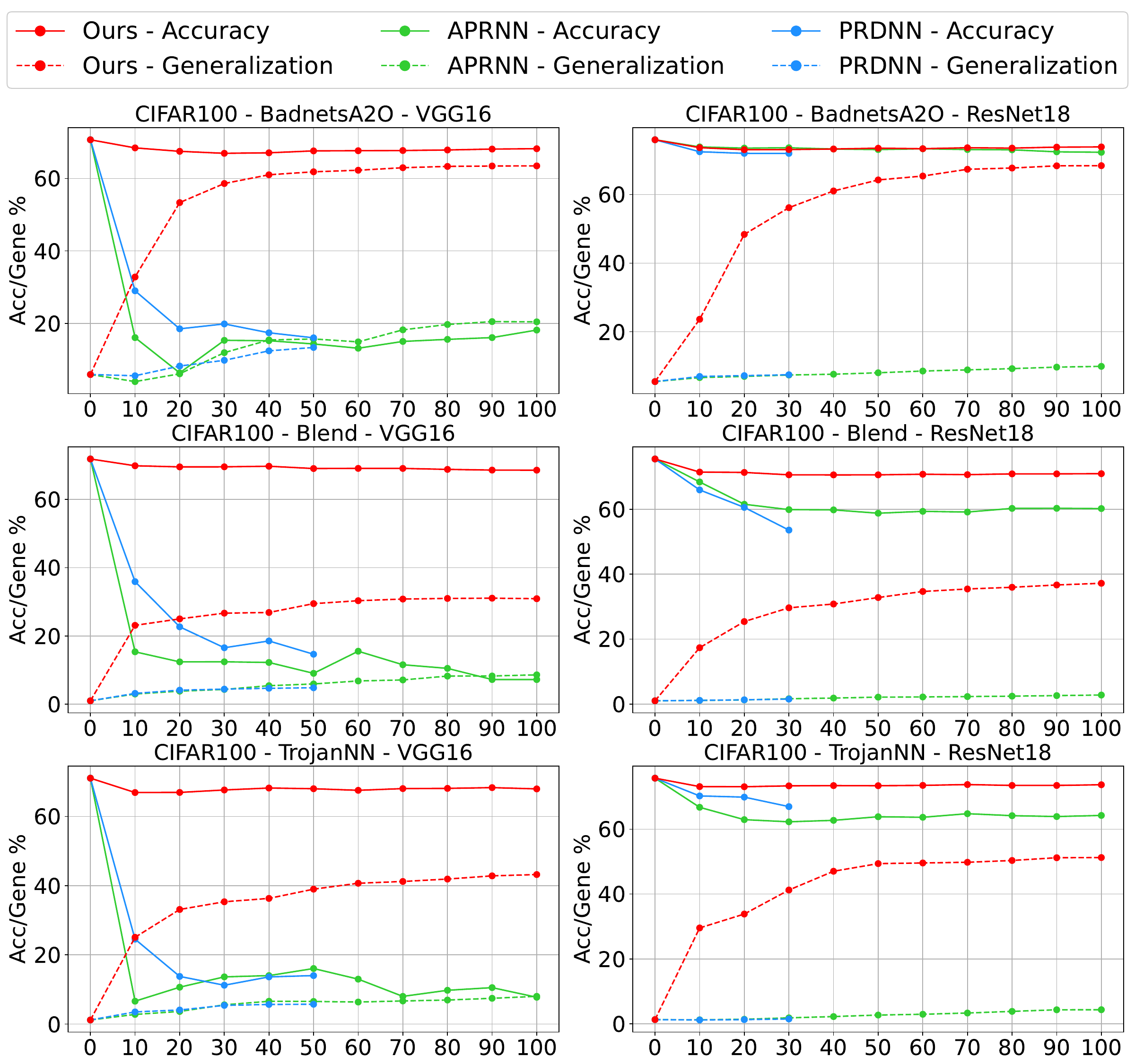}
\centering
\caption{Results of backdoor repair on CIFAR-100 with different numbers of properties (x-axis).}
\label{fig:back_num_cifar100}
\end{figure}

\noindent \textbf{Backdoor.}
We first impose constraints on the number of properties in the backdoor repair scenario, ranging from 10 to 100. 
Results are presented in Fig.\ref{fig:back_num_cifar100} (CIFAR-100) and Fig.\ref{fig:back_num} (full results, Appendix).
For clarity, we omit the PSR metric, as all methods achieve 100\% PSR when the repair is successful. 
However, as the number of properties increases, both APRNN and PRDNN encounter out-of-memory errors or infeasible LP issues (i.e., truncated curves in the figure), particularly on datasets with a large number of classes.
In terms of accuracy, the results highlight that \name exhibits exceptional stability in preserving the original knowledge of the model. 
In comparison, the results of APRNN and PRDNN are unstable, achieving satisfactory accuracy in a few scenarios.
Moreover, we find that their generalization improves as the number of properties increases. 
This may be attributed to the fact that more properties force them to change the wrong behavior of more activation status.
However, the number of activation status associated with the backdoor may far exceed the amount of data available to construct properties, which inherently limits their generalization.
In contrast, \name consistently demonstrates significantly higher generalization across all settings. 
Even with limited properties, it continues to exhibit considerable generalization.

For SEAM and I-REPAIR, we both vary the number of properties and additional training data. 
Specifically, we provide them with 1\%, 2\%, and 5\% of training data.
As shown in Fig.~\ref{fig:back-non-num} (Appendix), both I-REPAIR and our method show an acceptable decline in accuracy. However, I-REPAIR struggles with more covert attack strategies, as neither more properties nor clean data improve its PSR and generalization under the Blend attack. 
On the other hand, SEAM's performance generally improves with more clean data. 
Overall, our method does not rely on additional clean data and outperforms both I-REPAIR and SEAM in most configurations.

\noindent \textbf{Corruption.}
We conduct experiments on corruption repair with varying numbers of properties. 
The results of REASSURE are omitted as it shows 0\% accuracy drop and generalization.
Overall, Fig.~\ref{fig:corr_num} (Appendix) shows that \name remains effective, maintaining stable accuracy and consistently achieving best generalization.

\begin{figure}[!t]

\includegraphics[width=1.0\linewidth]{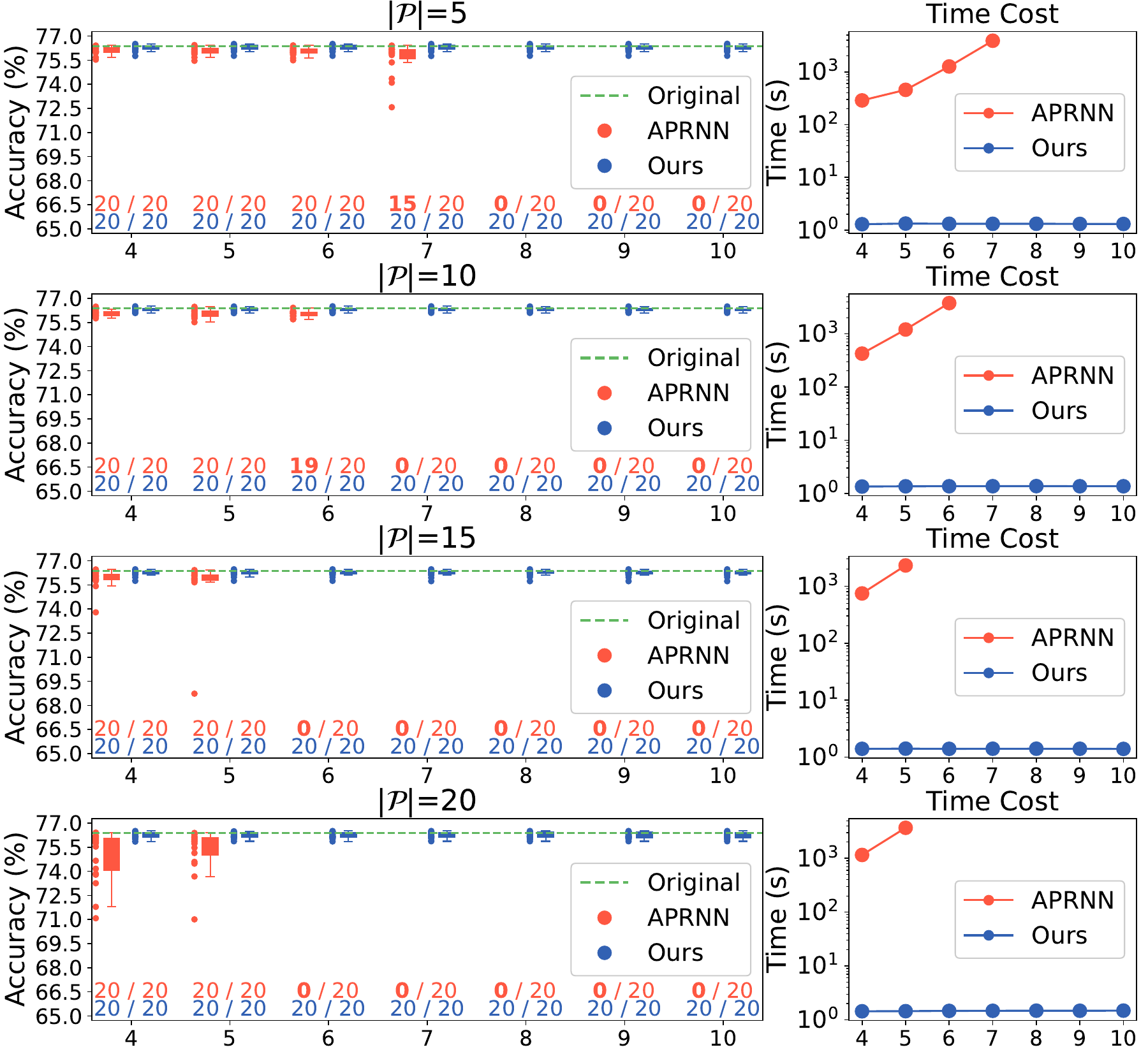}
\centering
\caption{Results of adversarial attack repair with multiple properties on CIFAR-10 (x-axis: space dimensionality).}
\label{fig:robustness-num-C}
\end{figure}

\noindent \textbf{Adversarial Attack.}
We conduct an evaluation on adversarial attack repair with a single property in Sec.~\ref{sec:main-exp-rob}, while in practice a repairer may require the network to exhibit robustness across multiple local spaces. 
So we increase the number of properties involved in the repair, the results are shown in Fig.~\ref{fig:robustness-num-C} and Fig.~\ref{fig:robustness-num-M} (Appendix).
As we can see, APRNN's performance gradually worsens as the number of properties increases. 
Specifically, when handling multiple local spaces, it struggles to scale beyond 5 (11, resp.) dimensions for CIFAR-10 (MNIST), whereas \name shows negligible accuracy loss and remarkable efficiency.

\begin{tcolorbox}[fonttitle = \bfseries, boxsep=1.2mm, top=0mm, bottom=0mm, left=2mm, right=2mm]
  \textbf{Remark 3:} \name continues to outperform baselines with various numbers of desired properties.
\end{tcolorbox}

\begin{figure}[t]
\includegraphics[width=1.0\linewidth]{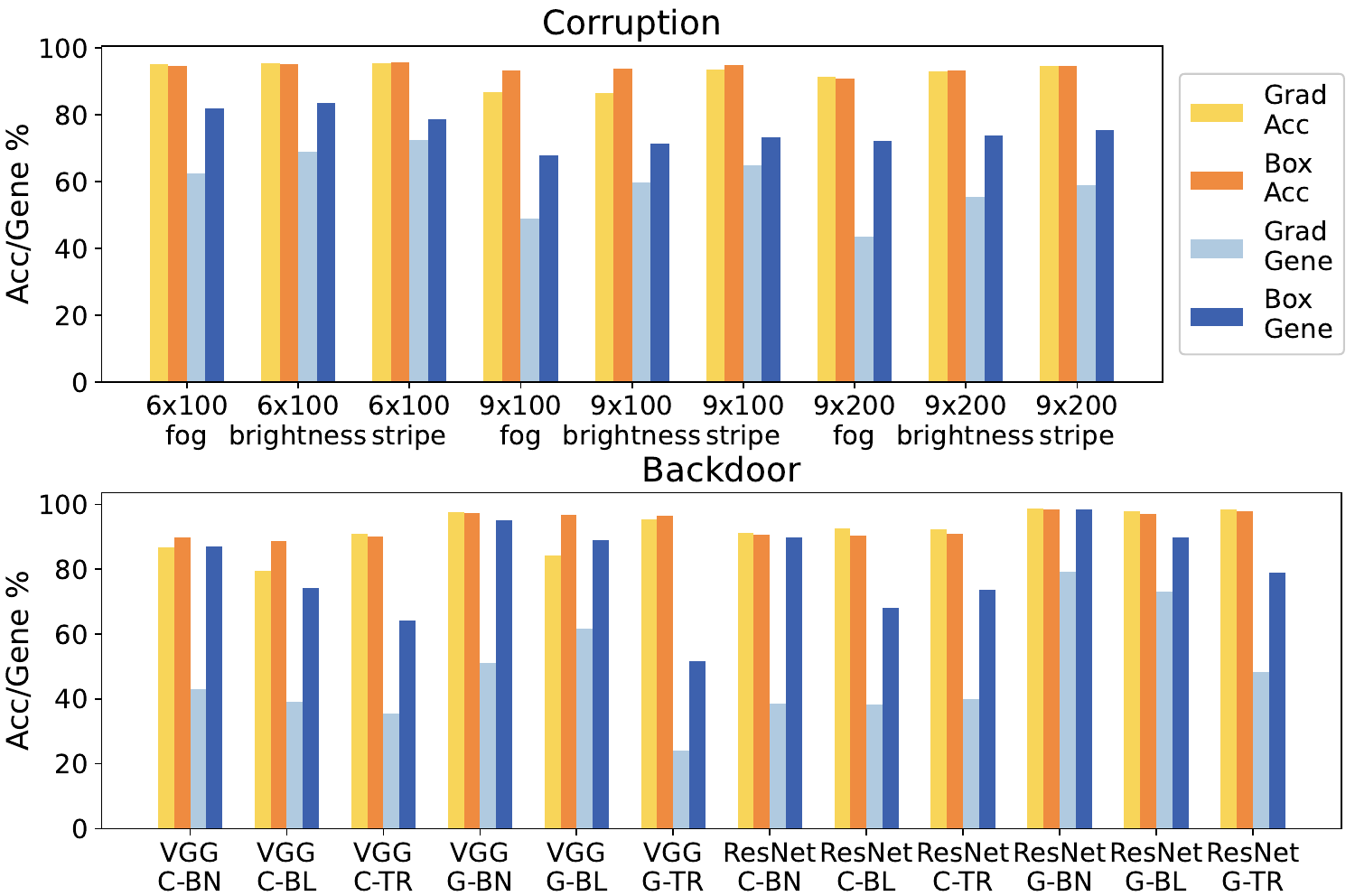}
\centering
\caption{Comparison of proxy box generate algorithm with gradient-based method. C, G, BN, BL and TR denote CIFAR10, GTSRB, BadnetsA2O, Blend and TrojanNN.}
\label{fig:ablation-box}
\end{figure}

\subsection{Ablation Study}
\label{sec:ablation}
\noindent \textbf{The Validity of Proxy Box Synthesis.}
We first investigate the impact of the box synthesis algorithm, which aims to generate proxy boxed to characterize the entire preimages.
Here we consider a gradient-based approach, 
where we optimize a single point \(\bm{h}\) to represent the whole preimage based on its gradients w.r.t. the objective in~\eqref{eq:box-opt}, i.e., 
\(\sum_{\psi \in \phi^{cons}} - \min\left(\bm{c}_{\psi}^\top \cdot f_{\mathrm{c}}(\bm{h})+d_{\psi}, 0\right)\). 
We conduct experiments on corruption and backdoor repair.
The results are shown in Fig.~\ref{fig:ablation-box}.
Overall, both two methods maintain high accuracy, as they preserve the preimage of the feature space. 
However, the box synthesis algorithm shows significantly better generalization across all scenarios. 
Here are two main reasons for this improvement. 
First, it analyzes the model's behavior across the entire box during each update, preventing certain dimensions from falling into local optima and halting updates. 
Second, the center \(\bm{h}_{\phi}^{*}\) returned by the box synthesis method promotes better generalization because the surrounding neighborhood (the entire box) stays within the preimage, which the gradient-based method cannot ensure.

\begin{figure}[t]
\includegraphics[width=1.0\linewidth]{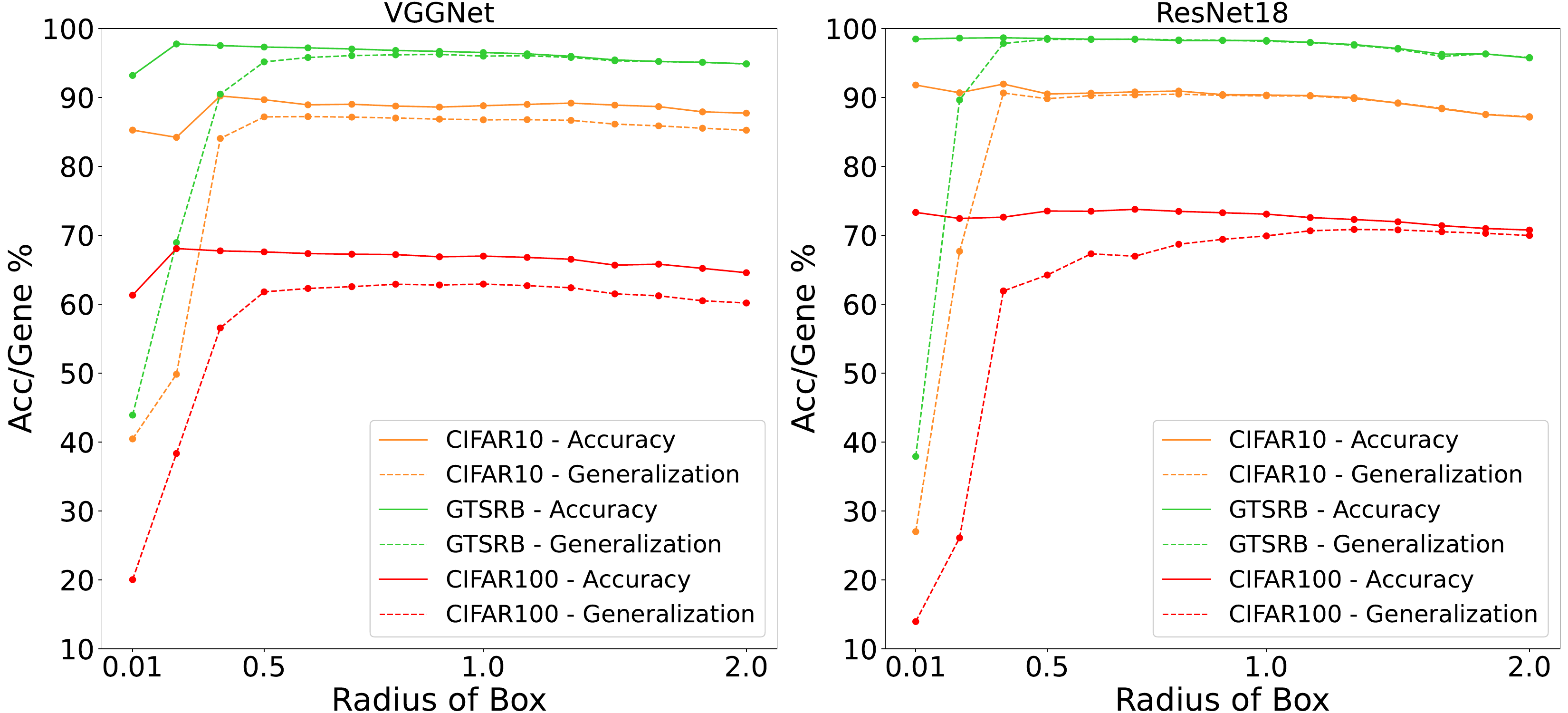}
\centering
\caption{Impact of box radius on repair results.}
\label{fig:ablation-r}
\end{figure}

\noindent \textbf{Proxy Box Radius.}
\label{sec:ablation-radius}
We study the impact of the proxy box radius $r$, the results are shown in Fig.~\ref{fig:ablation-r}.
As the radius of the box increases, we observe that the generalization of the repaired model improves consistently when the radius is relatively small. 
This is because a smaller radius may not be sufficient to jump out of local minima, which further supports the analysis in Section~\ref{sec:point-wise-repair}. 
Generally, the performance of \name remains robust at a moderate radius. 
However, when the radius becomes excessively large (e.g., 2.0), the results show a slight decline, since the larger box increases the relaxation error in the box synthesis process.

\noindent \textbf{The Validity of the Refine Score.}
We further evaluate the effectiveness of the \textit{Refine Score} in region-wise repair, comparing it against a baseline metric that selects refinement dimensions solely by the magnitude of the input range. 
On the global safety property repair task, the baseline repairs 34 out of 36 models with 357.7 steps and 7.5 seconds on average. In comparison, the \textit{Refine Score} repairs all 36 models with only 150.0 steps and 4.4 seconds, demonstrating clear superiority in both efficiency and effectiveness.

\subsection{Scalability}
\subsubsection{Other activation functions}
To evaluate the scalability of all repair methods, we conducted a set of backdoor repair experiments under the BadnetsA2O attack, where we train NNs with various activation functions.
Specifically, we consider Leaky ReLU, GeLU, and Tanh.
The former is a piecewise linear function, while the latter two are more complex non-linear functions. 
PRDNN and APRNN are excluded as their implementation only supports ReLU activation at this model scale.
As shown in Tab.~\ref{table:backdoor-other-act}, I-REPAIR exhibits unstable results across different datasets, while SEAM fails to maintain accuracy. 
In contrast, our method demonstrates impressive robustness when handling different activation functions.

\subsubsection{High dimensional region}
In Sec.~\ref{sec:main-exp-rob}, we assess APRNN and \name on repairing relatively low-dimensional input spaces.
Here we extend the evaluation to higher-dimensional regions, where APRNN faces significant challenges due to its reliance on enumerating the vertices of the input space.
For each dimensionality, we perform 50 repair tasks on CIFAR-10 and record the number of successes along with the average accuracy after repair. 
Tab.~\ref{table:high-dim-region} shows that \name maintains a high success rate and preserves model performance even in spaces with dimensions up to 18 times higher than those that APRNN can handle (up to 9 for CIFAR-10).

To further explore the scalability of \name, we conduct an additional experiment where all input dimensions are perturbed.
Specifically, on the MNIST dataset, we construct a 784-dimensional space by allowing all pixels to be perturbed within a radius of $10^{-3}$.
This setting is challenging due to increased approximation errors in such high-dimensional regions, which hinder both counterexample generation and verification.
Nevertheless, we run 50 repair tasks under this setting and observe that \name successfully repairs 49 of them.
In contrast, existing baselines such as APRNN are hardly feasible under these settings, as they rely on explicit vertex enumeration and scale only to 10–20 dimensions.

\begin{tcolorbox}[fonttitle = \bfseries, boxsep=1.5mm, top=1mm, bottom=1mm, left=2mm, right=2mm]
  \textbf{Remark 4:} \name greatly benefits from the box synthesis method and the refine score; it is robust to the proxy box radius and can scale to various activation functions and higher-dimensional spaces (18-fold improvement over SOTA).
\end{tcolorbox}


\begin{table}[t]
\setlength{\tabcolsep}{4pt}
\caption{Repair with different activation functions.}
\label{table:backdoor-other-act}
\centering
\footnotesize
\begin{tabular}{cc|cccc|cccc}
\toprule
\multirow{2}{*}{\makecell{Activation \\ Function}} &  \multirow{2}{*}{Method} & \multicolumn{4}{c|}{GTSRB}  & \multicolumn{4}{c}{CIFAR-10}  \\ 
& & PSR & Acc & Gene & T  & PSR & Acc & Gene & T  \\
\midrule
\multirow{4}{*}{\makecell{Leaky \\ ReLU}} &  Original & 0 & 97.32 & 7.63 & -& 0 & 91.87 & 12.79 & -\\
& {SEAM} & 96 & 64.12 & 50.32 & 1.6 & 58 & 71.80 & \underline{47.36} & \underline{2.7} \\
& {IREPAIR} & \textbf{100} & \underline{96.84} & \underline{67.09} & \underline{1.3} & 36 & \underline{87.59} & 27.01 & 6.0 \\
& {Ours} & \textbf{100} & \textbf{97.48} & \textbf{92.24} & \textbf{0.7} & \textbf{100} & \textbf{90.00} & \textbf{88.82} & \textbf{0.7} \\
\cmidrule{2-10}
\multirow{4}{*}{GeLU} &  Original & 0 & 97.50 & 7.69 & -& 0 & 90.83 & 13.21 & -\\
& {SEAM} & 92 & 53.81 & 49.11 & 1.8 & 56 & 69.36 & \underline{52.34} & \textbf{0.5} \\
& {IREPAIR} & \textbf{100} & \underline{92.89} & \underline{75.08} & \underline{1.5} & 22 & \underline{83.67} & 19.69 & 6.9 \\
& {Ours} & \textbf{100} & \textbf{96.07} & \textbf{92.16} & \textbf{1.1} & \textbf{100} & \textbf{88.96} & \textbf{74.02} & \underline{1.3} \\
\cmidrule{2-10}

 \multirow{4}{*}{Tanh} &  Original & 0 & 97.35 & 7.72 & -& 0 & 89.32 & 13.04 & -\\
& {SEAM} & 92 & 61.66 & 54.29 & \underline{0.7} & 56 & 54.58 & 36.13 & \textbf{0.5} \\
& {IREPAIR} & \textbf{100} & \textbf{97.51} & \underline{66.41} & \textbf{0.4} & \textbf{100} & \underline{82.90} & \underline{72.64} & 2.2 \\
& {Ours} & \textbf{100} & \underline{95.95} & \textbf{95.58} & 0.8 & \textbf{100} & \textbf{85.46} & \textbf{82.28} & \underline{1.1} \\
\bottomrule
\end{tabular}
\end{table}

\begin{table}[t]
\setlength{\tabcolsep}{4pt}
\caption{Results of repair with high dimensional region.}
\label{table:high-dim-region}
\centering
\footnotesize
\begin{tabular}{c|ccr|c|ccr}
\toprule
Dim & \#Success & Accuracy &  T  & Dim &\#Success & Accuracy &  T \\
\midrule
20 &50 / 50 & 76.270$\pm$0.039 & 1.3 & 40 &50 / 50 & 76.265$\pm$0.039 & 1.3 \\
60 &49 / 50 & 76.275$\pm$0.035 & 1.3 & 80 &49 / 50 & 76.272$\pm$0.039 & 1.7 \\
100 &48 / 50 & 76.269$\pm$0.040 & 1.5 & 120 &48 / 50 & 76.271$\pm$0.038 & 1.8 \\
140 &41 / 50 & 76.265$\pm$0.041 & 0.9 & 160 &41 / 50 & 76.250$\pm$0.046 & 2.1 \\
\bottomrule
\end{tabular}
\end{table}

\section{Discussion}
\noindent \textbf{Limitations.}
\name integrates a sound verifier to provide guarantees for region-wise repair and exhibits significantly improved scalability compared to prior provable methods.
However, its scalability is fundamentally limited by the verifier itself, as it requires multiple invocations for verification. 
For example, when repairing high-dimensional spaces that are challenging for the existing verifier, \name may engage in excessive property refinement but still fail to pass verification (few cases in Tab.~\ref{table:high-dim-region}). 
To further explore this issue, we incorporate different linear relaxation method based verifiers into \name for detailed analysis.
Tab.~\ref{table:more-verifier} (Appendix) shows that integrating stronger verifier significantly enhances it's capability to handle higher-dimensional spaces. 
We believe that using advanced verifiers will help \name scale further.

\noindent \textbf{Future work.}
While we apply \name to four different repair tasks, additional types of failures, such as fairness issue, remain to be explored.
Moreover, this work limits the input space for the desired properties to be a bounded box. 
It is worth exploring how \name can be extended to handle properties defined over other types of input specifications such as $\ell_2$-balls.
Incorporating verifiers specialized for such regions (e.g., SDP-CROWN~\cite{chiusdp}) may yield tighter bounds and directly benefit the repair process.
On the other hand, how to refine such regions becomes a core challenge, as traditional space partitioning does not directly apply.
One possible approach is to iteratively add linear constraints through the center to partition the ball.
We leave these extensions to future work to further enhance \name's versatility.

\section{Conclusion}
In this paper, we propose \name, a provable neural network repair framework.
The core idea is to synthesize small proxy boxes to effectively characterize the preimages and further guide the repair.
\name further employs an adaptive approach that includes counterexample generation and property refinement to address more practical tasks.
Thorough evaluation across diverse repair tasks demonstrate that \name exhibits remarkable effectiveness, efficiency and scalability compared to existing repair methods.

\begin{acks}
This work is supported by the National Natural Science Foundation of China (U21B2001), the Key R\&D Program of Zhejiang (2022C01018), and the CCF-Huawei Populus Grove Fund (CCF-HuaweiFM2024003).
\end{acks}

\bibliographystyle{ACM-Reference-Format}
\balance
\bibliography{main}


\begin{thebibliography}{71}


\ifx \showCODEN    \undefined \def \showCODEN     #1{\unskip}     \fi
\ifx \showISBNx    \undefined \def \showISBNx     #1{\unskip}     \fi
\ifx \showISBNxiii \undefined \def \showISBNxiii  #1{\unskip}     \fi
\ifx \showISSN     \undefined \def \showISSN      #1{\unskip}     \fi
\ifx \showLCCN     \undefined \def \showLCCN      #1{\unskip}     \fi
\ifx \shownote     \undefined \def \shownote      #1{#1}          \fi
\ifx \showarticletitle \undefined \def \showarticletitle #1{#1}   \fi
\ifx \showURL      \undefined \def \showURL       {\relax}        \fi
\providecommand\bibfield[2]{#2}
\providecommand\bibinfo[2]{#2}
\providecommand\natexlab[1]{#1}
\providecommand\showeprint[2][]{arXiv:#2}

\bibitem[Barni et~al\mbox{.}(2019)]%
        {barni2019new}
\bibfield{author}{\bibinfo{person}{Mauro Barni}, \bibinfo{person}{Kassem Kallas}, {and} \bibinfo{person}{Benedetta Tondi}.} \bibinfo{year}{2019}\natexlab{}.
\newblock \showarticletitle{A new backdoor attack in cnns by training set corruption without label poisoning}. In \bibinfo{booktitle}{\emph{2019 IEEE International Conference on Image Processing (ICIP)}}. IEEE, \bibinfo{pages}{101--105}.
\newblock


\bibitem[Batten et~al\mbox{.}(2021)]%
        {batten2021efficient}
\bibfield{author}{\bibinfo{person}{Ben Batten}, \bibinfo{person}{Panagiotis Kouvaros}, \bibinfo{person}{Alessio Lomuscio}, {and} \bibinfo{person}{Yang Zheng}.} \bibinfo{year}{2021}\natexlab{}.
\newblock \showarticletitle{Efficient neural network verification via layer-based semidefinite relaxations and linear cuts}. In \bibinfo{booktitle}{\emph{International Joint Conference on Artificial Intelligence}}. \bibinfo{pages}{2184--2190}.
\newblock


\bibitem[Boyd and Vandenberghe(2004)]%
        {boyd2004convex}
\bibfield{author}{\bibinfo{person}{Stephen~P Boyd} {and} \bibinfo{person}{Lieven Vandenberghe}.} \bibinfo{year}{2004}\natexlab{}.
\newblock \bibinfo{booktitle}{\emph{Convex optimization}}.
\newblock \bibinfo{publisher}{Cambridge university press}.
\newblock


\bibitem[Chen et~al\mbox{.}(2024a)]%
        {chen2024isolation}
\bibfield{author}{\bibinfo{person}{Jialuo Chen}, \bibinfo{person}{Jingyi Wang}, \bibinfo{person}{Youcheng Sun}, \bibinfo{person}{Peng Cheng}, {and} \bibinfo{person}{Jiming Chen}.} \bibinfo{year}{2024}\natexlab{a}.
\newblock \showarticletitle{Isolation-Based Debugging for Neural Networks}. In \bibinfo{booktitle}{\emph{Proceedings of the 33rd ACM SIGSOFT International Symposium on Software Testing and Analysis}}. \bibinfo{pages}{338--349}.
\newblock


\bibitem[Chen et~al\mbox{.}(2017)]%
        {chen2017targeted}
\bibfield{author}{\bibinfo{person}{Xinyun Chen}, \bibinfo{person}{Chang Liu}, \bibinfo{person}{Bo Li}, \bibinfo{person}{Kimberly Lu}, {and} \bibinfo{person}{Dawn Song}.} \bibinfo{year}{2017}\natexlab{}.
\newblock \showarticletitle{Targeted backdoor attacks on deep learning systems using data poisoning}.
\newblock \bibinfo{journal}{\emph{arXiv preprint arXiv:1712.05526}} (\bibinfo{year}{2017}).
\newblock


\bibitem[Chen et~al\mbox{.}(2024b)]%
        {chen2024interpretability}
\bibfield{author}{\bibinfo{person}{Zuohui Chen}, \bibinfo{person}{Jun Zhou}, \bibinfo{person}{Youcheng Sun}, \bibinfo{person}{Jingyi Wang}, \bibinfo{person}{Qi Xuan}, {and} \bibinfo{person}{Xiaoniu Yang}.} \bibinfo{year}{2024}\natexlab{b}.
\newblock \showarticletitle{Interpretability Based Neural Network Repair}. In \bibinfo{booktitle}{\emph{Proceedings of the 33rd ACM SIGSOFT International Symposium on Software Testing and Analysis}}. \bibinfo{pages}{908--919}.
\newblock


\bibitem[Chi et~al\mbox{.}(2025)]%
        {chi2025patch}
\bibfield{author}{\bibinfo{person}{Zhiming Chi}, \bibinfo{person}{Jianan Ma}, \bibinfo{person}{Pengfei Yang}, \bibinfo{person}{Cheng-Chao Huang}, \bibinfo{person}{Renjue Li}, \bibinfo{person}{Jingyi Wang}, \bibinfo{person}{Xiaowe Huang}, {and} \bibinfo{person}{Lijun Zhang}.} \bibinfo{year}{2025}\natexlab{}.
\newblock \showarticletitle{Patch Synthesis for Property Repair of Deep Neural Networks}. In \bibinfo{booktitle}{\emph{2025 IEEE/ACM 47th International Conference on Software Engineering (ICSE)}}. IEEE, \bibinfo{pages}{1191--1203}.
\newblock


\bibitem[Chiu et~al\mbox{.}({[n.\,d.]})]%
        {chiusdp}
\bibfield{author}{\bibinfo{person}{Hong-Ming Chiu}, \bibinfo{person}{Hao Chen}, \bibinfo{person}{Huan Zhang}, {and} \bibinfo{person}{Richard~Y Zhang}.} \bibinfo{year}{[n.\,d.]}\natexlab{}.
\newblock \showarticletitle{SDP-CROWN: Efficient Bound Propagation for Neural Network Verification with Tightness of Semidefinite Programming}. In \bibinfo{booktitle}{\emph{Forty-second International Conference on Machine Learning}}.
\newblock


\bibitem[Cullen et~al\mbox{.}(2024)]%
        {cullen2024s}
\bibfield{author}{\bibinfo{person}{Andrew~C Cullen}, \bibinfo{person}{Paul Montague}, \bibinfo{person}{Shijie Liu}, \bibinfo{person}{Sarah~M Erfani}, {and} \bibinfo{person}{Benjamin~IP Rubinstein}.} \bibinfo{year}{2024}\natexlab{}.
\newblock \showarticletitle{It’s Simplex! Disaggregating Measures to Improve Certified Robustness}. In \bibinfo{booktitle}{\emph{2024 IEEE Symposium on Security and Privacy (SP)}}. IEEE, \bibinfo{pages}{2886--2900}.
\newblock


\bibitem[Dong et~al\mbox{.}(2021)]%
        {dong2021towards}
\bibfield{author}{\bibinfo{person}{Guoliang Dong}, \bibinfo{person}{Jun Sun}, \bibinfo{person}{Xingen Wang}, \bibinfo{person}{Xinyu Wang}, {and} \bibinfo{person}{Ting Dai}.} \bibinfo{year}{2021}\natexlab{}.
\newblock \showarticletitle{Towards repairing neural networks correctly}. In \bibinfo{booktitle}{\emph{2021 IEEE 21st International Conference on Software Quality, Reliability and Security (QRS)}}. IEEE, \bibinfo{pages}{714--725}.
\newblock


\bibitem[Ehlers(2017)]%
        {ehlers2017formal}
\bibfield{author}{\bibinfo{person}{Ruediger Ehlers}.} \bibinfo{year}{2017}\natexlab{}.
\newblock \showarticletitle{Formal verification of piece-wise linear feed-forward neural networks}. In \bibinfo{booktitle}{\emph{International Symposium on Automated Technology for Verification and Analysis}}. Springer, \bibinfo{pages}{269--286}.
\newblock


\bibitem[Eykholt et~al\mbox{.}(2018)]%
        {eykholt2018robust}
\bibfield{author}{\bibinfo{person}{Kevin Eykholt}, \bibinfo{person}{Ivan Evtimov}, \bibinfo{person}{Earlence Fernandes}, \bibinfo{person}{Bo Li}, \bibinfo{person}{Amir Rahmati}, \bibinfo{person}{Chaowei Xiao}, \bibinfo{person}{Atul Prakash}, \bibinfo{person}{Tadayoshi Kohno}, {and} \bibinfo{person}{Dawn Song}.} \bibinfo{year}{2018}\natexlab{}.
\newblock \showarticletitle{Robust physical-world attacks on deep learning visual classification}. In \bibinfo{booktitle}{\emph{Proceedings of the IEEE conference on computer vision and pattern recognition}}. \bibinfo{pages}{1625--1634}.
\newblock


\bibitem[Fu and Li(2022)]%
        {reassure}
\bibfield{author}{\bibinfo{person}{Feisi Fu} {and} \bibinfo{person}{Wenchao Li}.} \bibinfo{year}{2022}\natexlab{}.
\newblock \showarticletitle{Sound and Complete Neural Network Repair with Minimality and Locality Guarantees}. In \bibinfo{booktitle}{\emph{The Tenth International Conference on Learning Representations, {ICLR} 2022, Virtual Event, April 25-29, 2022}}. \bibinfo{publisher}{OpenReview.net}.
\newblock


\bibitem[Gehr et~al\mbox{.}(2018)]%
        {gehr2018ai2}
\bibfield{author}{\bibinfo{person}{Timon Gehr}, \bibinfo{person}{Matthew Mirman}, \bibinfo{person}{Dana Drachsler-Cohen}, \bibinfo{person}{Petar Tsankov}, \bibinfo{person}{Swarat Chaudhuri}, {and} \bibinfo{person}{Martin Vechev}.} \bibinfo{year}{2018}\natexlab{}.
\newblock \showarticletitle{Ai2: Safety and robustness certification of neural networks with abstract interpretation}. In \bibinfo{booktitle}{\emph{2018 IEEE symposium on security and privacy (SP)}}. IEEE, \bibinfo{pages}{3--18}.
\newblock


\bibitem[Gong et~al\mbox{.}(2023)]%
        {gong2023redeem}
\bibfield{author}{\bibinfo{person}{Xueluan Gong}, \bibinfo{person}{Yanjiao Chen}, \bibinfo{person}{Wang Yang}, \bibinfo{person}{Qian Wang}, \bibinfo{person}{Yuzhe Gu}, \bibinfo{person}{Huayang Huang}, {and} \bibinfo{person}{Chao Shen}.} \bibinfo{year}{2023}\natexlab{}.
\newblock \showarticletitle{Redeem myself: Purifying backdoors in deep learning models using self attention distillation}. In \bibinfo{booktitle}{\emph{2023 IEEE Symposium on Security and Privacy (SP)}}. IEEE, \bibinfo{pages}{755--772}.
\newblock


\bibitem[Goodfellow et~al\mbox{.}(2014)]%
        {goodfellow2014explaining}
\bibfield{author}{\bibinfo{person}{Ian~J Goodfellow}, \bibinfo{person}{Jonathon Shlens}, {and} \bibinfo{person}{Christian Szegedy}.} \bibinfo{year}{2014}\natexlab{}.
\newblock \showarticletitle{Explaining and harnessing adversarial examples}.
\newblock \bibinfo{journal}{\emph{arXiv preprint arXiv:1412.6572}} (\bibinfo{year}{2014}).
\newblock


\bibitem[Gu et~al\mbox{.}(2019)]%
        {gu2019badnets}
\bibfield{author}{\bibinfo{person}{Tianyu Gu}, \bibinfo{person}{Kang Liu}, \bibinfo{person}{Brendan Dolan-Gavitt}, {and} \bibinfo{person}{Siddharth Garg}.} \bibinfo{year}{2019}\natexlab{}.
\newblock \showarticletitle{Badnets: Evaluating backdooring attacks on deep neural networks}.
\newblock \bibinfo{journal}{\emph{IEEE Access}}  \bibinfo{volume}{7} (\bibinfo{year}{2019}), \bibinfo{pages}{47230--47244}.
\newblock


\bibitem[Guo et~al\mbox{.}(2022)]%
        {guo2022improving}
\bibfield{author}{\bibinfo{person}{Yong Guo}, \bibinfo{person}{David Stutz}, {and} \bibinfo{person}{Bernt Schiele}.} \bibinfo{year}{2022}\natexlab{}.
\newblock \showarticletitle{Improving robustness by enhancing weak subnets}. In \bibinfo{booktitle}{\emph{European Conference on Computer Vision}}. Springer, \bibinfo{pages}{320--338}.
\newblock


\bibitem[He et~al\mbox{.}(2016)]%
        {resnet}
\bibfield{author}{\bibinfo{person}{Kaiming He}, \bibinfo{person}{Xiangyu Zhang}, \bibinfo{person}{Shaoqing Ren}, {and} \bibinfo{person}{Jian Sun}.} \bibinfo{year}{2016}\natexlab{}.
\newblock \showarticletitle{Deep residual learning for image recognition}. In \bibinfo{booktitle}{\emph{Proceedings of the IEEE conference on computer vision and pattern recognition}}. \bibinfo{pages}{770--778}.
\newblock


\bibitem[Hendrycks and Dietterich(2019)]%
        {hendrycks2019benchmarking}
\bibfield{author}{\bibinfo{person}{Dan Hendrycks} {and} \bibinfo{person}{Thomas Dietterich}.} \bibinfo{year}{2019}\natexlab{}.
\newblock \showarticletitle{Benchmarking neural network robustness to common corruptions and perturbations}.
\newblock \bibinfo{journal}{\emph{arXiv preprint arXiv:1903.12261}} (\bibinfo{year}{2019}).
\newblock


\bibitem[Hendrycks et~al\mbox{.}(2019)]%
        {hendrycks2019augmix}
\bibfield{author}{\bibinfo{person}{Dan Hendrycks}, \bibinfo{person}{Norman Mu}, \bibinfo{person}{Ekin~D Cubuk}, \bibinfo{person}{Barret Zoph}, \bibinfo{person}{Justin Gilmer}, {and} \bibinfo{person}{Balaji Lakshminarayanan}.} \bibinfo{year}{2019}\natexlab{}.
\newblock \showarticletitle{Augmix: A simple data processing method to improve robustness and uncertainty}.
\newblock \bibinfo{journal}{\emph{arXiv preprint arXiv:1912.02781}} (\bibinfo{year}{2019}).
\newblock


\bibitem[Henriksen et~al\mbox{.}(2022)]%
        {henriksen2022repairing}
\bibfield{author}{\bibinfo{person}{Patrick Henriksen}, \bibinfo{person}{Francesco Leofante}, {and} \bibinfo{person}{Alessio Lomuscio}.} \bibinfo{year}{2022}\natexlab{}.
\newblock \showarticletitle{Repairing misclassifications in neural networks using limited data}. In \bibinfo{booktitle}{\emph{Proceedings of the 37th ACM/SIGAPP Symposium on Applied Computing}}. \bibinfo{pages}{1031--1038}.
\newblock


\bibitem[Huang et~al\mbox{.}(2017)]%
        {huang2017safety}
\bibfield{author}{\bibinfo{person}{Xiaowei Huang}, \bibinfo{person}{Marta Kwiatkowska}, \bibinfo{person}{Sen Wang}, {and} \bibinfo{person}{Min Wu}.} \bibinfo{year}{2017}\natexlab{}.
\newblock \showarticletitle{Safety verification of deep neural networks}. In \bibinfo{booktitle}{\emph{Computer Aided Verification: 29th International Conference, CAV 2017, Heidelberg, Germany, July 24-28, 2017, Proceedings, Part I 30}}. Springer, \bibinfo{pages}{3--29}.
\newblock


\bibitem[Julian et~al\mbox{.}(2019)]%
        {julian2019verifying}
\bibfield{author}{\bibinfo{person}{Kyle~D Julian}, \bibinfo{person}{Shivam Sharma}, \bibinfo{person}{Jean-Baptiste Jeannin}, {and} \bibinfo{person}{Mykel~J Kochenderfer}.} \bibinfo{year}{2019}\natexlab{}.
\newblock \showarticletitle{Verifying aircraft collision avoidance neural networks through linear approximations of safe regions}.
\newblock \bibinfo{journal}{\emph{arXiv preprint arXiv:1903.00762}} (\bibinfo{year}{2019}).
\newblock


\bibitem[Katz et~al\mbox{.}(2017a)]%
        {katz2017reluplex}
\bibfield{author}{\bibinfo{person}{Guy Katz}, \bibinfo{person}{Clark Barrett}, \bibinfo{person}{David~L Dill}, \bibinfo{person}{Kyle Julian}, {and} \bibinfo{person}{Mykel~J Kochenderfer}.} \bibinfo{year}{2017}\natexlab{a}.
\newblock \showarticletitle{Reluplex: An efficient SMT solver for verifying deep neural networks}. In \bibinfo{booktitle}{\emph{Computer Aided Verification: 29th International Conference, CAV 2017, Heidelberg, Germany, July 24-28, 2017, Proceedings, Part I 30}}. Springer, \bibinfo{pages}{97--117}.
\newblock


\bibitem[Katz et~al\mbox{.}(2017b)]%
        {reluplex}
\bibfield{author}{\bibinfo{person}{Guy Katz}, \bibinfo{person}{Clark~W. Barrett}, \bibinfo{person}{David~L. Dill}, \bibinfo{person}{Kyle Julian}, {and} \bibinfo{person}{Mykel~J. Kochenderfer}.} \bibinfo{year}{2017}\natexlab{b}.
\newblock \showarticletitle{Reluplex: An Efficient {SMT} Solver for Verifying Deep Neural Networks}. In \bibinfo{booktitle}{\emph{CAV 2017}}. \bibinfo{publisher}{Springer}, \bibinfo{address}{Heidelberg, Germany}, \bibinfo{pages}{97--117}.
\newblock


\bibitem[Katz et~al\mbox{.}(2019)]%
        {marabou}
\bibfield{author}{\bibinfo{person}{Guy Katz}, \bibinfo{person}{Derek~A Huang}, \bibinfo{person}{Duligur Ibeling}, \bibinfo{person}{Kyle Julian}, \bibinfo{person}{Christopher Lazarus}, \bibinfo{person}{Rachel Lim}, \bibinfo{person}{Parth Shah}, \bibinfo{person}{Shantanu Thakoor}, \bibinfo{person}{Haoze Wu}, \bibinfo{person}{Aleksandar Zelji{\'c}}, {et~al\mbox{.}}} \bibinfo{year}{2019}\natexlab{}.
\newblock \showarticletitle{The marabou framework for verification and analysis of deep neural networks}. In \bibinfo{booktitle}{\emph{Computer Aided Verification: 31st International Conference, CAV 2019, New York City, NY, USA, July 15-18, 2019, Proceedings, Part I 31}}. Springer, \bibinfo{pages}{443--452}.
\newblock


\bibitem[Kotha et~al\mbox{.}(2023)]%
        {kotha2023provably}
\bibfield{author}{\bibinfo{person}{Suhas Kotha}, \bibinfo{person}{Christopher Brix}, \bibinfo{person}{J~Zico Kolter}, \bibinfo{person}{Krishnamurthy Dvijotham}, {and} \bibinfo{person}{Huan Zhang}.} \bibinfo{year}{2023}\natexlab{}.
\newblock \showarticletitle{Provably bounding neural network preimages}.
\newblock \bibinfo{journal}{\emph{Advances in Neural Information Processing Systems}}  \bibinfo{volume}{36} (\bibinfo{year}{2023}), \bibinfo{pages}{80270--80290}.
\newblock


\bibitem[Krizhevsky et~al\mbox{.}(2009)]%
        {cifar}
\bibfield{author}{\bibinfo{person}{Alex Krizhevsky}, \bibinfo{person}{Geoffrey Hinton}, {et~al\mbox{.}}} \bibinfo{year}{2009}\natexlab{}.
\newblock \showarticletitle{Learning multiple layers of features from tiny images}.
\newblock  (\bibinfo{year}{2009}).
\newblock


\bibitem[Levi and Kontorovich(2024)]%
        {levi2024splitting}
\bibfield{author}{\bibinfo{person}{Matan Levi} {and} \bibinfo{person}{Aryeh Kontorovich}.} \bibinfo{year}{2024}\natexlab{}.
\newblock \showarticletitle{Splitting the difference on adversarial training}. In \bibinfo{booktitle}{\emph{33rd USENIX Security Symposium (USENIX Security 24)}}. \bibinfo{pages}{3639--3656}.
\newblock


\bibitem[Liang et~al\mbox{.}(2025)]%
        {liang2025birdnn}
\bibfield{author}{\bibinfo{person}{Zhen Liang}, \bibinfo{person}{Taoran Wu}, \bibinfo{person}{Changyuan Zhao}, \bibinfo{person}{Wanwei Liu}, \bibinfo{person}{Bai Xue}, \bibinfo{person}{Wenjing Yang}, \bibinfo{person}{Ji Wang}, {and} \bibinfo{person}{Wanrong Huang}.} \bibinfo{year}{2025}\natexlab{}.
\newblock \showarticletitle{BIRDNN: Behavior-Imitation Based Repair for Deep Neural Networks}.
\newblock \bibinfo{journal}{\emph{Neural Networks}}  \bibinfo{volume}{183} (\bibinfo{year}{2025}), \bibinfo{pages}{106949}.
\newblock


\bibitem[Liu et~al\mbox{.}(2019)]%
        {liu2019abs}
\bibfield{author}{\bibinfo{person}{Yingqi Liu}, \bibinfo{person}{Wen-Chuan Lee}, \bibinfo{person}{Guanhong Tao}, \bibinfo{person}{Shiqing Ma}, \bibinfo{person}{Yousra Aafer}, {and} \bibinfo{person}{Xiangyu Zhang}.} \bibinfo{year}{2019}\natexlab{}.
\newblock \showarticletitle{Abs: Scanning neural networks for back-doors by artificial brain stimulation}. In \bibinfo{booktitle}{\emph{Proceedings of the 2019 ACM SIGSAC Conference on Computer and Communications Security}}. \bibinfo{pages}{1265--1282}.
\newblock


\bibitem[Liu et~al\mbox{.}(2018)]%
        {liu2018trojaning}
\bibfield{author}{\bibinfo{person}{Yingqi Liu}, \bibinfo{person}{Shiqing Ma}, \bibinfo{person}{Yousra Aafer}, \bibinfo{person}{Wen-Chuan Lee}, \bibinfo{person}{Juan Zhai}, \bibinfo{person}{Weihang Wang}, {and} \bibinfo{person}{Xiangyu Zhang}.} \bibinfo{year}{2018}\natexlab{}.
\newblock \showarticletitle{Trojaning attack on neural networks}. In \bibinfo{booktitle}{\emph{25th Annual Network And Distributed System Security Symposium (NDSS 2018)}}.
\newblock


\bibitem[Ma et~al\mbox{.}(2024)]%
        {ma2024vere}
\bibfield{author}{\bibinfo{person}{Jianan Ma}, \bibinfo{person}{Pengfei Yang}, \bibinfo{person}{Jingyi Wang}, \bibinfo{person}{Youcheng Sun}, \bibinfo{person}{Cheng-Chao Huang}, {and} \bibinfo{person}{Zhen Wang}.} \bibinfo{year}{2024}\natexlab{}.
\newblock \showarticletitle{VeRe: Verification Guided Synthesis for Repairing Deep Neural Networks}. In \bibinfo{booktitle}{\emph{Proceedings of the 46th IEEE/ACM International Conference on Software Engineering}}. \bibinfo{pages}{1--13}.
\newblock


\bibitem[Manfredi and Jestin(2016)]%
        {manfredi2016introduction}
\bibfield{author}{\bibinfo{person}{Guido Manfredi} {and} \bibinfo{person}{Yannick Jestin}.} \bibinfo{year}{2016}\natexlab{}.
\newblock \showarticletitle{An introduction to ACAS Xu and the challenges ahead}. In \bibinfo{booktitle}{\emph{2016 IEEE/AIAA 35th Digital Avionics Systems Conference (DASC)}}. IEEE, \bibinfo{pages}{1--9}.
\newblock


\bibitem[Marston and Baca(2015)]%
        {marston2015acas}
\bibfield{author}{\bibinfo{person}{Mike Marston} {and} \bibinfo{person}{Gabe Baca}.} \bibinfo{year}{2015}\natexlab{}.
\newblock \bibinfo{booktitle}{\emph{ACAS-Xu initial self-separation flight tests}}.
\newblock \bibinfo{type}{{T}echnical {R}eport}.
\newblock


\bibitem[Matoba and Fleuret(2020)]%
        {matoba2020exact}
\bibfield{author}{\bibinfo{person}{Kyle Matoba} {and} \bibinfo{person}{Fran{\c{c}}ois Fleuret}.} \bibinfo{year}{2020}\natexlab{}.
\newblock \showarticletitle{Exact preimages of neural network aircraft collision avoidance systems}. In \bibinfo{booktitle}{\emph{Proceedings of the Machine Learning for Engineering Modeling, Simulation, and Design Workshop at Neural Information Processing Systems}}. \bibinfo{pages}{1--9}.
\newblock


\bibitem[Mo et~al\mbox{.}(2024)]%
        {mo2024robust}
\bibfield{author}{\bibinfo{person}{Xiaoxing Mo}, \bibinfo{person}{Yechao Zhang}, \bibinfo{person}{Leo~Yu Zhang}, \bibinfo{person}{Wei Luo}, \bibinfo{person}{Nan Sun}, \bibinfo{person}{Shengshan Hu}, \bibinfo{person}{Shang Gao}, {and} \bibinfo{person}{Yang Xiang}.} \bibinfo{year}{2024}\natexlab{}.
\newblock \showarticletitle{Robust backdoor detection for deep learning via topological evolution dynamics}. In \bibinfo{booktitle}{\emph{2024 IEEE Symposium on Security and Privacy (SP)}}. IEEE Computer Society, \bibinfo{pages}{171--171}.
\newblock


\bibitem[Mu and Gilmer(2019)]%
        {mu2019mnist}
\bibfield{author}{\bibinfo{person}{Norman Mu} {and} \bibinfo{person}{Justin Gilmer}.} \bibinfo{year}{2019}\natexlab{}.
\newblock \showarticletitle{Mnist-c: A robustness benchmark for computer vision}.
\newblock \bibinfo{journal}{\emph{arXiv preprint arXiv:1906.02337}} (\bibinfo{year}{2019}).
\newblock


\bibitem[Narodytska et~al\mbox{.}(2018)]%
        {narodytska2018verifying}
\bibfield{author}{\bibinfo{person}{Nina Narodytska}, \bibinfo{person}{Shiva Kasiviswanathan}, \bibinfo{person}{Leonid Ryzhyk}, \bibinfo{person}{Mooly Sagiv}, {and} \bibinfo{person}{Toby Walsh}.} \bibinfo{year}{2018}\natexlab{}.
\newblock \showarticletitle{Verifying properties of binarized deep neural networks}. In \bibinfo{booktitle}{\emph{Proceedings of the AAAI Conference on Artificial Intelligence}}, Vol.~\bibinfo{volume}{32}.
\newblock


\bibitem[Netzer et~al\mbox{.}(2011)]%
        {netzer2011reading}
\bibfield{author}{\bibinfo{person}{Yuval Netzer}, \bibinfo{person}{Tao Wang}, \bibinfo{person}{Adam Coates}, \bibinfo{person}{Alessandro Bissacco}, \bibinfo{person}{Bo Wu}, {and} \bibinfo{person}{Andrew~Y Ng}.} \bibinfo{year}{2011}\natexlab{}.
\newblock \showarticletitle{Reading digits in natural images with unsupervised feature learning}.
\newblock  (\bibinfo{year}{2011}).
\newblock


\bibitem[Simonyan and Zisserman(2014)]%
        {vgg}
\bibfield{author}{\bibinfo{person}{Karen Simonyan} {and} \bibinfo{person}{Andrew Zisserman}.} \bibinfo{year}{2014}\natexlab{}.
\newblock \showarticletitle{Very deep convolutional networks for large-scale image recognition}.
\newblock \bibinfo{journal}{\emph{arXiv preprint arXiv:1409.1556}} (\bibinfo{year}{2014}).
\newblock


\bibitem[Singh(2019)]%
        {erangithub}
\bibfield{author}{\bibinfo{person}{Gagandeep Singh}.} \bibinfo{year}{2019}\natexlab{}.
\newblock \bibinfo{title}{ETH Robustness Analyzer for Neural Networks (ERAN)}.
\newblock
\urldef\tempurl%
\url{https://github.com/eth-sri/eran}
\showURL{%
\tempurl}


\bibitem[Singh et~al\mbox{.}(2018)]%
        {singh2018fast}
\bibfield{author}{\bibinfo{person}{Gagandeep Singh}, \bibinfo{person}{Timon Gehr}, \bibinfo{person}{Matthew Mirman}, \bibinfo{person}{Markus P{\"u}schel}, {and} \bibinfo{person}{Martin Vechev}.} \bibinfo{year}{2018}\natexlab{}.
\newblock \showarticletitle{Fast and effective robustness certification}.
\newblock \bibinfo{journal}{\emph{Advances in neural information processing systems}}  \bibinfo{volume}{31} (\bibinfo{year}{2018}).
\newblock


\bibitem[Singh et~al\mbox{.}(2019)]%
        {singh2019abstract}
\bibfield{author}{\bibinfo{person}{Gagandeep Singh}, \bibinfo{person}{Timon Gehr}, \bibinfo{person}{Markus P{\"u}schel}, {and} \bibinfo{person}{Martin Vechev}.} \bibinfo{year}{2019}\natexlab{}.
\newblock \showarticletitle{An abstract domain for certifying neural networks}.
\newblock \bibinfo{journal}{\emph{Proceedings of the ACM on Programming Languages}} \bibinfo{volume}{3}, \bibinfo{number}{POPL} (\bibinfo{year}{2019}), \bibinfo{pages}{1--30}.
\newblock


\bibitem[Sohn et~al\mbox{.}(2023)]%
        {sohn2023arachne}
\bibfield{author}{\bibinfo{person}{Jeongju Sohn}, \bibinfo{person}{Sungmin Kang}, {and} \bibinfo{person}{Shin Yoo}.} \bibinfo{year}{2023}\natexlab{}.
\newblock \showarticletitle{Arachne: Search-based repair of deep neural networks}.
\newblock \bibinfo{journal}{\emph{ACM Transactions on Software Engineering and Methodology}} \bibinfo{volume}{32}, \bibinfo{number}{4} (\bibinfo{year}{2023}), \bibinfo{pages}{1--26}.
\newblock


\bibitem[Sotoudeh and Thakur(2021)]%
        {sotoudeh2021provable}
\bibfield{author}{\bibinfo{person}{Matthew Sotoudeh} {and} \bibinfo{person}{Aditya~V Thakur}.} \bibinfo{year}{2021}\natexlab{}.
\newblock \showarticletitle{Provable repair of deep neural networks}. In \bibinfo{booktitle}{\emph{Proceedings of the 42nd ACM SIGPLAN International Conference on Programming Language Design and Implementation}}. \bibinfo{pages}{588--603}.
\newblock


\bibitem[Stallkamp et~al\mbox{.}(2012)]%
        {stallkamp2012man}
\bibfield{author}{\bibinfo{person}{Johannes Stallkamp}, \bibinfo{person}{Marc Schlipsing}, \bibinfo{person}{Jan Salmen}, {and} \bibinfo{person}{Christian Igel}.} \bibinfo{year}{2012}\natexlab{}.
\newblock \showarticletitle{Man vs. computer: Benchmarking machine learning algorithms for traffic sign recognition}.
\newblock \bibinfo{journal}{\emph{Neural networks}}  \bibinfo{volume}{32} (\bibinfo{year}{2012}), \bibinfo{pages}{323--332}.
\newblock


\bibitem[Sun et~al\mbox{.}(2022)]%
        {sun2022causality}
\bibfield{author}{\bibinfo{person}{Bing Sun}, \bibinfo{person}{Jun Sun}, \bibinfo{person}{Long~H Pham}, {and} \bibinfo{person}{Jie Shi}.} \bibinfo{year}{2022}\natexlab{}.
\newblock \showarticletitle{Causality-based neural network repair}. In \bibinfo{booktitle}{\emph{Proceedings of the 44th International Conference on Software Engineering}}. \bibinfo{pages}{338--349}.
\newblock


\bibitem[Tao et~al\mbox{.}(2023)]%
        {tao2023architecture}
\bibfield{author}{\bibinfo{person}{Zhe Tao}, \bibinfo{person}{Stephanie Nawas}, \bibinfo{person}{Jacqueline Mitchell}, {and} \bibinfo{person}{Aditya~V Thakur}.} \bibinfo{year}{2023}\natexlab{}.
\newblock \showarticletitle{Architecture-preserving provable repair of deep neural networks}.
\newblock \bibinfo{journal}{\emph{Proceedings of the ACM on Programming Languages}} \bibinfo{volume}{7}, \bibinfo{number}{PLDI} (\bibinfo{year}{2023}), \bibinfo{pages}{443--467}.
\newblock


\bibitem[Tramer and Boneh(2019)]%
        {tramer2019adversarial}
\bibfield{author}{\bibinfo{person}{Florian Tramer} {and} \bibinfo{person}{Dan Boneh}.} \bibinfo{year}{2019}\natexlab{}.
\newblock \showarticletitle{Adversarial training and robustness for multiple perturbations}.
\newblock \bibinfo{journal}{\emph{Advances in neural information processing systems}}  \bibinfo{volume}{32} (\bibinfo{year}{2019}).
\newblock


\bibitem[Usman et~al\mbox{.}(2021)]%
        {usman2021nn}
\bibfield{author}{\bibinfo{person}{Muhammad Usman}, \bibinfo{person}{Divya Gopinath}, \bibinfo{person}{Youcheng Sun}, \bibinfo{person}{Yannic Noller}, {and} \bibinfo{person}{Corina~S P{\u{a}}s{\u{a}}reanu}.} \bibinfo{year}{2021}\natexlab{}.
\newblock \showarticletitle{NN repair: constraint-based repair of neural network classifiers}. In \bibinfo{booktitle}{\emph{Computer Aided Verification: 33rd International Conference, CAV 2021, Virtual Event, July 20--23, 2021, Proceedings, Part I 33}}. Springer, \bibinfo{pages}{3--25}.
\newblock


\bibitem[Wan et~al\mbox{.}(2023)]%
        {wan2023bounceattack}
\bibfield{author}{\bibinfo{person}{Jie Wan}, \bibinfo{person}{Jianhao Fu}, \bibinfo{person}{Lijin Wang}, {and} \bibinfo{person}{Ziqi Yang}.} \bibinfo{year}{2023}\natexlab{}.
\newblock \showarticletitle{Bounceattack: A query-efficient decision-based adversarial attack by bouncing into the wild}. In \bibinfo{booktitle}{\emph{2024 IEEE Symposium on Security and Privacy (SP)}}. IEEE Computer Society, \bibinfo{pages}{68--68}.
\newblock


\bibitem[Wang et~al\mbox{.}(2019)]%
        {wang2019neural}
\bibfield{author}{\bibinfo{person}{Bolun Wang}, \bibinfo{person}{Yuanshun Yao}, \bibinfo{person}{Shawn Shan}, \bibinfo{person}{Huiying Li}, \bibinfo{person}{Bimal Viswanath}, \bibinfo{person}{Haitao Zheng}, {and} \bibinfo{person}{Ben~Y Zhao}.} \bibinfo{year}{2019}\natexlab{}.
\newblock \showarticletitle{Neural cleanse: Identifying and mitigating backdoor attacks in neural networks}. In \bibinfo{booktitle}{\emph{2019 IEEE symposium on security and privacy (SP)}}. IEEE, \bibinfo{pages}{707--723}.
\newblock


\bibitem[Wang et~al\mbox{.}(2024)]%
        {wang2024mm}
\bibfield{author}{\bibinfo{person}{Hang Wang}, \bibinfo{person}{Zhen Xiang}, \bibinfo{person}{David~J Miller}, {and} \bibinfo{person}{George Kesidis}.} \bibinfo{year}{2024}\natexlab{}.
\newblock \showarticletitle{Mm-bd: Post-training detection of backdoor attacks with arbitrary backdoor pattern types using a maximum margin statistic}. In \bibinfo{booktitle}{\emph{2024 IEEE Symposium on Security and Privacy (SP)}}. IEEE, \bibinfo{pages}{1994--2012}.
\newblock


\bibitem[Wang et~al\mbox{.}(2021a)]%
        {wang2021augmax}
\bibfield{author}{\bibinfo{person}{Haotao Wang}, \bibinfo{person}{Chaowei Xiao}, \bibinfo{person}{Jean Kossaifi}, \bibinfo{person}{Zhiding Yu}, \bibinfo{person}{Anima Anandkumar}, {and} \bibinfo{person}{Zhangyang Wang}.} \bibinfo{year}{2021}\natexlab{a}.
\newblock \showarticletitle{Augmax: Adversarial composition of random augmentations for robust training}.
\newblock \bibinfo{journal}{\emph{Advances in neural information processing systems}}  \bibinfo{volume}{34} (\bibinfo{year}{2021}), \bibinfo{pages}{237--250}.
\newblock


\bibitem[Wang et~al\mbox{.}(2021b)]%
        {wang2021beta}
\bibfield{author}{\bibinfo{person}{Shiqi Wang}, \bibinfo{person}{Huan Zhang}, \bibinfo{person}{Kaidi Xu}, \bibinfo{person}{Xue Lin}, \bibinfo{person}{Suman Jana}, \bibinfo{person}{Cho-Jui Hsieh}, {and} \bibinfo{person}{J~Zico Kolter}.} \bibinfo{year}{2021}\natexlab{b}.
\newblock \showarticletitle{Beta-crown: Efficient bound propagation with per-neuron split constraints for neural network robustness verification}.
\newblock \bibinfo{journal}{\emph{Advances in Neural Information Processing Systems}}  \bibinfo{volume}{34} (\bibinfo{year}{2021}), \bibinfo{pages}{29909--29921}.
\newblock


\bibitem[Weng et~al\mbox{.}(2018)]%
        {fastlin}
\bibfield{author}{\bibinfo{person}{Lily Weng}, \bibinfo{person}{Huan Zhang}, \bibinfo{person}{Hongge Chen}, \bibinfo{person}{Zhao Song}, \bibinfo{person}{Cho-Jui Hsieh}, \bibinfo{person}{Luca Daniel}, \bibinfo{person}{Duane Boning}, {and} \bibinfo{person}{Inderjit Dhillon}.} \bibinfo{year}{2018}\natexlab{}.
\newblock \showarticletitle{Towards fast computation of certified robustness for relu networks}. In \bibinfo{booktitle}{\emph{International Conference on Machine Learning}}. PMLR, \bibinfo{pages}{5276--5285}.
\newblock


\bibitem[Xiang et~al\mbox{.}(2021)]%
        {xiang2021patchguard}
\bibfield{author}{\bibinfo{person}{Chong Xiang}, \bibinfo{person}{Arjun~Nitin Bhagoji}, \bibinfo{person}{Vikash Sehwag}, {and} \bibinfo{person}{Prateek Mittal}.} \bibinfo{year}{2021}\natexlab{}.
\newblock \showarticletitle{$\{$PatchGuard$\}$: A provably robust defense against adversarial patches via small receptive fields and masking}. In \bibinfo{booktitle}{\emph{30th USENIX Security Symposium (USENIX Security 21)}}. \bibinfo{pages}{2237--2254}.
\newblock


\bibitem[Xie et~al\mbox{.}(2020)]%
        {xie2020adversarial}
\bibfield{author}{\bibinfo{person}{Cihang Xie}, \bibinfo{person}{Mingxing Tan}, \bibinfo{person}{Boqing Gong}, \bibinfo{person}{Jiang Wang}, \bibinfo{person}{Alan~L Yuille}, {and} \bibinfo{person}{Quoc~V Le}.} \bibinfo{year}{2020}\natexlab{}.
\newblock \showarticletitle{Adversarial examples improve image recognition}. In \bibinfo{booktitle}{\emph{Proceedings of the IEEE/CVF conference on computer vision and pattern recognition}}. \bibinfo{pages}{819--828}.
\newblock


\bibitem[Xu et~al\mbox{.}(2020a)]%
        {xu2020automatic}
\bibfield{author}{\bibinfo{person}{Kaidi Xu}, \bibinfo{person}{Zhouxing Shi}, \bibinfo{person}{Huan Zhang}, \bibinfo{person}{Yihan Wang}, \bibinfo{person}{Kai-Wei Chang}, \bibinfo{person}{Minlie Huang}, \bibinfo{person}{Bhavya Kailkhura}, \bibinfo{person}{Xue Lin}, {and} \bibinfo{person}{Cho-Jui Hsieh}.} \bibinfo{year}{2020}\natexlab{a}.
\newblock \showarticletitle{Automatic perturbation analysis for scalable certified robustness and beyond}.
\newblock \bibinfo{journal}{\emph{Advances in Neural Information Processing Systems}}  \bibinfo{volume}{33} (\bibinfo{year}{2020}), \bibinfo{pages}{1129--1141}.
\newblock


\bibitem[Xu et~al\mbox{.}(2020b)]%
        {xu2020fast}
\bibfield{author}{\bibinfo{person}{Kaidi Xu}, \bibinfo{person}{Huan Zhang}, \bibinfo{person}{Shiqi Wang}, \bibinfo{person}{Yihan Wang}, \bibinfo{person}{Suman Jana}, \bibinfo{person}{Xue Lin}, {and} \bibinfo{person}{Cho-Jui Hsieh}.} \bibinfo{year}{2020}\natexlab{b}.
\newblock \showarticletitle{Fast and complete: Enabling complete neural network verification with rapid and massively parallel incomplete verifiers}.
\newblock \bibinfo{journal}{\emph{arXiv preprint arXiv:2011.13824}} (\bibinfo{year}{2020}).
\newblock


\bibitem[Xu et~al\mbox{.}(2021)]%
        {xu2021detecting}
\bibfield{author}{\bibinfo{person}{Xiaojun Xu}, \bibinfo{person}{Qi Wang}, \bibinfo{person}{Huichen Li}, \bibinfo{person}{Nikita Borisov}, \bibinfo{person}{Carl~A Gunter}, {and} \bibinfo{person}{Bo Li}.} \bibinfo{year}{2021}\natexlab{}.
\newblock \showarticletitle{Detecting ai trojans using meta neural analysis}. In \bibinfo{booktitle}{\emph{2021 IEEE Symposium on Security and Privacy (SP)}}. \bibinfo{pages}{103--120}.
\newblock


\bibitem[Yang et~al\mbox{.}(2021)]%
        {deepsrgrextended}
\bibfield{author}{\bibinfo{person}{Pengfei Yang}, \bibinfo{person}{Renjue Li}, \bibinfo{person}{Jianlin Li}, \bibinfo{person}{Cheng{-}Chao Huang}, \bibinfo{person}{Jingyi Wang}, \bibinfo{person}{Jun Sun}, \bibinfo{person}{Bai Xue}, {and} \bibinfo{person}{Lijun Zhang}.} \bibinfo{year}{2021}\natexlab{}.
\newblock \showarticletitle{Improving Neural Network Verification through Spurious Region Guided Refinement}. In \bibinfo{booktitle}{\emph{{TACAS} 2021}} \emph{(\bibinfo{series}{Lecture Notes in Computer Science}, Vol.~\bibinfo{volume}{12651})}. \bibinfo{publisher}{Springer}, \bibinfo{pages}{389--408}.
\newblock


\bibitem[Zhang et~al\mbox{.}(2024a)]%
        {zhang2024exploring}
\bibfield{author}{\bibinfo{person}{Kaiyuan Zhang}, \bibinfo{person}{Siyuan Cheng}, \bibinfo{person}{Guangyu Shen}, \bibinfo{person}{Guanhong Tao}, \bibinfo{person}{Shengwei An}, \bibinfo{person}{Anuran Makur}, \bibinfo{person}{Shiqing Ma}, {and} \bibinfo{person}{Xiangyu Zhang}.} \bibinfo{year}{2024}\natexlab{a}.
\newblock \showarticletitle{Exploring the Orthogonality and Linearity of Backdoor Attacks}. In \bibinfo{booktitle}{\emph{2024 IEEE Symposium on Security and Privacy (SP)}}. IEEE Computer Society, \bibinfo{pages}{225--225}.
\newblock


\bibitem[Zhang et~al\mbox{.}(2024b)]%
        {zhang2024text}
\bibfield{author}{\bibinfo{person}{Xinyu Zhang}, \bibinfo{person}{Hanbin Hong}, \bibinfo{person}{Yuan Hong}, \bibinfo{person}{Peng Huang}, \bibinfo{person}{Binghui Wang}, \bibinfo{person}{Zhongjie Ba}, {and} \bibinfo{person}{Kui Ren}.} \bibinfo{year}{2024}\natexlab{b}.
\newblock \showarticletitle{Text-crs: A generalized certified robustness framework against textual adversarial attacks}. In \bibinfo{booktitle}{\emph{2024 IEEE Symposium on Security and Privacy (SP)}}. IEEE, \bibinfo{pages}{2920--2938}.
\newblock


\bibitem[Zhang et~al\mbox{.}(2024c)]%
        {zhang2024provable}
\bibfield{author}{\bibinfo{person}{Xiyue Zhang}, \bibinfo{person}{Benjie Wang}, {and} \bibinfo{person}{Marta Kwiatkowska}.} \bibinfo{year}{2024}\natexlab{c}.
\newblock \showarticletitle{Provable preimage under-approximation for neural networks}. In \bibinfo{booktitle}{\emph{International Conference on Tools and Algorithms for the Construction and Analysis of Systems}}. Springer, \bibinfo{pages}{3--23}.
\newblock


\bibitem[Zhao et~al\mbox{.}(2021)]%
        {zhao2021ai}
\bibfield{author}{\bibinfo{person}{Yue Zhao}, \bibinfo{person}{Hong Zhu}, \bibinfo{person}{Kai Chen}, {and} \bibinfo{person}{Shengzhi Zhang}.} \bibinfo{year}{2021}\natexlab{}.
\newblock \showarticletitle{Ai-lancet: Locating error-inducing neurons to optimize neural networks}. In \bibinfo{booktitle}{\emph{Proceedings of the 2021 ACM SIGSAC Conference on Computer and Communications Security}}. \bibinfo{pages}{141--158}.
\newblock


\bibitem[Zheng et~al\mbox{.}(2022)]%
        {zheng2022pre}
\bibfield{author}{\bibinfo{person}{Runkai Zheng}, \bibinfo{person}{Rongjun Tang}, \bibinfo{person}{Jianze Li}, {and} \bibinfo{person}{Li Liu}.} \bibinfo{year}{2022}\natexlab{}.
\newblock \showarticletitle{Pre-activation distributions expose backdoor neurons}.
\newblock \bibinfo{journal}{\emph{Advances in Neural Information Processing Systems}}  \bibinfo{volume}{35} (\bibinfo{year}{2022}), \bibinfo{pages}{18667--18680}.
\newblock


\bibitem[Zhu et~al\mbox{.}(2023b)]%
        {zhu2023ai}
\bibfield{author}{\bibinfo{person}{Hong Zhu}, \bibinfo{person}{Shengzhi Zhang}, {and} \bibinfo{person}{Kai Chen}.} \bibinfo{year}{2023}\natexlab{b}.
\newblock \showarticletitle{Ai-guardian: Defeating adversarial attacks using backdoors}. In \bibinfo{booktitle}{\emph{2023 IEEE Symposium on Security and Privacy (SP)}}. IEEE, \bibinfo{pages}{701--718}.
\newblock


\bibitem[Zhu et~al\mbox{.}(2023a)]%
        {zhu2023selective}
\bibfield{author}{\bibinfo{person}{Rui Zhu}, \bibinfo{person}{Di Tang}, \bibinfo{person}{Siyuan Tang}, \bibinfo{person}{XiaoFeng Wang}, {and} \bibinfo{person}{Haixu Tang}.} \bibinfo{year}{2023}\natexlab{a}.
\newblock \showarticletitle{Selective amnesia: On efficient, high-fidelity and blind suppression of backdoor effects in trojaned machine learning models}. In \bibinfo{booktitle}{\emph{2023 IEEE Symposium on Security and Privacy (SP)}}. IEEE, \bibinfo{pages}{1--19}.
\newblock


\end{thebibliography}

\appendix

\section{Proofs of Theorems}

\subsection{Proof of Theorem~\ref{thm:point}}
\label{appendix:proof-1}
\begin{proof}
We first prove the inequality chain in Eq.~\eqref{eq:distance-bound}. Consider two cases:
\begin{enumerate}[left=0pt, label=(\roman*)]
    \item \label{case:inbox} 
    When $\tilde{f}_{\mathrm{e}}(\bm{x}_\phi) \in \mathcal{B}_{\phi}$, the left inequality holds trivially since $\mathcal{L}_{\mathrm{Proj}} = 0$ and $\lVert \bm{h}_{\phi}^* - \Pi_{\mathcal{B}_{\phi}}\tilde{f}_{\mathrm{e}}(\bm{x}_\phi) \rVert_p = \lVert \bm{h}_{\phi}^* - \tilde{f}_{\mathrm{e}}(\bm{x}_\phi) \rVert_p =\mathcal{L}$. 
    The right inequality becomes:
    $\mathcal{L} \leq r\cdot \mathrm{s}^{1/p}$, which holds by $\lVert \tilde{f}_{\mathrm{e}}(\bm{x}_\phi) - \bm{h}^*_\phi \rVert_p \leq \mathrm{s}^{1/p} \lVert \tilde{f}_{\mathrm{e}}(\bm{x}_\phi) - \bm{h}^*_\phi \rVert_\infty \leq \mathrm{s}^{1/p} \cdot r$.
    \item \label{case:outbox} 
    If $\tilde{f}_{\mathrm{e}}(\bm{x}_\phi) \notin \mathcal{B}_{\phi}$, the left inequality follows from the definition of projection operator~\cite{boyd2004convex} and the geometric
    nature of the proxy box: 
    \begin{itemize}[left=1pt]
        \item $\mathcal{L}_{\mathrm{Proj}} = \min_{\bm{h} \in \mathcal{B}_\phi} \lVert \tilde{f}_{\mathrm{e}}(\bm{x}_\phi) - \bm{h} \rVert_p \leq \lVert \tilde{f}_{\mathrm{e}}(\bm{x}_\phi) - \bm{h}^*_\phi \rVert_p = \mathcal{L}$;
        \item Let $\bm{v} = \tilde{f}_{\mathrm{e}}(\bm{x}_\phi) - \bm{h}^*_\phi$ and $\tilde{\bm{v}} = \Pi_{\mathcal{B}_{\phi}}\tilde{f}_{\mathrm{e}}(\bm{x}_\phi) - \bm{h}^*_\phi$. Since the box projection performs coordinate-wise truncation, we have:
        $|\tilde{v}_i| = \min(|v_i|, r) \leq |v_i|$.
        Therefore, for any $p \geq 1$:
        \[\lVert \tilde{\bm{v}} \rVert_p = \left( \sum_{i=1}^{\mathrm{s}} |\tilde{v}_i|^p \right)^{1/p} \leq \left( \sum_{i=1}^{\mathrm{s}} |v_i|^p \right)^{1/p} = \lVert \bm{v} \rVert_p
        \]
        which proves $\lVert \Pi_{\mathcal{B}_{\phi}}\tilde{f}_{\mathrm{e}}(\bm{x}_\phi) - \bm{h}^*_\phi \rVert_p \leq \mathcal{L}$.
    \end{itemize}
    To sum up, we have $\max \left(\lVert \bm{h}_{\phi}^* - \Pi_{\mathcal{B}_{\phi}}\tilde{f}_{\mathrm{e}}(\bm{x}_\phi) \rVert_p \, , \, \mathcal{L}_{\mathrm{Proj}}\right) \leq \mathcal{L} $.
    For the right inequality, by the Minkowski inequality we have 
    $\mathcal{L} \le \mathcal{L}_{\mathrm{Proj}} + \lVert \Pi_{\mathcal{B}_{\phi}}\tilde{f}_{\mathrm{e}}(\bm{x}_\phi) - \bm{h}^*_\phi\rVert_p$ where $\lVert \Pi_{\mathcal{B}_{\phi}}\tilde{f}_{\mathrm{e}}(\bm{x}_\phi) - \bm{h}^*_\phi\rVert_p \le r\cdot \mathrm{s}^{1/p}$ since $\tilde{f}_{\mathrm{e}}(\bm{x}_\phi)$ is on the surface of $\mathcal{B}_{\phi}$.
\end{enumerate}
Combining cases~\ref{case:inbox} and~\ref{case:outbox}, We conclude that Eq.~\eqref{eq:distance-bound} holds.

The repair guarantee follows as $\mathcal{L} \leq r$ implies $\lVert \tilde{f}_{\mathrm{e}}(\bm{x}_\phi) - \bm{h}^*_\phi \rVert_\infty \leq r$ by $\ell_p$-norm monotonicity, placing $\tilde{f}_{\mathrm{e}}(\bm{x}_\phi)$ in $\mathcal{B}_\phi$.
Since $\mathcal{B}_\phi \subset f_{\mathrm{c}}^{-1}(\mathbb{R}^{\mathrm{s}}, \phi^{out})$, we have $f_{\mathrm{c}}(\tilde{f}_{\mathrm{e}}(\bm{x}_\phi)) \in \phi^{out}$.
\end{proof}

\subsection{Proof of Theorem~\ref{thm:region}}
\label{appendix:proof-2}
\begin{proof}
Algorithm~\ref{algorithm:region-wise} returns a repaired model $\tilde{f}$ only at line 26, i.e., only if the unsatisfied constraints set $\mathrm{Vio}(\phi)$ for all properties $\phi \in \mathcal{P}$ is empty.
This condition implies: $lb_\psi \geq 0, \; \forall \phi \in \mathcal{P}, \psi \in \phi^{cons}$.
From the linear bound definition:
\[
lb_\psi = \min_{\bm{x} \in \phi^{in}} \left( \underline{\bm{w}}_\psi^\top \bm{x} + \underline{b}_\psi \right) \leq  \underline{\bm{w}}_\psi^\top \bm{x} + \underline{b}_\psi  \leq\bm{c}_\psi^\top \tilde{f}(\bm{x}) + d_\psi, \;\; \forall \bm{x} \in \phi^{in}
\]
we conclude $\bm{c}_\psi^\top f(\bm{x}) + d_\psi \geq 0$ for all $\bm{x} \in \phi^{in}$, $\psi \in \phi^{cons}$ and $\phi \in \mathcal{P}$.
Finally. we have $\forall \phi \in \mathcal{P}: \;\tilde{f} \models \phi $. \qedhere
\end{proof}

\section{Experimental Details}
\label{appendix:setup}
\noindent \textbf{Platform.} We conducted all the experiments on a machine with Dual AMD EPYC 7763 64-Core Processor with 256 GB of memory and RTX-4090 running Ubuntu 20.04.

\subsection{Details of Repair Tasks}

\subsubsection{Backdoor}
\label{appendix:setup-backdoor}
For this task, we train VGGNet~\cite{vgg} and ResNet-18~\cite{resnet} models on four datasets including CIFAR-10~\cite{cifar}, SVHN~\cite{netzer2011reading}, GTSRB~\cite{stallkamp2012man} and CIFAR-100~\cite{cifar}. 
We consider five common backdoor attacks, i.e., BadNets~\cite{gu2019badnets} using two attack strategies (All to One and All to All), TrojanNN\cite{liu2018trojaning}, Blend~\cite{chen2017targeted} and SIG~\cite{barni2019new}. 
For each dataset, we inject the backdoor trigger into its test set \(\mathcal{D}\) to obtain \textbf{the poisoned set} \(\Tilde{\mathcal{D}}\) and divide it into two disjoint sets: \textbf{the repair set \(\Tilde{\mathcal{D}}_\mathrm{r}\)} consisting of 1\,000 data and \textbf{the generalization set \(\tilde{\mathcal{D}}_\mathrm{g}\)} comprising the remaining data. We randomly select \textbf{a few buggy data} from \textbf{the repair set} to construct our desired property set \(\mathcal{P}=\{\phi_1, \dots, \phi_p\}\), i.e., each property \(\phi\) corresponds to a buggy data \(\bm{x}\), along with the output constraints \(\bigwedge_{j=1}^{n} f(\bm{x})_{cl(\bm{x})} - f(\bm{x})_j \geq 0 \), where \(cl(\bm{x})\) is the correct label of \(\bm{x}\). 
Finally, the repaired network is assessed on datasets \(\mathcal{D}\) and \(\tilde{\mathcal{D}}_\mathrm{g}\) to measure its \textit{Acc} and \textit{Gene}.

\subsubsection{Corruption} 
The MNIST-C dataset includes corrupted images from the original MNIST test set. 
We consider all 15 corruption types, such as \textit{fog}, \textit{brightness}, \textit{stripe}, etc. 
Each corruption set consists of 10,000 images, which we divide into a \textbf{repair set} (1\,000) and a \textbf{generalization set} (9\,000).
From \textbf{the repair set}, we further randomly selected a subset of buggy data to construct \textbf{the desired property set} \(\mathcal{P}\).
We evaluate the repaired model on the MNIST test set and the generalization set to measure its \textit{Acc} and \textit{Gene}.

\subsubsection{Adversarial Attack}
In this task, we first randomly select several buggy data from the original training set and then construct the specification for each buggy data \(\bm{x}\). 
Following APRNN, we perturb the first \(k\) non-zero pixels of the input to construct the desirable property for \(\bm{x}\) as follows:
\begin{equation*}
\begin{split}
\begin{aligned}
   &\phi^{in} =  \{ \bm{x^{\prime}} \mid 
   \forall p \in \mathrm{P}, |  \bm{x^{\prime}}_p-\bm{x}_p|  \leq \mathrm{R}, \, \forall q \in \mathrm{NP}, \bm{x^{\prime}}_q=\bm{x}_q \}\\
   & \phi^{cons} = \{f(\bm{x})_{cl(\bm{x})} - f(\bm{x})_{j} \geq 0 \mid 1 \leq j \leq n \}
\end{aligned}
\end{split}
\end{equation*}
where \(\mathrm{P}\) and \(\mathrm{NP}\) represent the set of indices of the first \(k\) non-zero pixels and the remaining pixels, respectively. \(\mathrm{R}\) is the perturbation radius.
Specifically, the perturbation radii are set to 0.1 for MNIST, and 2/255 for both GTSRB and CIFAR-10.

\subsubsection{Safety Property Violation.}
\label{appendix:setup-safety}
This task aims to correct the buggy model in the ACAS Xu benchmark so that it satisfies the predefined global safety property.
ACAS Xu system~\cite{manfredi2016introduction} is an aircraft collision avoidance system designed for unmanned aircraft. 
The input of DNNs for this system consists of five variables describing the speed ($v_{int}$ and $v_{own}$) and relative position (relative distance $\rho$ and angles $\theta, \psi$) of both the intruder and ownship, and the outputs are the scores of five advisories that can be given to the ownship: Clear-of-Conflict (COC), weak right (WR), strong right (SR), weak left (WL), or strong left (SL).
These models are subject to a set of safety-critical properties~\cite{reluplex}.
Following prior repair frameworks~\cite{sun2022causality, sotoudeh2021provable, liang2025birdnn}, we consider Property-2:
\begin{itemize}[left=0pt]
    \item \textbf{Description:} If the intruder is distant and is significantly slower than the ownship, the score of a COC advisory will be minimal (ACAS Xu system chooses the lowest score as the best action). 
    \item \textbf{Desired Property}: 
    \begin{equation*}
    \begin{aligned}
       &\phi^{in}= \{(\rho, \theta, \psi, v_{own}, v_{int}) \mid \rho \ge 55947.691, v_{own} \ge 1145, v_{int} \le 60\}\\
       & \phi^{cons}= \{f(\cdot)_{COC} \le f(\cdot)_j \mid j \in \{WR, SR, SL, WL\}\}
    \end{aligned}
    \end{equation*}
\end{itemize}

\subsection{Details of Models}
\noindent \textbf{Corruption.} The models used for corruption repair are sourced from ERAN~\cite{erangithub}, a robustness platform for DNNs. As in APRNN, we refer to the networks by the number of layers and the width of hidden layers, e.g., a ``6 × 100'' network has 6 hidden layers with 100 neurons per layer.

\noindent \textbf{Backdoor.} We use SGD as the optimizer to train the backdoored model with learning rate 0.1, momentum 0.9 and batch size 128 for 50 epochs on SVHN and GTSRB, 100 epochs on CIFAR-10 and CIFAR-100.

\noindent \textbf{Adversarial Attack.} The 3x100 model for MNIST is sourced from ERAN~\cite{erangithub}. Other models are trained using SGD with learning rate 0.1, momentum 0.9 and batch size 128 for 20 epochs on GTSRB and 100 epochs on CIFAR-10.

\noindent \textbf{Safety Property Violation.} 
As reported in~\cite{dong2021towards}, over 30 out of the 45 DNNs in the ACAS Xu system violate Property 2. 
We conduct a comprehensive evaluation of all these models.

\begin{table*}[!t]
\caption{Results of adversarial attack repair with various
verifiers. auto-LiRPA is the most precise verifier, followed by
CROWN-Forward, then IBP.}
\label{table:more-verifier}
\centering
\small
\begin{tabular}{c|c|c|c|c|c|c|c|c}
\toprule
Dim$\rightarrow$ & \multirow{2}{*}{20 \#Success}& \multirow{2}{*}{40 \#Success}&\multirow{2}{*}{60 \#Success}&\multirow{2}{*}{80 \#Success}&\multirow{2}{*}{100 \#Success}&\multirow{2}{*}{120 \#Success}&\multirow{2}{*}{140 \#Success}&\multirow{2}{*}{160 \#Success}\\
Verifier$\downarrow$ & &  &   &  &  &  & &  \\
\midrule
IBP  &49 / 50 &46 / 50 &39 / 50 &22 / 50 &11 / 50 &6 / 50 &1 / 50 & 0 / 50 \\
CROWN-Forward &\textbf{50 / 50} &\textbf{50 / 50} &\underline{48 / 50} &\underline{48 / 50} &\textbf{48 / 50} &\underline{45 / 50} &\underline{40 / 50} & \underline{34 / 50} \\
auto-LiRPA       &\textbf{50 / 50} &\textbf{50 / 50} &\textbf{49 / 50} &\textbf{49 / 50} &\textbf{48 / 50} &\textbf{48 / 50} &\textbf{41 / 50} & \textbf{41 / 50} \\

\bottomrule
\end{tabular}
\end{table*}

\subsection{Details of Backdoor Attacks}
\label{appendix:attacks}
We implement various backdoor attack as follows:
\begin{itemize}[left=0pt]
    \item \textbf{Badnets}~\cite{gu2019badnets}: For \textit{All to One attack}, the backdoor trigger is a 5 × 5 white square at the bottom right corner of images. 
    In the \textit{All to All attack}, the target labels for backdoored data with original labels \(y\) are set to \(y_t = (y + 1) \mod n\). Here $\mod$ is short for ``modulus''. The poisoning rate is set to be \(10\%\).
    \item \textbf{Blend}~\cite{chen2017targeted}: We randomly generate a trigger pattern \(\bm{t}\) by sampling pixel values from a uniform distribution in [0, 255]. Then we obtain the backdoored data by attaching the trigger \(\bm{t}\) to the sample \(\bm{x}\) and setting the blended ratio to 0.2, i.e., \(\Tilde{\bm{x}}=0.2 \cdot \bm{t} + 0.8 \cdot \bm{x} \). Given the target class, we randomly poison 10\% of training samples.
    \item \textbf{TrojanNN}~\cite{liu2018trojaning}: Following~\cite{zhang2024exploring}, we use a reversed watermark as the trigger pattern. The threshold for the mask used to attach the trigger is set to 0.3. Given the target label, we attach the trigger to 10\% of training samples for poisoning.
    \item \textbf{SIG}~\cite{barni2019new}: Following~\cite{barni2019new}, we overlay a sinusoidal signal over the inputs as the trigger. Given the target label, we attach the trigger to 80\% of samples from the target class, i.e., the actual poisoning rate is \(\frac{80}{n}\)\% for the entire dataset, where \(n\) is the number of classes.
\end{itemize}

\subsection{Details of Setup in Repair Methods}
\label{appendix:baseline}
Here we briefly introduce the setup of all methods, especially hyper-parameter settings. For fundamental hyper-parameters such as learning rate and number of epochs, we adhere to the settings in the original papers. For other parameters, we provide the following explanations:
\begin{itemize}[left=0pt]
    \item \textbf{lb \& ub}: In PRDNN and APRNN, this parameter restricts the modifications to the model (\(\pm \left\|\Delta \theta\right\|_{\infty}\)). Following APRNN, we set it to \(\pm3\) for corruption repair, \(\pm5\) for global safety property repair, and \(\pm10\) for both adversarial attack and backdoor repairs.
    \item \textbf{Rows}: In PRDNN and APRNN, this parameter is designed for large networks (such as VGG and ResNet), as repairing all parameters in a layer requires an unacceptable amount of memory. We set it according to the original setting to 400-800 for VGG and 0-200 for ResNet, where 400-800 indicates the number of rows in the parameter matrix to be repaired for one layer.
    \item \textbf{Layers}: In PRDNN, APRNN and our method, this parameter involves how to select the layers for repair (i.e., how to partition the network $f$ into \(f_{\mathrm{e}}\) and \(f_{\mathrm{c}}\)). For PRDNN and APRNN, we try our best to select the best-performing layer. For our method, we set parameters uniformly based on the model size. For small networks, we include one activation layer in \(f_{\mathrm{c}}\), while for larger networks, we include two activation layers.
\end{itemize}

Additionally, we provide further details on the configuration of SEAM~\cite{zhu2023selective}, as it is the latest state-of-the-art method for backdoor removal based on a certain amount of clean samples. It first utilizes a small set of clean samples \(\mathcal{D}_u\) (with wrong labels) for unlearning, and then uses a relatively larger set of clean samples \(\mathcal{D}_r\) (labeled correctly) for recovery. In our scenarios, we made additional modifications to SEAM to enhance its performance.
\begin{itemize}[left=0pt]
    \item \textbf{Unlearning}: We provide the buggy data that used to construct property set \(\mathcal{P}\) for SEAM to facilitate the unlearning process. Compared to using clean samples with incorrect labels, this operation allows for a more effective unlearning of the backdoor.
    \item \textbf{Recover}: During the recovery phase, we provide SEAM with additional clean samples, enabling it to utilize both the buggy data set (used in unlearning step) and the extra clean samples for recovery.
\end{itemize}


\begin{figure*}[t]
\includegraphics[width=1.0\textwidth]{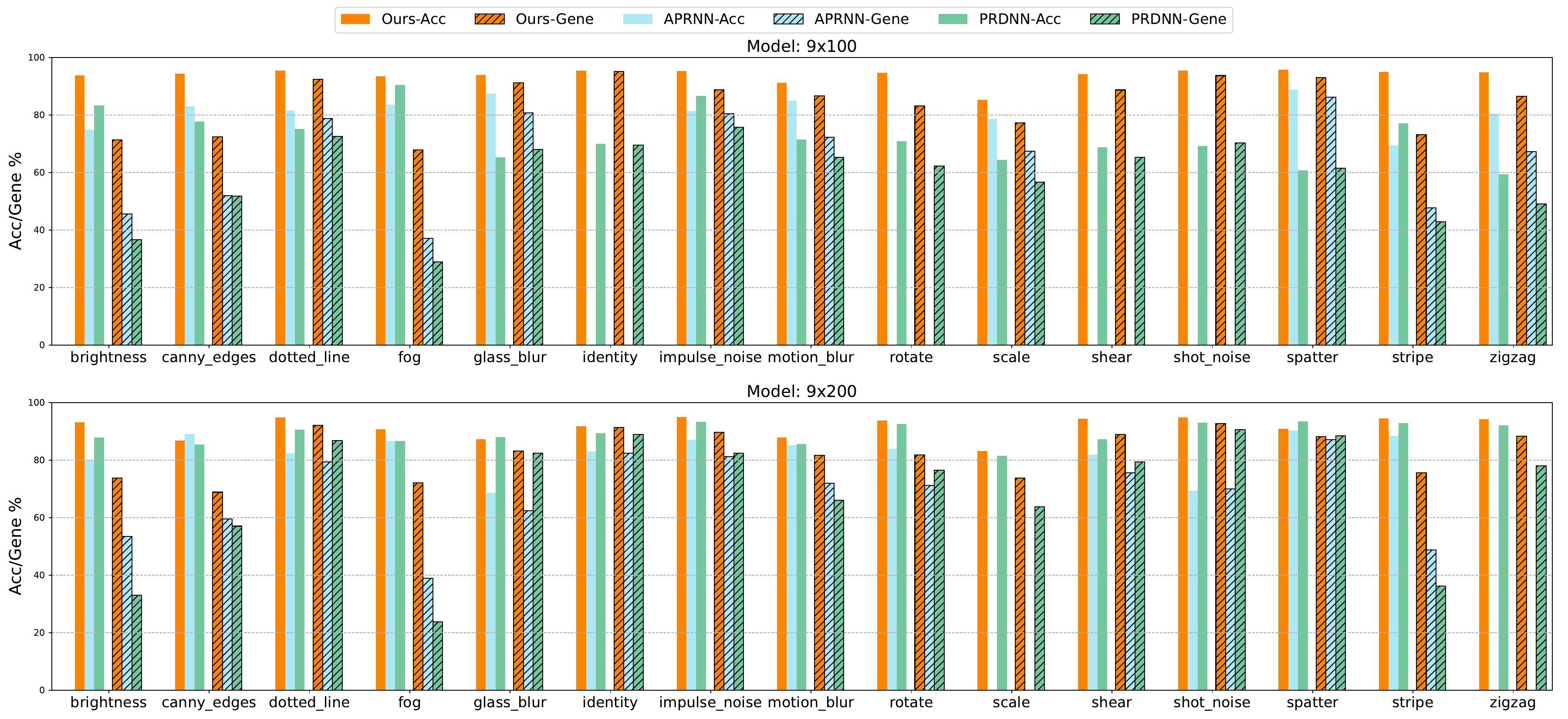}
\centering
\caption{Results of point-wise corruption repair on the 9×100 and 9x200 models under different corruptions. We omit the bar if the corresponding repair is infeasible (e.g., APRNN on the 9×200 model under the scale corruption).}
\label{fig:corr-9100200}
\end{figure*}

\begin{figure*}[t]
\includegraphics[width=1.0\textwidth]{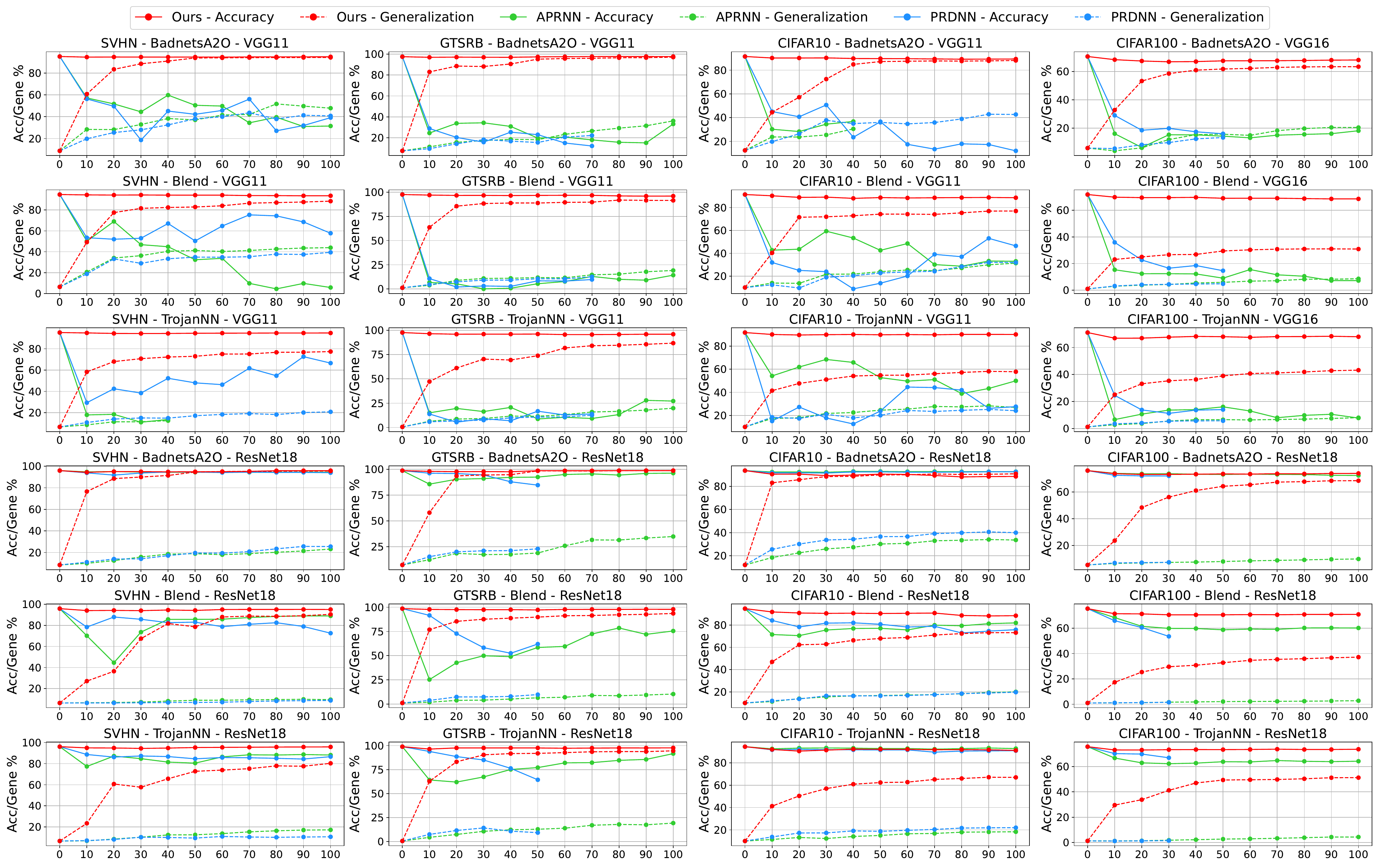}
\centering
\caption{Full results of backdoor repair with different numbers of desired properties (x-axis).}
\label{fig:back_num}
\end{figure*}

\begin{figure*}[t]
    \centering
    \begin{subfigure}[b]{\textwidth}
        \centering
        \captionsetup{skip=3pt}
        \includegraphics[width=\textwidth]{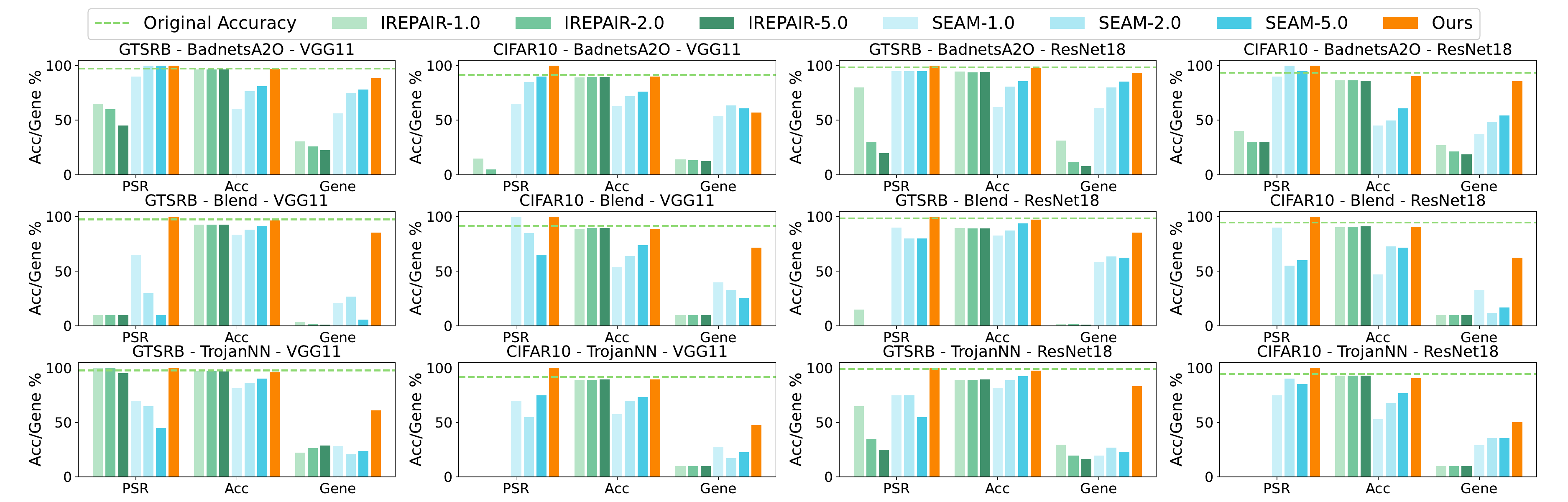} 
        \caption{\(\left |  \mathcal{P}\right | = 20\)}
    \end{subfigure}

    \begin{subfigure}[b]{\textwidth}
        \centering
        \captionsetup{skip=3pt}
        \includegraphics[width=\textwidth]{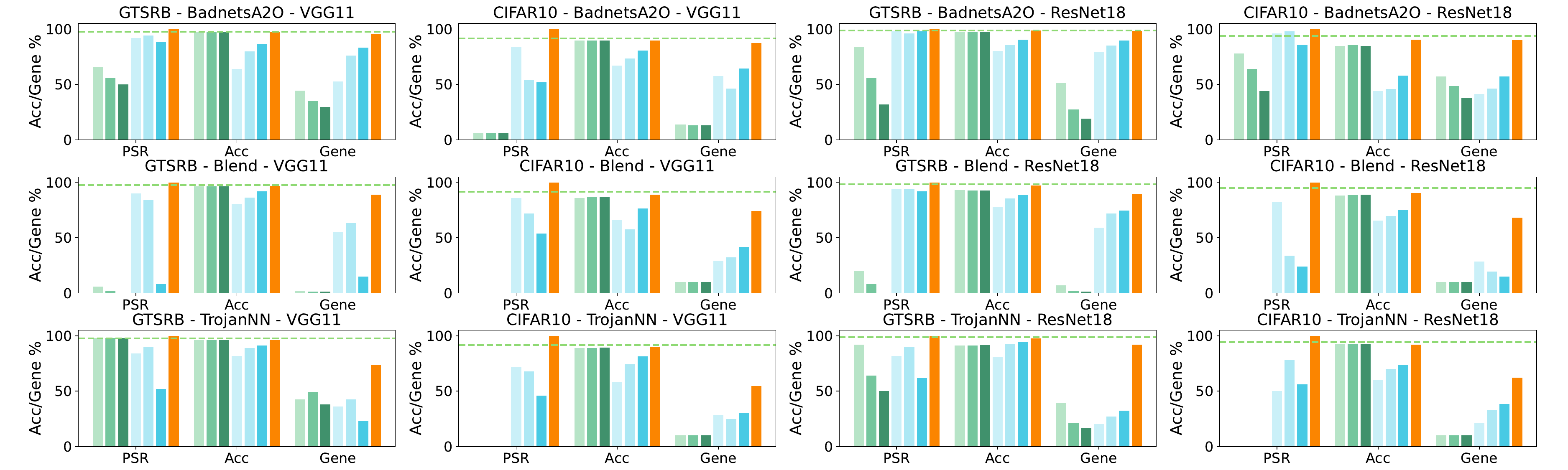} 
        \caption{\(\left |  \mathcal{P}\right | = 50\)}
    \end{subfigure}

    \begin{subfigure}[b]{\textwidth}
        \centering
        \captionsetup{skip=3pt}
        \includegraphics[width=\textwidth]{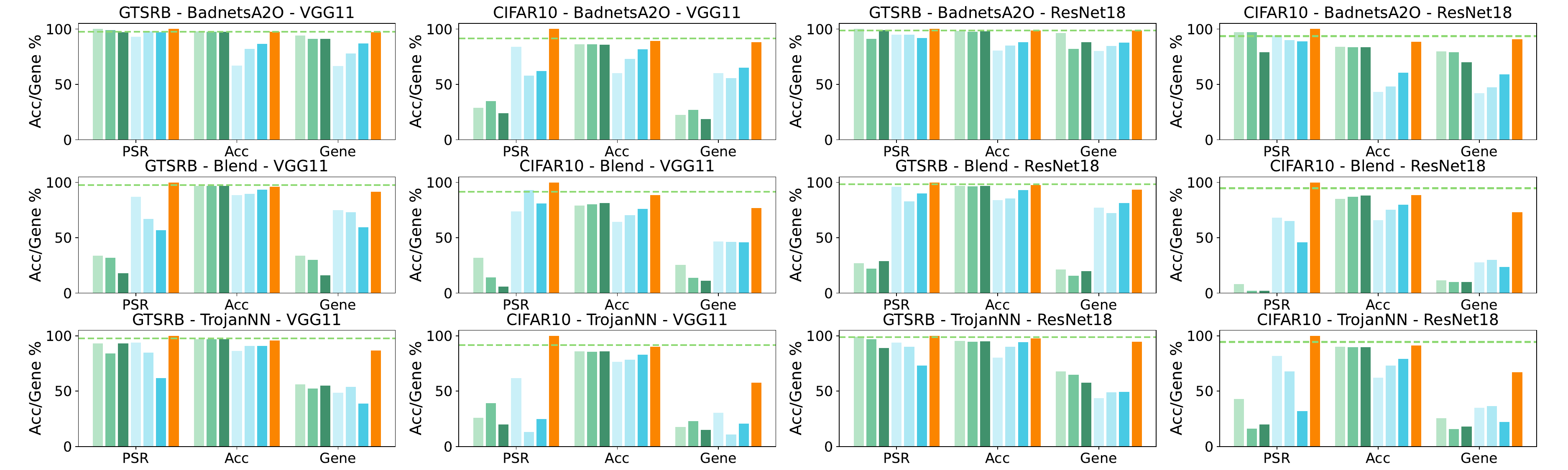} 
        \caption{\(\left |  \mathcal{P}\right | = 100\)}
    \end{subfigure}

    \caption{Comparison of \name with SEAM-X and IREPAIR-X with different number of  properties. X denotes the percentage of additional data accessible, e.g., SEAM-1.0 indicates that 500 additional clean data is available for CIFAR-10. }
    \label{fig:back-non-num}
\end{figure*}

\begin{figure*}[t]
\includegraphics[width=1.0\textwidth]{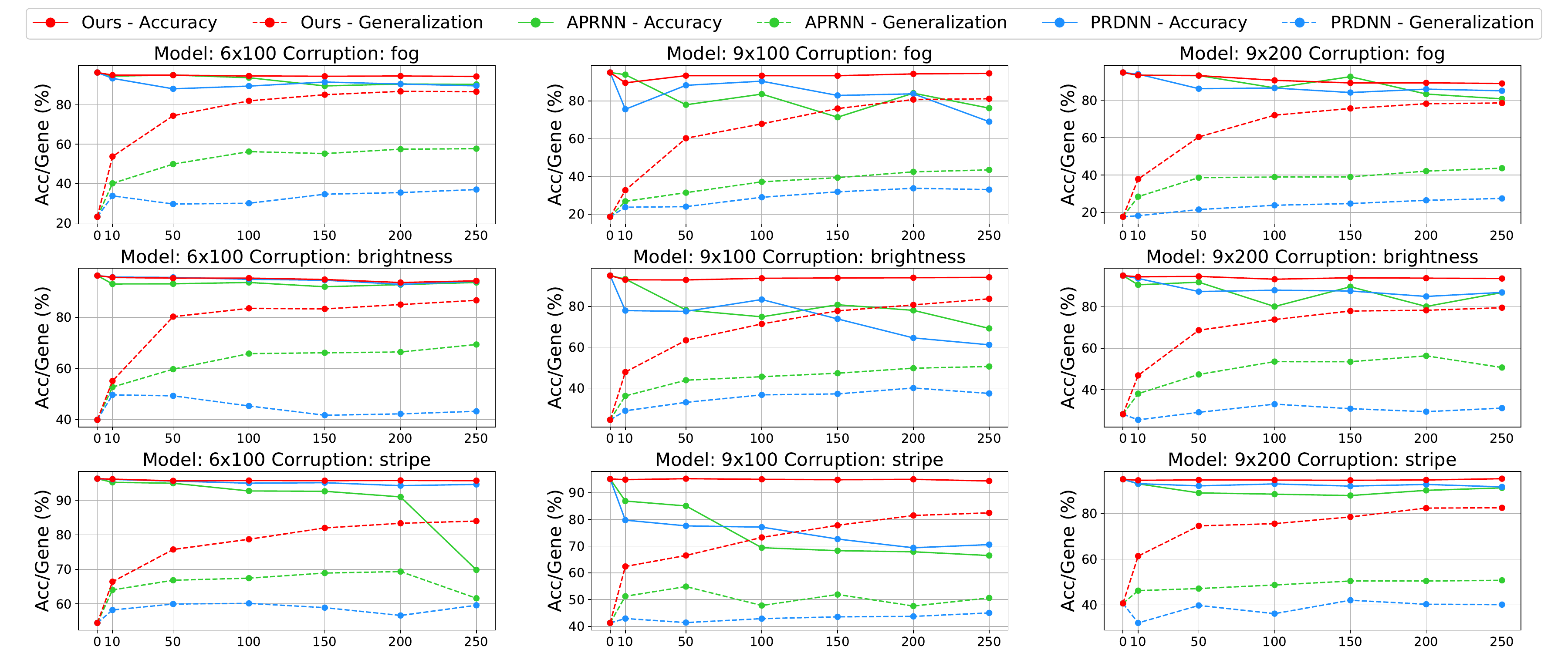}
\centering
\caption{Results of corruption repair with different numbers of desirable properties (x-axis).}
\label{fig:corr_num}
\end{figure*}

\begin{figure*}[t]
\includegraphics[width=1.0
\textwidth]{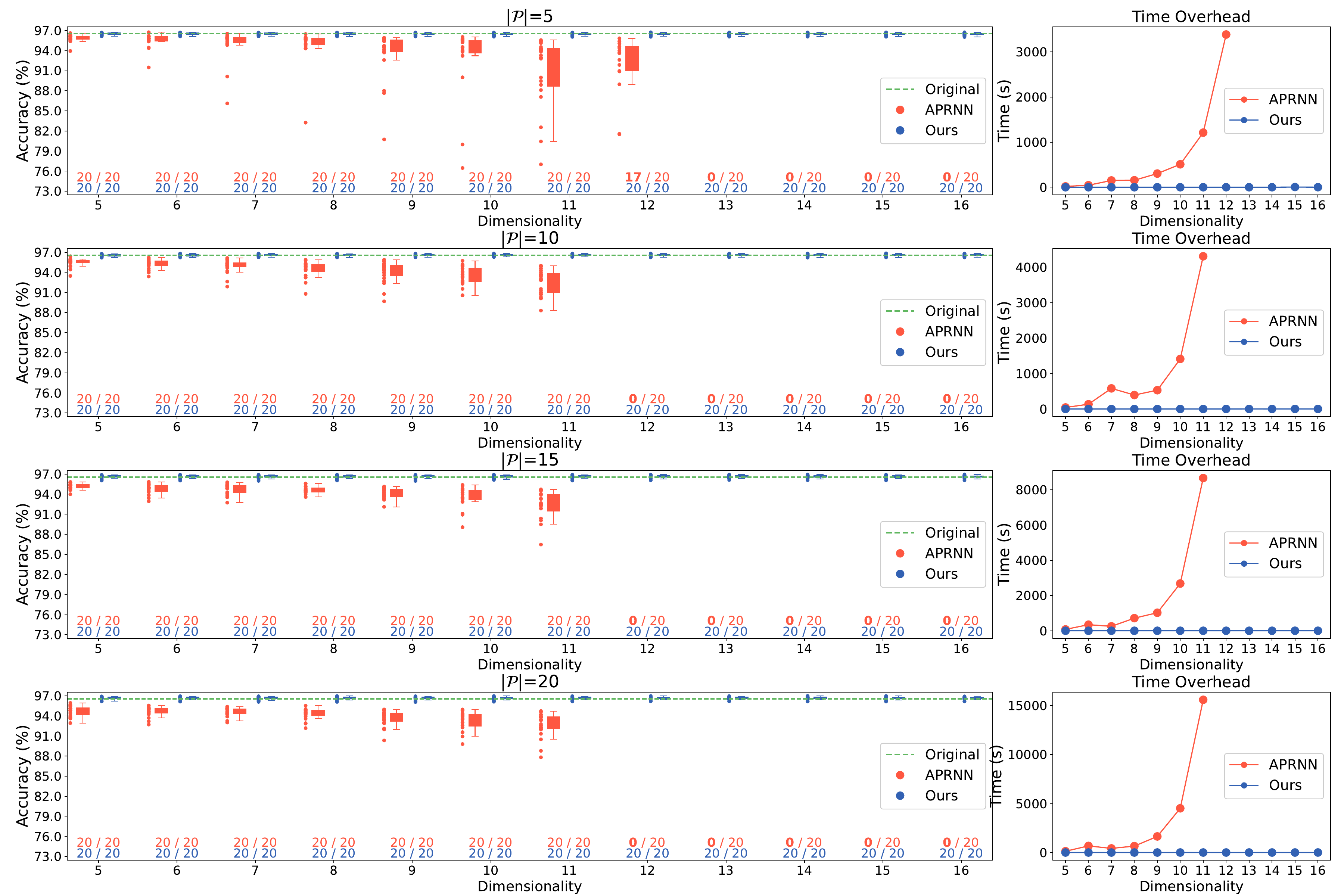}
\centering
\caption{Results of adversarial attack repair with multiple properties on MNIST dataset.}
\label{fig:robustness-num-M}
\end{figure*}

\end{document}